\documentclass[aps,prd,twocolumn,nofootinbib,preprintnumbers]{revtex4-2}
\usepackage{hyperref}

\usepackage[utf8]{inputenc}
\usepackage{comment}
\usepackage{physics}
\usepackage{amsfonts}
\usepackage{graphicx}
\usepackage{dcolumn}
\usepackage{bm}
\usepackage{multibib}
\usepackage{placeins}
\usepackage{xcolor}
\usepackage{orcidlink}
\usepackage{siunitx} 
\usepackage{cleveref} 
\crefname{figure}{Figure}{Figures}
\Crefname{figure}{Figure}{Figures}
\crefname{table}{Table}{Tables}
\Crefname{table}{Table}{Tables}
\crefname{section}{Section}{Sections} 
\Crefname{section}{Section}{Sections} 
\crefname{appendix}{Appendix}{Appendixes}
\Crefname{appendix}{Appendix}{Appendixes}
\crefname{equation}{Eq.}{Eqs.}
\Crefname{equation}{Eq.}{Eqs.}

\usepackage{booktabs}

\AddToHook{cmd/appendix/before}{\crefalias{section}{appendix}}

\newcommand{\myquote}[1]{``#1''}

\newcommand{\myindent}{\hspace*{2em}}
\newcommand{\rhohat}{\hat{\rho}^\mathrm{O(3)}}
\newcommand{\omegaone}{  \Omega_\mathrm{GP}^{(1)}}
\newcommand{\omegatwo}{  \Omega_\mathrm{GP}^{(2)}}
\newcommand{\omegathree}{\Omega_\mathrm{GP}^{(3)}}
\newcommand{\omegafour}{ \Omega_\mathrm{GP}^{(4)}}
\newcommand{\omegafive}{ \Omega_\mathrm{GP}^{(5)}}

\newcites{SM}{Supplementary Material Bibliography}

\begin{document}

\setlength\abovedisplayskip{10pt}
\setlength\belowdisplayskip{10pt}

\setlength{\parskip}{14pt}
\setlength{\parindent}{0pt}

\preprint{MITP-26-031}
\title{Operator Learning for Spectral Reconstruction from Euclidean Correlators}

\author{Alessandro~\surname{De~Santis}~\orcidlink{0000-0002-2674-4222}}
\email{desantia@uni-mainz.de}
\affiliation{Helmholtz-Institut Mainz, Johannes Gutenberg-Universit{\"a}t Mainz}
\affiliation{GSI Helmholtz Centre for Heavy Ion Research, 64291 Darmstadt, Germany}

\begin{abstract}
In this work, we propose a novel supervised machine-learning-based strategy for extracting smeared spectral functions from Euclidean correlation functions. The strategy revisits the numerically ill-posed spectral reconstruction problem within the framework of \textit{Operator Learning} through the use of DeepONet-like architectures. To illustrate the method, we construct an \textit{ensemble of neural networks} trained on mock data generated from Gaussian Processes containing different levels of prior knowledge. This ensemble is then employed to estimate the systematic uncertainty associated with the fact that a neural network provides only an approximation to the target operator. The procedure is fully validated on previously unseen noisy mock data. To demonstrate the potential of the method for phenomenological applications, we reconstruct the inclusive rate above the four-particle threshold and up to high energies in the $1+1$-dimensional O(3) non-linear $\sigma$ model, starting from correlation functions computed in Monte Carlo simulations. The final result is consistent with the known analytic spectral density and, compared with the state-of-the-art Hansen-Lupo-Tantalo algorithm, exhibits a significant reduction in the total uncertainty even under mild physical assumptions encoded in the training set. The proposed strategy can be naturally extended to more phenomenologically relevant observables in lattice QCD.
\end{abstract}

\maketitle

\section{\label{sec:introduction}Introduction}

Spectral reconstruction in Lattice Quantum Chromodynamics (QCD) is the task of extracting information about the spectral densities associated with Euclidean correlation functions, which are the primary observables computed in first-principles Monte Carlo simulations. This topic is of considerable phenomenological relevance in hadron physics, since many physical observables, including both inclusive and exclusive scattering amplitudes measured at colliders, can be directly related to spectral densities. The ability to perform this reconstruction therefore provides a clear theoretical framework for probing the Standard Model in the non-perturbative regime. As is well known, spectral reconstruction is as challenging as it is important since, in practice, it requires the numerical inversion of a Laplace transform, an ill-conditioned problem in the presence of statistical noise affecting the Euclidean correlators and of a limited amount of available information. Interest in this problem has increased dramatically in recent years, and a variety of methods aimed at achieving stable results with controlled statistical and systematic uncertainties have been proposed.

Among the many existing methods, the Hansen-Lupo-Tantalo (HLT) method, introduced in Ref.~\cite{HLT1} and recently revisited in Ref.~\cite{HLT2}, occupies a particularly prominent position. The HLT method inherits its regularization strategy from the Backus-Gilbert method \cite{backus1968resolving} and is specifically designed to extract in a model-independent way spectral functions convolved with arbitrary smearing kernels while maintaining full control over both statistical and systematic uncertainties. Its success has been established through a thorough validation in the $1+1$-dimensional O(3) non-linear $\sigma$ model \cite{O3} and by several high-impact phenomenological applications \cite{Rratio,tau1,tau2, Ds1, Ds2, Bs, Spharelon1,Spharelon2,Barone:2023tbl}. For this reason, we regard it as the state-of-the-art spectral reconstruction algorithm and use it as the benchmark against which we compare the new strategy proposed in this work. Other methods, whose description is beyond the scope of this work, are based on Chebyshev polynomials \cite{Barata:1990rn, Bailas:2020qmv}, Bayesian frameworks \cite{MEM, BR, DelDebbio:2024lwm,Horak:2021syv,PhysRevD.107.076019,Medrano:2025cmg,Dutrieux:2024rem}, and  Refs.~\cite{Tsuji:2026zku, Abbott:2026wdw, Bruno:2024fqc, Bergamaschi:2023xzx,jay2026kerneltransformationsboundssmeared,Giusti:2026mcy,cdjf-65yf,Dutrieux:2025jed,Karpie:2021pap,Hashimoto:2026emk} for other approaches and advancements in the field.

From the list of references, we have omitted methods based on machine learning, since it constitutes the main focus of this work. Indeed, the spectral reconstruction operation has also been approached from a machine-learning perspective in Refs.~\cite{Buzzicotti:2023qdv, DelDebbio:2026abr, Kades:2019wtd, Karpie:2019eiq, Wang:2021cqw, Chen:2021jey, PhysRevD.106.L051502,chen2022machinelearningspectralfunctions, PhysRevD.104.076011, Lechien:2022ieg, PhysRevLett.124.056401,Andratschke:2026jpq}. In particular, in Ref.~\cite{Buzzicotti:2023qdv}, together with M.~Buzzicotti and N.~Tantalo, we made a significant step forward towards turning neural networks into reliable tools for phenomenological predictions by devising a novel supervised strategy whose main features are model independence and a robust procedure for estimating systematic uncertainties based on an \textit{ensemble of neural networks}. After validating the method on mock data, we compared the results obtained from actual lattice correlators with those produced by the HLT method and found full consistency together with a comparable level of precision. Unfortunately, the method is impractical since, as is common to other machine-learning approaches explored so far, the neural networks are trained to map Euclidean correlators directly onto the desired smeared spectral function at fixed input and output parameters. Consequently, applying the method to a correlator obtained from a different Monte Carlo simulation, or even to the same correlator but for different output parameters, requires generating the entire ensemble of neural networks from scratch, resulting in a substantial consumption of computational resources.

In this work we develop a new machine-learning strategy that overcomes this limitation. To achieve this goal, we adopt a different paradigm and reformulate the problem within the framework of \textit{Operator Learning}, the branch of machine learning devoted to the representation of operators, mapping one function space into another, through artificial neural networks. A major breakthrough in this field was achieved in Ref.~\cite{DeepONet}, where the Deep Operator Network architecture (DeepONet, for short) was introduced. DeepONet has been shown to provide a universal approximation of linear and non-linear operators and has been successfully applied to a broad range of problems. The spectral reconstruction operation is an operator and, building on this observation, we develop a new supervised strategy based on DeepONets to address the problem. To the best of our knowledge, this is the first application of DeepONets in the context of lattice field theory and, beyond spectral reconstruction, the strategy presented here can be applied to other relevant problems, such as the extraction of parton distribution functions, which is an ill-posed inverse problem as well (see, for instance, Ref.~\cite{Medrano:2025cmg}).

Unlike the approach of Ref.~\cite{Buzzicotti:2023qdv}, we do not pursue a fully model-independent strategy. Instead, we train our neural networks on mock data generated within a class of functions that incorporates prior knowledge of the physical process under investigation. We then carry out a study about the dependence of the final result on the amount of prior knowledge, by progressively relaxing the physical assumptions. We discuss how full model independence may be, at least in principle, recovered within this framework. At the same time, we retain the concept of neural-network ensembles to estimate the systematic uncertainties associated with the network predictions. After validating the strategy on noisy mock data, we reconstruct, as a benchmark, the inclusive rate in the two-dimensional O(3) non-linear $\sigma$ model using the same Euclidean correlators computed in Ref.~\cite{O3}. A comparison with the HLT results reported in Ref.~\cite{O3} shows that our method provides access to a significantly more precise determination of the inclusive rate. The method presented here opens the way to a potentially more powerful approach for extracting spectral information from four-dimensional lattice QCD simulations.

The remainder of this paper is organized as follows. In \Cref{sec:framework}, we formulate the spectral reconstruction problem and introduce the DeepONet architecture. In \Cref{sec:dataset}, we present the benchmark model and describe the procedure used to generate the training set. In \Cref{sec:training}, we discuss the neural-network architectures considered in this work together with the strategy used to estimate the systematic uncertainties associated with their predictions. A closure test is also presented. Finally, in \Cref{sec:benchmark}, we apply the proposed strategy to the benchmark model and, after numerically performing all the relevant limits required to recover the physical result, compare our reconstruction with both the known analytic spectral function and the result obtained using the HLT method.

\section{\label{sec:framework}Framework}

\subsection{\label{sec::spectral_reconstruction} Spectral Reconstruction}
In this section, we formulate the mathematical framework underlying spectral reconstruction and introduce some of the notation used throughout this work. Since the topic is well established, we keep the focus on the aspects needed in the following sections. For a more detailed discussion, we refer the reader to Refs.~\cite{HLT1,HLT2,Buzzicotti:2023qdv}.

In this work we restrict ourselves to correlation functions with only one temporal dependence and generalisation to higher-dimensional cases is also possible. Euclidean correlation functions are related to their spectral density through the Laplace transform, 
\begin{flalign}\label{eq:laplace_continuum}
C(t) = \int_{0}^{\infty}\dd{\omega} e^{-\omega t}\rho(\omega).
\end{flalign}
In the lattice field theory framework, the spacetime is discretized within a box and the quantity that is actually computed is a finite-volume Euclidean correlator at discrete times. In this case \cref{eq:laplace_continuum} becomes
\begin{flalign}\label{eq:laplace}
C_{L}(an) = \int_{0}^{\infty}\dd{\omega} e^{-\omega a n}\rho_{L}(\omega)\,.
\end{flalign}
In the above equation, $a$ denotes the lattice spacing in physical units, while the physical Euclidean time is $t_n=an$ with $n=0,1,\cdots, N_T/2$. In writing \cref{eq:laplace}, we have assumed periodic boundary conditions in time and a temporal extent sufficiently large to neglect wrap-around effects. The subscript $L$ emphasizes that both the correlator and the associated spectral density depend on the spatial lattice extent $L=aN_L$. Making this dependence explicit is crucial because a comparison with experimentally measured quantities related to the spectral density is only possible after taking the infinite-volume limit, for which a suitable regularization procedure is required. Indeed, finite-volume spectral densities are defined only on the discrete spectrum of the finite-volume Hamiltonian,
\begin{flalign}\label{eq:rho_L}
\rho_L(\omega)=\sum_{n=1}^{\infty}c_n(L) \delta\big(\omega-\omega_n(L)\big)\,,
\end{flalign}
and the $L\to \infty$ limit of this quantity is not mathematically well defined. Therefore, one should not aim at extracting $\rho_L(\omega)$ directly from $C_L(a\tau)$ but rather a \textit{smeared} version of the spectral density,
\begin{flalign}
\rho_L[K]=\int_{0}^{\infty}\dd{\omega}K(\omega)\rho_L(\omega)\,.
\end{flalign}
Here, the \textit{smearing kernel} $K(\omega)$ may be any Schwartz function. The smeared spectral density is a well-defined quantity that admits a well-defined infinite-volume limit. The choice of the smearing kernel can be motivated either by theoretical considerations or by practical advantages. In the former case, the physical observable of interest can be obtained only for a specific choice of $K(\omega)$. This is the case, for example, in the study of semileptonic inclusive rates (see, for instance, Refs.~\cite{tau1,Ds2,Bs}) or exclusive scattering amplitudes \cite{Patella:2024cto,PhysRevD.100.034521}. In the latter case, the spectral density is directly measured experimentally, and one may exploit this fact by choosing a smearing kernel that is theoretically convenient in terms of the achievable precision. This is the case, for instance, in the study of the $R$-ratio presented in Ref.~\cite{Rratio}. Notice that the correlator itself is the result of applying the smearing kernel $e^{-t\omega}$ to the associated spectral function.

The formalism based on smeared spectral densities also enables the reconstruction of the unsmeared spectral density itself through a procedure first proposed in Ref.~\cite{PhysRevD.96.094513}. The basic idea is to choose as smearing kernel a \textit{resolution function} $K_\sigma(E-\omega)$, dependent on a \textit{smearing parameter} $\sigma$ and on the energy $E$, with unit area and satisfying the following property
\begin{flalign}
&\lim_{\sigma\to 0}K_\sigma(E-\omega)=\delta(E-\omega)\,.
\end{flalign}
The resulting smeared spectral density,
\begin{flalign}
\rho_{\sigma,L}(E)=\int_{0}^{\infty}\dd{\omega} K_\sigma(E-\omega)\rho_L(\omega)\,,
\end{flalign}
is a smooth function of the energy $E$, whose degree of smoothness is controlled by the smearing parameter $\sigma$. Being a smeared quantity, it admits a well-defined infinite-volume limit. Moreover, exploiting the defining property of the resolution function, the unsmeared spectral density can be recovered through the double limit
\begin{flalign}\label{eq:double_limit}
\rho(E)=\lim_{\sigma \to 0}\lim_{L\to\infty}\rho_{\sigma,L}(E)\,.
\end{flalign}
The two limits do not commute and must therefore be taken in the order shown above. A convenient choice of resolution function is e.g. the normalized Gaussian with center $E$ and variance $\sigma^2$,
\begin{flalign}\label{eq:gaussian}
K_\sigma(E-\omega)=\frac{1}{\sqrt{2\pi}\sigma}e^{-\frac{(\omega-E)^2}{2\sigma^2}}\,.
\end{flalign}
Replacing the spectral density with its smeared counterpart does not remedy the numerical difficulty of the reconstruction problem, which remains ill-conditioned because of the unavoidable statistical uncertainties affecting the correlator $C_L(a\tau)$. The exponential nature of the Laplace transform makes the inverse operation particularly sensitive even to tiny distortions of the correlator. For this reason, any algorithm designed to solve the spectral reconstruction problem must incorporate a suitable regularization procedure to suppress the effects of statistical noise. The HLT method of Ref.~\cite{HLT1}, which is a well-established technique for extracting smeared spectral densities from Euclidean correlators, makes the dependence on the regularization explicit and provides a theoretically clean procedure to assess it. Despite its theoretical robustness, the HLT method suffers from the fact that the solution is expressed as a linear combination of the correlator values with coefficients chosen to approximate the smearing kernel on the exponential basis. As a consequence, systematic uncertainties can be made arbitrarily small only at the price of increasingly large statistical uncertainties. For sharply peaked smearing kernels (for instance, those associated with small values of the smearing parameter), this limits the achievable precision, and small uncertainties can be obtained only through a computationally expensive increase in the correlator statistics.

On the other hand, neural networks can approximate operators through highly non-linear transformations of the input correlator. As a consequence, the impact of statistical noise is handled in a fundamentally different way, leaving open the possibility of achieving better performance than HLT. This motivates us to revisit the machine-learning-based extraction of smeared spectral densities from noisy correlators. In order to place our new method in a concrete setting, we focus in the remainder of this paper on the extraction of spectral functions smeared with the Gaussian kernel of \Cref{eq:gaussian}. This choice is also motivated by the fact that, in order to benchmark our method against HLT using actual lattice data, we ultimately aim to perform the double limit in \Cref{eq:double_limit} and recover the spectral function associated with the inclusive rate in the two-dimensional O(3) model. As in the HLT method, however, the choice of smearing kernel in our strategy is entirely arbitrary. Having motivated the focus on smeared spectral functions, we now drop the subscript $L$ for the sake of notation and proceed to the formulation of the problem in the context of Operator Learning.

\subsection{\label{sec::operator_learning} Operator Learning}

In this section, we formulate our goal, the extraction of a smeared spectral density from a noisy Euclidean correlator, within the framework of Operator Learning. To better understand how this approach differs from the machine-learning applications discussed in the introduction, let us briefly review the strategy developed in Ref.~\cite{Buzzicotti:2023qdv}. There, the objective was the same. From a practical perspective, the neural network was trained to associate a Euclidean correlator, defined on a fixed set of $N$ Euclidean times and therefore represented as a vector $\vec{C}$, with the corresponding smeared spectral density evaluated at a fixed value of the smearing parameter $\sigma$ and on a grid of $N_E$ equally spaced energies. The output of the network is therefore another vector, denoted by $\vec{\rho}_\sigma$. From this perspective, the neural network learns a map from a vector to a vector,
\begin{flalign}\label{eq:O_vector}
\hat{\mathcal{O}}: \mathbb{R}^N \mapsto \mathbb{R}^{N_E}\,.
\end{flalign}
As a consequence, spectral reconstruction can only be performed for correlators defined on the same set of Euclidean times used during training, while the prediction is valid only for the specific values of the energies and smearing parameter used to define the output space $\mathbb{R}^{N_E}$. The strategy was shown to generalize very well to previously unseen data and to provide reliable uncertainty estimates. Its main drawback, however, is immediately evident: if one wishes to change the set of Euclidean times defining the correlator, or if one is interested in obtaining the spectral density at a different set of energies or values of $\sigma$, the entire training procedure must be repeated from scratch.

\begin{figure*}[t]
\centering
\includegraphics[width=0.7\linewidth]{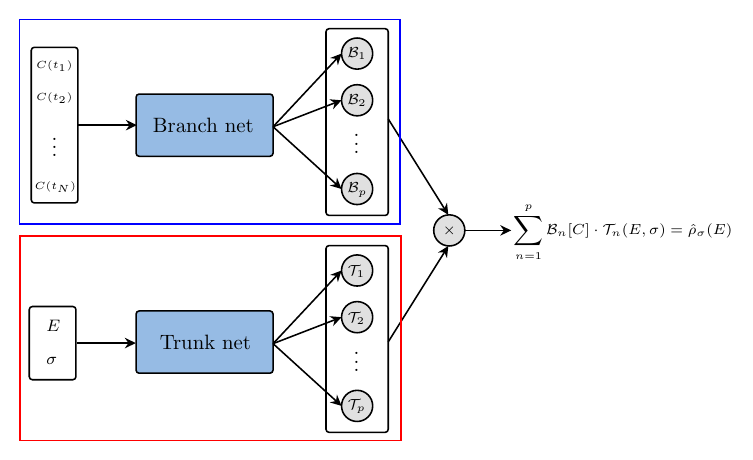}
\caption{\label{fig:DeepONet} DeepONet architecture employed in this work and inspired by Ref.~\cite{DeepONet}. The architecture consists of two separate neural networks, the \textit{Branch Net} and the \textit{Trunk Net}. The Trunk Net processes the domain of the output spectral functions, which in this case is parameterized by the energy $E$ and the smearing parameter $\sigma$, and maps it onto a latent representation. The Branch Net instead compresses the input function, the Euclidean correlator represented as a vector indexed by the Euclidean time slice, into a latent representation of the same dimension as that produced by the Trunk Net. The reconstructed smeared spectral function is then obtained as the dot product of the two latent representations. }
\end{figure*}
In the Operator Learning framework, an artificial neural network is trained to reproduce an operator $\hat{\mathcal{O}}$ that maps a space of functions onto another space of functions, 
\begin{flalign}\label{eq:O_space}
\hat{\mathcal{O}}:  \mathcal{C} \mapsto \mathcal{S}\,.
\end{flalign}
In our case, $\mathcal{C}$ is the space where Euclidean correlators are defined,
\begin{flalign}
\mathcal{C} = \big\{C(t)\; | \; t\in\Omega_t\subset \mathbb{R}^+\big\}\,,
\end{flalign}
while $\mathcal{{S}}$ is the space of the smeared spectral functions,
\begin{flalign}
\mathcal{S}=\big\{\rho_\sigma(E)\;|\; (E,\sigma)\in \Omega_{E,\sigma}\subset \mathbb{R}\times \mathbb{R}^+\big\}.
\end{flalign}
Notice that smeared spectral functions can, both in principle and in practice, be evaluated at negative energies, whereas the smearing parameter is restricted to non-negative values. In light of the limitations discussed for the vector-to-vector map in \Cref{eq:O_vector}, a neural-network representation of the operator in \Cref{eq:O_space} is considerably more powerful, since the domains of both the input correlator and the output prediction are not constrained to the fixed grids used during training but can be accessed continuously.

The existence of a neural-network representation of a linear or non-linear operator is supported by the so-called Universal Approximation Theorem (UAT), whose proof is given in Ref.~\cite{UAT} and whose statement is reported in \Cref{app:UAT} for convenience. The practical relevance of the UAT is limited by the fact that it provides no guidance on which architecture should be employed to approximate a given operator. In practice, a trained neural network always provides only an approximation to the target map. We therefore adopt the viewpoint that the relevant question is not whether the operator can be represented exactly, but rather whether the systematic uncertainty associated with its neural-network approximation can be reliably estimated. We provide numerical evidence that this is indeed the case. The strategy, discussed in \Cref{sec::systematic}, is based on combining several approximations obtained from an ensemble of independently trained neural networks. This approach is a reinterpretation of that proposed in Ref.~\cite{Buzzicotti:2023qdv}, where it was shown to be highly effective, and constitutes a central ingredient of the methodology developed in the present work.

\subsection{\label{sec::DeepOnet}DeepONet architecture}
We now explain how the neural network is in practice realized. In the following, we denote by a hat $\hat{\;}$ any estimator of the solution, regardless of whether it is obtained from the neural network output or from an alternative reconstruction method.  The most common neural network architecture that has been extensively employed in several contexts to approximate operators is the Deep Operator Network (DeepONet) introduced in Ref.~\cite{DeepONet} to solve partial and ordinary differential equations and whose structure is sketched in \cref{fig:DeepONet}. In the DeepONet architecture, the estimator of the solution to our problem is given by the linear combination
\begin{flalign}\label{eq:master_equation_DeepONet}
\hat{\rho}_\sigma(E)=\sum_{n=1}^{p}\mathcal{B}_n[C]\cdot \mathcal{T}_n(E,\sigma)\,.
\end{flalign}
The architecture consists of two neural networks whose sets of trainable parameters are independent. The \textit{Branch Net} takes as input a vector $\vec{C}\in\mathbb{R}^N$ whose components are the values of the Euclidean correlator evaluated on a fixed grid of Euclidean times, namely $C_n=C(t_n)$ with $t_n\in\Omega_t$ and $n=1,\cdots,N$. The task of the Branch Net is to parametrize a non-linear map of the correlator onto a latent representation living in the space $\mathbb{R}^p$, whose components are denoted by $\mathcal{B}_n[C]$ with $n=1,\cdots,p$. The \textit{Trunk Net}, on the other hand, is designed to map the domain $\Omega_{E,\sigma}$ of the smeared spectral function, represented by the two-dimensional vector $(E,\sigma)$, onto a vector in $\mathbb{R}^p$ whose components are denoted by $\mathcal{T}_n(E,\sigma)$ with $n=1,\cdots,p$. The outputs of the Branch and Trunk Nets are then combined according to \Cref{eq:master_equation_DeepONet}. The underlying interpretation is that the Branch Net maps the input correlator onto a new basis, whose mathematical properties depend on the neural-network parametrization, such as the choice of activation functions, the number of layers, and the number of neurons. The Trunk Net, in turn, maps the domain $\Omega_{E,\sigma}$  non-linearly onto a set of coefficients that, when combined with the Branch representation, provide an approximation to the target smeared spectral function. The optimization of the network parameters is performed in a standard supervised-learning setting. A correlator $\vec{C}$ is fed to the Branch Net while, simultaneously, the Trunk Net receives a pair $(E,\sigma)$ randomly sampled from the domain $\Omega_{E,\sigma}$. The prediction $\hat{\rho}_\sigma(E)$ is then computed and compared with the corresponding exact value $\rho_\sigma^\mathrm{true}(E)$. The optimal network parameters are obtained by minimizing the resulting loss function over a large set of mock data through standard back-propagation algorithms until convergence is achieved. The construction of the training set is described in \Cref{sec:dataset}, while the details of the architectures and the training procedure are presented in \Cref{sec:training}. If the training set contains a sufficiently rich variety of functions and the domain $\Omega_{E,\sigma}$ is sampled uniformly, the resulting trained architecture provides an approximation to the target operator where, unlike the map in \cref{eq:O_vector}, $E$ and $\sigma$ are not discrete indices but continuous coordinates. This operator can be employed to make predictions on previously unseen data belonging to the same function space used for training.

What is particularly interesting about this architecture is the decoupling between the input correlator and the parameters of the output smeared spectral density. For fixed values of $E$ and $\sigma$, the output of the Trunk Net is independent of the correlator. Conversely, a given correlator is associated with a unique latent representation that does not depend on the values of $E$ and $\sigma$. A further advantage is that the dimension $p$ of the latent space can be chosen independently of the number of Euclidean time slices at which the correlator is known and can, in principle, be made arbitrarily large. To appreciate the significance of this point, it is useful to compare the DeepONet estimator with the estimator provided by the HLT method,
\begin{flalign}\label{eq:master_equation_HLT}
\hat{\rho}_\sigma^\mathrm{HLT}(E)=\sum_{n=1}^{N}C(an)\cdot g_n(E,\sigma)\,.
\end{flalign}
Without entering into the details, the coefficients $g_n$ are obtained by minimizing the functional
\begin{flalign}
A[\vec{g}]=\int_{0}^{\infty} \dd{\omega}\bigg[K_\sigma(E-\omega)-\sum_{n=1}^{N}g_ne^{-\omega an}\bigg]^2\,,
\end{flalign}
which measures, in the $L_2$ norm, the difference between the target smearing kernel and its approximation in the basis of exponentials. In \Cref{eq:master_equation_HLT}, the sum is limited by the number $N$ of available Euclidean time slices, and the coefficients $g_n$ are constrained to approximate the target smearing kernel in the basis of exponential functions. As a consequence, these coefficients are usually highly oscillatory in sign and very large in magnitude. For this reason, a regularization procedure is required to control the amplification of the statistical noise affecting the correlator.  By contrast, the basis underlying \Cref{eq:master_equation_DeepONet} is not fixed \textit{a priori} and is instead learned during training. Moreover, the dimension of the latent space is not constrained by the number of available time slices and can be chosen independently. In addition, the magnitude of the coefficients $\mathcal{T}_n$ can be controlled through an appropriate choice of activation functions and network architecture, effectively introducing a built-in mechanism that mitigates the effects of the ill-posedness of the reconstruction problem. The accuracy of the operator approximation can then be systematically assessed by increasing the dimension of the latent space. In addition, physical constraints, such as the positivity of the spectral density, can be naturally incorporated into \Cref{eq:master_equation_DeepONet}, thereby preventing unphysical predictions. By contrast, the implementation of such constraints within the HLT framework is considerably less straightforward.

The last important point to address concerns the choice of the domain $\Omega_t$ on which the input correlator is defined. As anticipated, $C(t)$ enters the Branch Net as a vector and, consequently, $\Omega_t$ is a discrete set of time points. This is not merely a consequence of the fact that lattice correlators are known only at discrete times corresponding to integer multiples of the lattice spacing. Rather, it reflects the fact that neural networks are mathematical objects that naturally operate on vectors and matrices. Therefore, even if the correlator $C(t)$ were known continuously over the entire interval $[0,T]$, a discretization of both the interval and the correlator would still be required. A natural choice would be to define $\Omega_t$ so that it coincides with the set of times at which the lattice correlator of interest is measured. However, this poses a problem if one wishes to reuse a trained neural network for correlators obtained from simulations with different lattice spacings. Two possible strategies can be adopted to avoid retraining:
\begin{enumerate}
\item The set of time points $\Omega_t$ is fixed \textit{a priori} in physical units. Any new lattice correlator must then be interpolated (or extrapolated) in order to be evaluated at the selected values of $\Omega_t$;

\item The Branch Net is modified so that it becomes aware not only of the values $C(t_n)$ but also of the physical meaning of the coordinates $t_n$. In this case, the training set must contain not only correlators generated from different spectral functions, but also correlators defined on different time grids. Correlators originating from different lattice simulations can then be fed directly to the network without requiring any additional preprocessing.
\end{enumerate}
We extensively explored the second strategy, as it is conceptually more appealing and potentially more powerful. In particular, we modified the Branch Net by incorporating the Deep Sets strategy of Ref.~\cite{Deepset}, which is specifically designed to process data defined on varying sets of coordinates. While we found that this approach can, in principle, work, its performance in terms of reconstruction accuracy was significantly worse than that achieved with the first strategy. It is certainly possible that alternative architectures could provide better results in this regard. However, in the present work we adopt the first strategy and keep the set of times $\Omega_t$ fixed in physical units (see \Cref{sec::sampling}).

As a final remark, we emphasize that the present work focuses on training a neural network, through supervised learning, to reconstruct a smeared spectral density. Alternative approaches, see for example Ref.~\cite{Andratschke:2026jpq}, propose to parameterize directly the unsmeared spectral density $\rho(E)$ and train a neural network to determine its values on a discrete set of energies under the constraint, supplemented by a suitable regularization, that the Laplace transform of the reconstructed spectral density reproduces the input correlator. Although we do not espouse this strategy, since, as discussed in \Cref{sec::spectral_reconstruction}, the unsmeared spectral density cannot be directly reconstructed in a mathematically well-defined way without introducing a smearing procedure, we note that it could be naturally implemented within the DeepONet framework. The advantage would be that the energy variable $E$ would become a continuous rather than a discrete index. The importance of obtaining neural-network representations with continuous indices has recently been emphasized, in a completely different context, in Ref.~\cite{Cipriani:2025ini}.

\section{\label{sec:dataset}Benchmark model and dataset}

After this lengthy, but necessary, introduction to the problem, we now turn to the implementation details of the proposed strategy and its validation. As anticipated in \cref{sec:introduction}, we do not pursue in this work the fully model-independent approach of Ref.~\cite{Buzzicotti:2023qdv}. While such a strategy remains in principle feasible, demonstrating the new methodology within a completely model-independent framework would require a substantial additional computational effort. We discuss in \cref{app:GP} how the problem could be approached in that case. Nonetheless, we provide in this work a sensitivity study on the choice of the priors towards the direction of model independence.

\subsection{\label{sec::benchmark_model} Benchmark model}
\begin{table}[]
\begin{tabular}{cccccc}
\toprule
ID & $(L/a)\times(T/a)$ & $\beta$ & $am$ & $mL$ & $mT$ \\
\midrule
A1 & $640\times320$    & 1.63 & 0.0447967(66)  &  29 & 14 \\
A2 & $1280\times640 $  & 1.72 & 0.0257692(35)  &  33 & 17 \\
A3 & $1920\times960$   & 1.78 & 0.0176061(39) &   34 & 17 \\
A4 & $2880\times1440$  & 1.85 & 0.0112591(21)  &  32 & 16 \\ \midrule
B1 & $5760\times1440$  & 1.85 & 0.0112601(75)  &  65 & 16 \\ 
B2 & $2880\times2880$  & 1.85 & 0.0112463(82)  &  32 & 32 \\
\bottomrule
\end{tabular}
\caption{\label{tab:ensembles} Action parameters of the ensembles used in this work. Ensembles A1--A4 have approximately the same physical volume and are used to perform the continuum extrapolation. Ensembles B1 and B2 are used to estimate finite-volume effects. The scale setting has been recalculated by numerically solving the condition $m^{-1}C(m^{-1}) = 0.046615$ on each ensemble. Note that the original correlators for the ensemble A3 are currently unavailable and have therefore been generated from scratch for this work, using the same simulation and algorithmic parameters as in Ref.~\cite{O3}, to which we refer for additional details, but with $60\%$ of the original statistics.}
\end{table}
As a final benchmark for validating our strategy on lattice correlators, we reconstruct the inclusive rate in the $1+1$-dimensional O(3) non-linear $\sigma$ model above the four-particle threshold, where the Lellouch-L\"uscher formalism Ref.~\cite{Lellouch:2000pv} does not apply. The very same model was used in Ref.~\cite{O3} as a benchmark to validate the HLT method. The Euclidean correlators employed in this work are the same as those generated in Ref.~\cite{O3} and were kindly provided by the authors upon request. We refer the reader to Ref.~\cite{O3} for a detailed discussion of both the model and the lattice simulations, and keep the presentation here brief to avoid unnecessary repetition.

The O(3) model is asymptotically free and exhibits a dynamically generated mass gap $m$ that we use as unit to express all the dimensionful quantities in the following sections. Most importantly, the O(3) model is integrable. As a consequence, the exact inclusive spectral density can be computed analytically without relying on lattice simulations, thereby providing an ideal benchmark for validating the reconstruction procedure in presence of realistic data. The Euclidean action is
\begin{flalign}
S[\sigma]=\frac{1}{2g^2}\int \dd[2]{x}\partial_\mu\bar\sigma(x)\cdot \partial_\mu\bar\sigma(x)\,,
\end{flalign}
where $\bar\sigma^a(x)$ is a three-component real field of unit length. The model admits a conserved Noether current $j_\mu(x)$, which enters the spectral density
\begin{flalign}\label{eq:rho_O3}
\rho(E)=2\pi\bra{0}\hat{j}^a_1(0)\delta^2(\hat{P}-p)\hat{j}_1^a(0)\ket{0}\,.
\end{flalign}
This spectral density is related to the current-current correlator in the time-momentum representation through
\begin{flalign}
C(t)=\int\dd{\bm{x}}\bra{0} \hat{j}^a_1(0,\bm{x})&e^{-\hat{H}t}\hat{j}^a_1(0)\ket{0}\\
&=\int_{0}^{\infty}\dd{\omega}e^{-\omega t}\rho(\omega)\,.
\end{flalign}
The Euclidean correlator $C(t)$ has been measured on the lattice ensembles listed in \Cref{tab:ensembles}. The table contains six ensembles: A1-A4, which have approximately the same physical volume but different lattice spacings, and B1 and B2, which have the same lattice spacing as A4 but larger spatial and temporal extents, respectively.

Our goal is to employ the strategy presented in \Cref{sec::DeepOnet} to reconstruct the smeared spectral density for several values of $\sigma$ from these correlators and subsequently perform the infinite-volume, continuum, and $\sigma\to 0$ limits. This procedure yields the unsmeared spectral density, which can then be compared with the analytically known result. This is precisely the same program carried out in Ref.~\cite{O3} using the HLT method. The exact spectral density can be written as a sum of contributions labelled by the number $n$ of asymptotic particles propagating between the two currents,
\begin{flalign}\label{eq:rho_O3}
\rho^\mathrm{O(3)}(E)=\sum_{\mathrm{even} \;n\ge 2}\rho_n^\mathrm{O(3)}(E)\,.
\end{flalign}
The threshold of each multi-particle contribution is $n\cdot m$, so that $\rho^\mathrm{O(3)}(E)=0$ for $E\le 2m$. Additional details on the determination of the analytic spectral density are provided in \Cref{app:analytic_rho}. The spectral density in \Cref{eq:rho_O3} is associated with an inclusive scattering rate analogous to the phenomenologically relevant $R$-ratio, $e^+e^-\to\mathrm{hadrons}$, which corresponds to the spectral density of the vector-vector current in lattice QCD. The $R$-ratio has already been studied using the HLT method in Refs.~\cite{Rratio,Margari:2026him} and, as a future application, it can be probed using the strategy presented in this work.

\subsection{\label{sec::dataset} Dataset generation}

We now describe how to generate a physics-informed training set without relying on a specific parametric family of spectral functions, as is commonly done when a particular functional form, such as a Breit-Wigner distribution, is assumed \textit{a priori}. Neural networks trained on datasets generated from a restricted class of models are, by construction, biased toward that class. As a consequence, there is generally no reason to expect reliable predictions when the true spectral function lies outside the family of functions used to generate the training data. To overcome this limitation while still restricting the space of admissible functions according to prior knowledge of the underlying spectral density, derived for instance from theoretical considerations or phenomenological expectations, we introduce a novel strategy based on Gaussian Processes (GPs) (see Refs.~\cite{valentine2020gaussian,DelDebbio:2024lwm,williams2006gaussian} for introductions to the subject). The basic idea behind a GP is to generalize a multivariate Gaussian distribution, characterized by a mean vector $\vec{\mu}$ and a covariance matrix $\hat{\Sigma}$,
\begin{flalign}
\vec{x} \sim \mathcal{N}(\vec{\mu}, \hat{\Sigma})\,,
\end{flalign}
to a probability distribution over functions,
\begin{flalign}
\rho(E) \sim \mathcal{GP}\big(\mu(E), \mathcal{K}(E,E')\big)\,,
\end{flalign}
where $\mu(E)$ is a \textit{mean function} of the energy and $\mathcal{K}(E,E')$ is a \textit{covariance kernel}. The latter determines how the values of the spectral density at two different energies, $E$ and $E'$, are correlated. In this way, the covariance kernel controls the characteristic smoothness and correlation length of the functions generated by the process. A GP should be viewed as a probability distribution in the space of functions rather than in the space of individual points, as is the case for the multivariate Gaussian distribution $\mathcal{N}$. Once the mean function and covariance kernel have been specified, one can directly sample functions from the corresponding probability distribution. The key advantage of this framework is that prior physical knowledge can be incorporated naturally through suitable choices of the mean function and covariance kernel, thereby constraining the generated functions without imposing a specific parametric form.
\begin{figure}
\centering
\includegraphics[width=\linewidth]{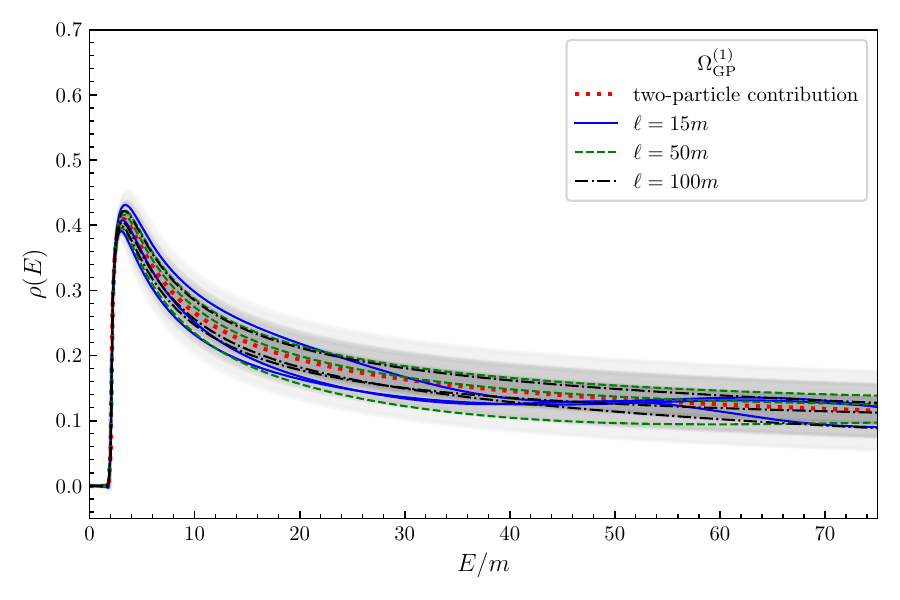}
\includegraphics[width=\linewidth]{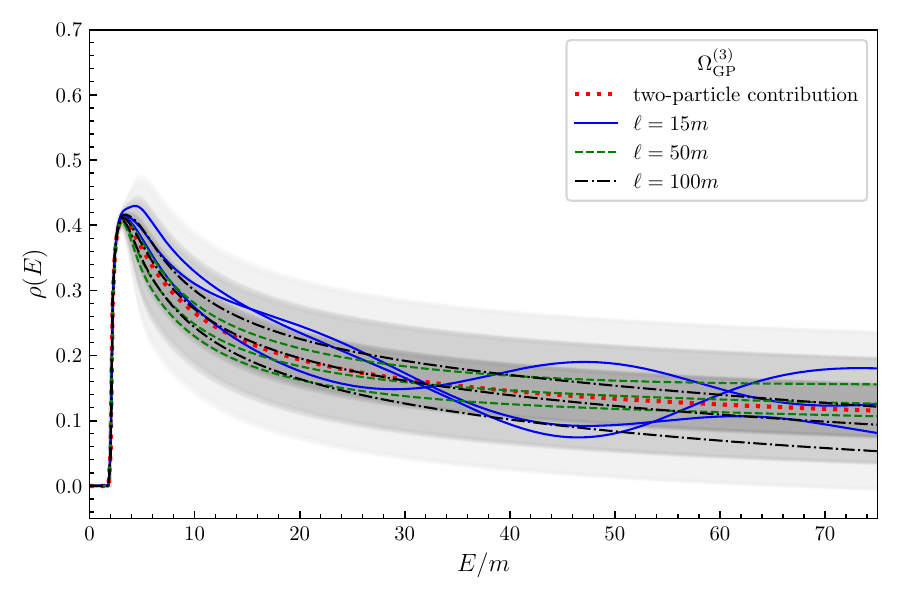}
\includegraphics[width=\linewidth]{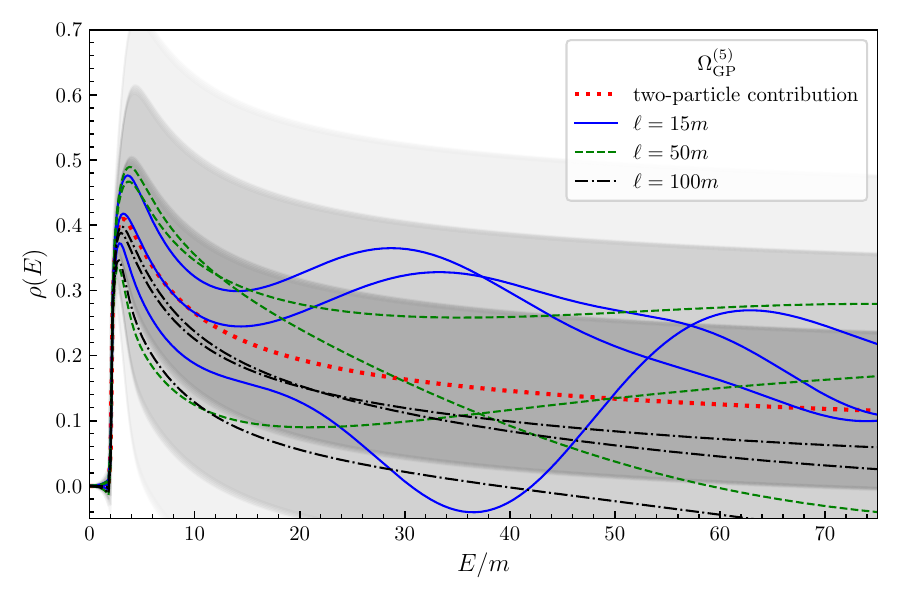}
\caption{\label{fig:function_space} Space of functions that are generated from a GP to assemble the training set for the supervised learning. From top to bottom the figures refer to the choices $(\sigma_\mathrm{GP},E_\star)=(0.02,3m)$, $(\sigma_\mathrm{GP},E_\star)=(0.04,4m)$ and $(\sigma_\mathrm{GP},E_\star)=(0.12,3m)$. The red dotted line is the analytical two-particle contribution defined in \cref{eq:rho2}, and it is used as mean of the GP. The three grey bands, from the darkest to the lightest, represent the $1\sigma$, $2\sigma$ and $3\sigma$ GP prior uncertainty $\sqrt{\mathcal{K}(E,E)}$. The other curves are examples of functions drawn randomly from the prior distribution by setting $\ell=15m$ (solid blue), $\ell=50m$ (dashed green) and $\ell=100m$ (dash-dotted black).}
\end{figure}
The physical information about the O(3) model that we exploit in the generation of the training set is the following. First, the two-particle contribution provides the dominant contribution to the spectral density, while multiparticle states become progressively less important as the number of particles increases (see \Cref{app:analytic_rho}). Second, above the four-particle threshold, the spectral density is a smooth function of the energy\footnote{The smoothness of the spectral density was already exploited in Ref.~\cite{O3} to probe the high-energy regime. In that work, the smearing parameter was allowed to increase linearly with the energy, since small values of $\sigma$ at large energies are prohibitive from the point of view of the achievable precision within the HLT method for the available correlator statistics.}. To generate a dataset of spectral functions compatible with this prior knowledge, we choose as mean function the two-particle contribution to the O(3) inclusive rate, which can be written in closed form,
\begin{flalign}\label{eq:rho2}
\mu(E) = \rho_{2}^\mathrm{O(3)}&(E)=\\\nonumber
=&\frac{3\pi^3}{8\theta^2}\frac{\theta^2+\pi^2}{\theta^2+4\pi^2}\tanh^3\frac{\theta}{2}\bigg\rvert_{\theta=2\cosh^{-1}\frac{E}{2m}}\,.
\end{flalign}
The covariance kernel is instead the  non-stationary symmetric function
\begin{flalign}\label{eq:covariance_kernel}
\mathcal{K}(E,E')=\xi(E)\xi(E')\,\exp\left(-\frac{(E-E')^2}{2\ell^2}\right) \;,
\end{flalign}
with
\begin{flalign}\label{eq:xi}
\xi(E)=\frac{\sigma_\mathrm{GP}}{1+e^{-(E-E_\star)/\varepsilon}}\;.
\end{flalign}

We set the parameter $\varepsilon=0.5m$. To perform a study on the sensitivity of the reconstruction to the prior parameters, we consider the five following combinations of hyper-parameters,
\begin{flalign}
\Omega_\mathrm{GP}^{(1)}: \quad\sigma_\mathrm{GP} =0.02 \qquad E_\star = 3m\;,\nonumber\\[6pt]
\Omega_\mathrm{GP}^{(2)}: \quad\sigma_\mathrm{GP} =0.04 \qquad E_\star = 3m\;,\nonumber\\[6pt]
\Omega_\mathrm{GP}^{(3)}: \quad\sigma_\mathrm{GP} =0.04 \qquad E_\star = 4m\;,\label{eq:GP_parameters} \\[6pt]
\Omega_\mathrm{GP}^{(4)}: \quad\sigma_\mathrm{GP} =0.08 \qquad E_\star = 3m\;,\nonumber\\[6pt]
\Omega_\mathrm{GP}^{(5)}: \quad\sigma_\mathrm{GP} =0.12 \qquad E_\star = 3m\;.\nonumber
\end{flalign}
We label these choices by $\Omega_\mathrm{GP}^{(1)}$, $\Omega_\mathrm{GP}^{(2)}$, $\Omega_\mathrm{GP}^{(3)}$, $\Omega_\mathrm{GP}^{(4)}$ and $\Omega_\mathrm{GP}^{(5)}$, respectively. The superscript numbers reflect the level of prior assumptions, from strong (1), through medium (3), to low (5). In the remainder of the text, it is understood that when we refer to $\Omega_\mathrm{GP}^{(i)}$ we intend that the result under discussion has been obtained by training the neural networks with spectral functions extracted from $\Omega_\mathrm{GP}^{(i)}$. The reconstruction of $\rho^{\mathrm{O(3)}}(E)$ is done independently for each choice of the hyper-parameters in order to assess their impact and, at the end, the five reconstructions are combined together for a final comparison with the HLT method.
The probability distribution over functions generated by some of the resulting GPs is illustrated in \Cref{fig:function_space}. The choice of $\rho_2^{\mathrm{O(3)}}$ is natural, since it determines the overall range of admissible spectral functions and, in particular, captures the rapid variation of the spectral density between $2m$ and $4m$, where the two-particle contribution turns on. The choice of the covariance kernel requires a more detailed discussion, as it introduces several parameters that control the properties of the generated spectral functions. The function $\xi(E)$ determines the local variance of the GP. In particular, at a given energy $E$, spectral functions sampled from the process fluctuate around the mean function with variance $\mathcal{K}(E,E)=\xi^2(E)$. The choice of \Cref{eq:xi} is such that this variance remains small for $E\ll E_\star$, i.e. below the four-particle threshold for the values of $E_\star$ given in \Cref{eq:GP_parameters}, where the spectral density is expected to be well approximated by the two-particle contribution. The variance then increases smoothly, with a transition scale controlled by $\varepsilon$, starting at $E_\star$ and eventually reaching a plateau for $E\gg4m$, where $\mathcal{K}(E,E)=\sigma_\mathrm{GP}^2$.  This completely specifies the local statistical uncertainty associated with the GP, given by $\sqrt{\mathcal{K}(E,E)}$. In \Cref{fig:function_space}, this uncertainty is represented by the gray bands corresponding to one, two, and three standard deviations around the mean function. \\
\myindent The parameter $\ell$ entering \Cref{eq:covariance_kernel} plays the role of a correlation length and therefore controls the smoothness of the sampled spectral functions. Smaller values of $\ell$ allow for increasingly rapid fluctuations, and in the limit $\ell\to0$ the sampled functions recover the distributional nature of finite-volume spectral functions. Conversely, larger values of $\ell$ suppress short-distance fluctuations and produce smoother spectral functions. This behavior can be appreciated in \Cref{fig:function_space}, where samples generated with $\ell=15m$ (blue curves), $\ell=50m$ (green curves), and $\ell=100m$ (black curves) are compared. While the analytic spectral density $\rho^{\mathrm{O(3)}}(E)$ is a very smooth function of the energy, which would naturally suggest choosing a large correlation length, one should keep in mind that the Euclidean correlators measured in lattice simulations are affected by both cutoff and finite-volume effects. These effects may induce less smooth features in the corresponding spectral densities. Therefore, in order to avoid imposing an excessive degree of smoothness that could bias the reconstruction and potentially miss the true solution, we generate mock data from several instances of the GP by sampling the correlation length $\ell$ uniformly in the interval $[15m,100m]$. In addition, we vary the mass entering the mean function in \Cref{eq:rho2} uniformly in the range $[0.95m,1.05m]$. This choice is intended to account for small distortions of the low-energy region induced by finite-volume effects, thereby further enlarging the space of admissible spectral functions represented in the training set. The resulting distribution of functions is illustrated in \cref{fig:function_space}. Reconstructing the spectral function for the O(3) model using five different choices of the GP hyper-parameters helps quantify the overall impact of this choice on the final result. The choice $\Omega_\mathrm{GP}^{(1)}$ generates a space of spectral functions which are very close to the mean function, while the choice $\Omega_\mathrm{GP}^{(5)}$ generates functions that are substantially different from the mean function (and even become negative), both in amplitude and shape. On the practical side, we can consider the space of functions generated by $\Omega_\mathrm{GP}^{(5)}$ as a proxy for model independence above the four-particle threshold. We extend the discussion about GP and model independence in \cref{app:GP}.

\subsection{\label{sec::sampling}Function sampling}

We provide now the practical recipe to generate mock data entering the training set. First, we fix the domains $\Omega_t$ and $\Omega_{E,\sigma}$. As discussed in \Cref{sec::DeepOnet}, we choose a fixed grid of Euclidean times expressed in physical units and set $\Omega_t=\{1,2,\cdots,540\}am$, where $am$ corresponds to the finest lattice spacing among the ensembles listed in \Cref{tab:ensembles}, namely that of ensemble A4. With this choice, the correlators measured on ensembles A4, B1, and B2 require no interpolation. For later convenience, we denote the correlator measured on ensemble ID by $C^\mathrm{ID}$. The largest Euclidean time included in the analysis is $540am\sim6$, which is well below $mT/2$. This allows us to safely neglect the wrap-around effects induced by the periodic boundary conditions. As for the domain $\Omega_{E,\sigma}=\Omega_E\times\Omega_\sigma$, our goal is to reconstruct $\rho^\mathrm{O(3)}(E)$ above the inelastic four-particle threshold and up to energies where the spectral density approximately matches the perturbative prediction (see Ref.~\cite{O3}). We therefore choose $\Omega_E=[4m,40m]$ and $\Omega_\sigma=[0.5m,7m]$. The range $\Omega_\sigma$ extends from very small values of $\sigma$, which are required to perform the $\sigma\to0$ extrapolation, to relatively large values, allowing us to probe different levels of sensitivity to the underlying spectral density.

To generate a spectral function from the GP for given choice of the hyper-parameters we proceed as follows. We first define a discrete set of energies $E_n=n\cdot\Delta E$ with $n=0,1,\cdots,N_E-1$. At each point of this grid, we evaluate the mean function, obtaining an $N_E$-dimensional vector $\vec{\mu}$ with components $\mu_n=\rho_2^\mathrm{O(3)}(E_n)$. Likewise, we construct the $N_E\times N_E$ covariance matrix $\hat{\mathcal{K}}$, whose elements are given by $\mathcal{K}_{n,m}=\mathcal{K}(E_n,E_m)$. We then use the \texttt{numpy.random.multivariate\_normal} routine in Python to draw random samples from the multivariate Gaussian distribution $\vec{\rho}\sim\mathcal{N}(\vec{\mu},\hat{\mathcal{K}})$. Once a sample $\vec{\rho}$ has been generated, the corresponding correlator is computed as
\begin{flalign}\label{eq:C_mock}
C(t)=\Delta E \sum_{n=0}^{N_E-1}\rho_n e^{-tE_n}\qquad \forall t\in\Omega_t\,,
\end{flalign}
while, after drawing a pair $(E,\sigma)$ from $\Omega_E\times\Omega_\sigma$, the corresponding true smeared spectral density is obtained as 
\begin{flalign}\label{eq:rho_mock}
\rho_\sigma^\mathrm{true}(E)=\Delta E \sum_{n=0}^{N_E-1}\rho_n K_\sigma(E-E_n)\,.
\end{flalign}
The value of $N_E$ is then fixed such that the largest energy at which $\rho(E)$ is sampled is $200m$, well above the largest ultraviolet cutoff of the ensembles listed in \Cref{tab:ensembles}. \Cref{eq:C_mock,eq:rho_mock} are precisely what one would obtain by assuming that the underlying spectral density is a finite-volume spectral function of the form given in \Cref{eq:rho_L}, with energy levels separated by $\Delta E$. In general, the energy levels are not equally spaced. However, the smallest value of the smearing parameter in $\Omega_\sigma$ is $\sigma=\Delta E$, which is not sufficiently small for the finite-volume spectrum to become resolved in the smeared spectral density. On the other hand, as shown in the appendix of Ref.~\cite{O3}, the L\"uscher quantization condition \cite{Luscher:1990ck} together with the Lellouch-L\"uscher relation \cite{Lellouch:2000pv} can be used to determine the two-particle energy levels contributing to $\rho^\mathrm{O(3)}(E)$. For a volume $mL=30$, comparable to those of the ensembles considered here, the largest separation between two consecutive peaks is approximately $0.4m$, which is even smaller than $\Delta E$. From now on, whenever we refer to a mock correlator, we mean the vector $\vec{C}$ constructed according to the procedure described above.

Following the procedure described above, for each of the five choices of the GP hyper-parameters given in \cref{eq:GP_parameters} we generated several training datasets containing $N_\rho=1\times10^5$, $2\times10^5$, $4\times10^5$, and $8\times10^5$ samples, respectively. By a sample, we mean a Euclidean correlator represented by the vector $\vec{C}$ and fed as input to the Branch Net, together with a pair $(E,\sigma)$ randomly drawn from $\Omega_{E,\sigma}$ and fed to the Trunk Net, as well as the corresponding target value $\rho_\sigma^\mathrm{true}(E)$ used to evaluate the prediction error and optimize the network parameters. The purpose of generating multiple datasets is related to our training strategy based on an ensemble of neural networks for the estimation of systematic uncertainties, as discussed in \Cref{sec::ensemble_networks}. We also produce a set containing $0.5\times10^5$ samples to be used as a validation test set during the training of the neural networks. Each input correlator in the training and validation sets is corrupted with noise according to the procedure explained in the next section. Of course, different noisy replicas of the same correlator can be used as an augmentation technique to expand the training set.

\subsection{\label{sec::preprocessing} Noise generation and preprocessing}
Correlators computed on Monte Carlo gauge configurations are naturally affected by statistical noise and, like any other reconstruction method, neural networks must be regularized against it. It is well known that neural networks can be very robust in handling noisy inputs. The most common strategy is to include noisy correlators directly in the training set and let the network learn to distinguish statistical fluctuations from the underlying physical information relevant to the prediction task. In lattice simulations, statistical fluctuations are generally correlated across Euclidean times and the noise-to-signal ratio typically grows exponentially at large time separations. Incorporating these features is particularly important when the ultimate goal is to apply the trained network to actual lattice correlators. To generate realistic noise, we exploit the measurements of the correlator $C^\mathrm{A4}$ as follows. Given a mock correlator generated according to the procedure described in \Cref{sec::sampling}, a realization of a noisy bootstrap sample is obtained through
\begin{flalign}\label{eq:noise}
C_i(t) = C(t)\big[1+\lambda \delta_i(t)\big]\,,
\end{flalign}
where
\begin{flalign}
\delta_i(t)= \frac{C_i^\mathrm{A4}(t)-\expval{C^\mathrm{A4}(t)}}{\expval{C^\mathrm{A4}(t)}}\,.
\end{flalign}
In the above equation, $C_i^\mathrm{A4}(t)$ denotes the $i$-th bootstrap sample of the lattice correlator $C^\mathrm{A4}(t)$. By construction, the relative fluctuation $\delta_i(t)$ has vanishing expectation value and therefore
\begin{flalign}
\expval{C_i(t)} = C(t)\,.
\end{flalign}
It is straightforward to show that the covariance matrix $\hat{\Sigma}$ of the mock correlator is a rescaled version of the covariance matrix of the lattice correlator,
\begin{flalign}
\Sigma[C](t,t')=\lambda^2\frac{C(t)C(t')}{\expval{C^\mathrm{A4}(t)}\expval{C^\mathrm{A4}(t')}}\hat{\Sigma}[C^\mathrm{A4}](t,t')\,.
\end{flalign}
According to this procedure, starting from $N_b$ bootstrap samples of the correlator $C^\mathrm{A4}$, one can generate $N_b$ bootstrap samples of the mock correlator that preserve both the temporal correlations and the noise-to-signal ratio of the original lattice data. In particular, we fixed a set of 5000 bootstrap samples from which to extract the quantities $C_i^\mathrm{A4}(t)$. The same set is used both for the training and validation sets and for all the training sets used in this work. The closure tests (see \cref{sec::closure_test}), however, are performed using a new set of 5000 bootstrap samples, independent of the first one. The subsequent analysis of the real lattice correlators is done using 500 bootstrap samples for each ensemble, independent of the two sets. The parameter $\lambda$, which we refer to as the \textit{noise level}, controls the amount of noise injected into the mock correlators. For $\lambda=0$, no noise is added, while for $\lambda=1$ the noise-to-signal ratio matches that of the lattice correlator. The motivation for introducing this parameter stems from the fact that the lattice correlators measured on ensembles A1-A4 have approximately the same statistics, whereas the correlators measured on ensembles B1 and B2, which are used to estimate finite-size effects, were generated with significantly fewer gauge configurations. As a consequence, their noise-to-signal ratio is approximately $3.5$ times larger than that of $C^\mathrm{A4}$. To account for this difference and to assess the performance of our strategy as a function of the noise level, we generate one ensemble of neural networks for $\Omega_\mathrm{GP}^{(3)}$ with $\lambda=3.5$, to be used exclusively for $C^{\mathrm{B1}}$ and $C^{\mathrm{B2}}$, which are only used to estimate finite-size effects (see \cref{sec::infinite_volume_limit}). We then generate an ensemble of neural networks with $\lambda=1$ for each of the choices in \cref{eq:GP_parameters}, to be used for the $C^\mathrm{A1-A4}$ correlators. In this work, we have therefore generated six independent ensembles of neural networks.

The last step before discussing the training strategy is the preprocessing of the correlators before they are fed to the Branch Nets. This step is necessary because a typical correlator spans several orders of magnitude as the Euclidean time increases due to the exponential kernel appearing in the Laplace transform. It has long been known that neural networks have difficulty treating on an equal footing input features that differ substantially in magnitude, often assigning little weight to the smallest components and, in extreme cases, even ignoring them. To address this issue, we adopt the standardization procedure. For each Euclidean time slice, we subtract from every correlator the mean value computed over all samples in the dataset and divide the result by the corresponding standard deviation. The resulting dataset therefore has zero mean and unit variance at each time slice. The parameters defining this transformation are stored and subsequently applied both to mock data and to lattice correlators during the prediction stage. This preprocessing does not alter the correlations among different time slices. No preprocessing is instead applied to the domains $\Omega_{E,\sigma}$ and $\Omega_t$, nor to the output smeared spectral densities.

\section{\label{sec:training} Training and closure test}

In this section, we provide the implementation details of the DeepONet architectures used in this work, describe the strategy adopted to estimate systematic uncertainties, and present evidence of its effectiveness through a closure test performed on previously unseen mock data.

\subsection{\label{sec::ensemble_networks}Ensemble of neural networks}

For a given latent-space dimension $p$ (see \Cref{fig:DeepONet}), the Branch and Trunk networks must be designed as maps from $\mathbb{R}^{540}$ to $\mathbb{R}^{p}$ and from $\mathbb{R}^{2}$ to $\mathbb{R}^{p}$, respectively. Since the Branch and Trunk networks are completely independent, their architectures can in principle be entirely different. In this work, however, we adopt a simple and symmetric design and employ the same architecture for both networks. Specifically, we use a standard Feed-Forward architecture consisting of a sequence of fully connected layers, each containing the same number of hidden neurons. Alternative architectures, based for instance on convolutional layers or on so-called Fourier-feature networks \cite{tancik2020fourier}, have also been shown to be highly effective. Exploring such architectures, however, is beyond the scope of this work. As activation function in the intermediate layers, we use the non-polynomial Gaussian Error Linear Unit (GELU) \cite{GELU},
\begin{flalign}
g_\mathrm{GELU}(x)=\frac{x}{2}\big[1+\mathrm{erf}(x/\sqrt{2})\big]\,.
\end{flalign}
Even though constraints, such as positivity of the prediction, could be applied in the output layer, we decide here not to impose any restriction on the output solution and then the last layer of the Branch and Trunk Nets contains no activation functions.
\begin{table*}[t]
\begin{tabular}{cccccccc}
\toprule
& & & \multicolumn{2}{c}{Branch Net} & \multicolumn{2}{c}{Trunk Net} &  \\
\midrule
$k$ &  $N_\rho\times 10^{-5}$ & $p$ &\quad layers \quad&\quad neurons \quad&\quad layers \quad&\quad neurons \quad &\quad params \\
\midrule\midrule
 1  &    1  &    16   &   2  &    64   &     2 &   64   &  45216  \\
 2  &    1  &    32   &   2  &    64   &     2 &   64   &  47296  \\
 3  &    1  &    64   &   2  &    64   &     2 &   64   &  51456  \\
 4  &    1  &   128   &   2  &    64   &     2 &   64   &  59776  \\ \midrule
 5  &    1  &    64   &   2  &   128   &     2 &  128   & 119168  \\
 6  &    1  &    64   &   2  &   256   &     2 &  256   & 303744  \\ \midrule 
 7  &    2  &    64   &   2  &    64   &     2 &   64   &  51456  \\
 8  &    4  &    64   &   2  &    64   &     2 &   64   &  51456  \\
 9  &    8  &    64   &   2  &    64   &     2 &   64   &  51456  \\ \midrule  
10  &    1  &    16   &   3  &    64   &     3 &   64   &  53536  \\ 
11  &    1  &    32   &   3  &    64   &     3 &   64   &  55616  \\ 
12  &    1  &    64   &   3  &    64   &     3 &   64   &  59776  \\ 
13  &    1  &   128   &   3  &    64   &     3 &   64   &  68096  \\ \midrule   
14  &    1  &    64   &   1  &    64   &     1 &   64   &  43136  \\
15  &    1  &   128   &   1  &    64   &     1 &   64   &  51456  \\ \bottomrule
\end{tabular}
\caption{\label{tab:networks} Summary of the neural networks used in this work. The Branch and Trunk Nets are stacks of standard fully-connected layers where the individual nodes also contain biases. The columns report, in order, the neural network index $k$, the number of samples in the training set (in units of $10^{5}$), the dimension of the latent space $p$, the number of layers and neurons in the Branch and Trunk Nets, respectively, the total number of trainable parameters. }
\end{table*}
\begin{table*}[t]
{
\setlength{\tabcolsep}{10pt} 
\begin{tabular}{ccccccccccccc}
\toprule
& \multicolumn{2}{c}{$\Omega_\mathrm{GP}^{(1)}$ ($\lambda=1$)} 
& \multicolumn{2}{c}{$\Omega_\mathrm{GP}^{(2)}$ ($\lambda=1$)}
& \multicolumn{2}{c}{$\Omega_\mathrm{GP}^{(3)}$ ($\lambda=1$)}
& \multicolumn{2}{c}{$\Omega_\mathrm{GP}^{(3)}$ ($\lambda=3.5$)}
& \multicolumn{2}{c}{$\Omega_\mathrm{GP}^{(4)}$ ($\lambda=1$)}
& \multicolumn{2}{c}{$\Omega_\mathrm{GP}^{(5)}$ ($\lambda=1$)}\\
\midrule
$k$ & 
$\mathcal{L}_\mathrm{val}^{(k)}$ & $w^{(k)}$ &
$\mathcal{L}_\mathrm{val}^{(k)}$ & $w^{(k)}$ &
$\mathcal{L}_\mathrm{val}^{(k)}$ & $w^{(k)}$ &
$\mathcal{L}_\mathrm{val}^{(k)}$ & $w^{(k)}$ &
$\mathcal{L}_\mathrm{val}^{(k)}$ & $w^{(k)}$ &
$\mathcal{L}_\mathrm{val}^{(k)}$ & $w^{(k)}$ \\

\midrule\midrule
1   &  1.4    &   0.073  &  4.5    &   0.071  &  3.8   &  0.072   &  7.2   &   0.070  &   16.9   &   0.067   &   36.0   &   0.067   \\
2   &  1.5    &   0.072  &  4.4    &   0.072  &  3.8   &  0.072   &  7.0   &   0.071  &   16.1   &   0.067   &   37.1   &   0.067   \\
3   &  1.3    &   0.074  &  4.6    &   0.070  &  3.7   &  0.073   &  6.8   &   0.073  &   17.4   &   0.066   &   38.0   &   0.066   \\
4   &  1.4    &   0.074  &  4.2    &   0.074  &  4.0   &  0.071   &  7.6   &   0.067  &   15.3   &   0.068   &   38.1   &   0.066   \\ \midrule
5   &  1.6    &   0.072  &  5.2    &   0.066  &  4.8   &  0.065   &  7.4   &   0.068  &   8.5    &   0.073   &   43.2   &   0.063   \\
6   &  7.6    &   0.039  &  5.0    &   0.068  &  4.7   &  0.066   &  9.1   &   0.058  &   18.0   &   0.066   &   43.2   &   0.063   \\
\midrule
7   &  1.2    &   0.075  &  3.8    &   0.077  &  3.3   &  0.076   &  5.8   &   0.080  &   13.0   &   0.069   &   27.0   &   0.074   \\
8   &  1.0    &   0.076  &  2.9    &   0.084  &  2.6   &  0.081   &  5.3   &   0.084  &   9.8    &   0.072   &   19.6   &   0.080   \\
9   &  0.9    &   0.077  &  2.5    &   0.087  &  2.3   &  0.083   &  4.9   &   0.087  &   8.1    &   0.073   &   16.5   &   0.082   \\
\midrule
10  &  1.5    &   0.072  &  5.8    &   0.063  &  4.6   &  0.066   &  7.7   &   0.066  &  20.9    &   0.064   &   43.0   &   0.063   \\
11  &  1.6    &   0.072  &  5.5    &   0.065  &  4.3   &  0.069   &  8.0   &   0.064  &  21.2    &   0.064   &   45.2   &   0.062   \\
12  &  1.5    &   0.073  &  5.4    &   0.065  &  4.5   &  0.067   &  7.7   &   0.066  &  20.5    &   0.064   &   44.0   &   0.062   \\
13  &  1.5    &   0.073  &  5.2    &   0.066  &  4.2   &  0.069   &  7.5   &   0.067  &  21.9    &   0.063   &   46.6   &   0.061   \\
\midrule
14  &  7.6    &   0.040  &  11.1   &   0.037  &  9.5   &  0.041   &  14.0  &   0.035  &  23.8    &   0.062   &   43.9   &   0.062   \\
15  &  8.2    &   0.037  &  11.4   &   0.036  &  12.1  &  0.031   &  12.2  &   0.042  &  25.2    &   0.061   &   43.4   &   0.063   \\
\bottomrule
\end{tabular}
}
\caption{\label{tab:performances} 
Summary of the performances of all the training runs carried out in this work. The first column reports the architecture index according to \cref{tab:networks}. For each choice of the GP hyper-parameters and $\lambda$, the table reports the minimum validation loss (in units of $10^{-6}$) and the corresponding normalized weight.}
\end{table*}
In order to obtain reliable predictions and estimate the systematic uncertainty associated with the fact that a neural network provides only an approximation to the target operator, we trained several architectures by varying both the network complexity and the size of the training set. In particular, we considered architectures with different numbers of layers and neurons at fixed training-set size, as well as architectures with different training-set sizes at fixed network structure. Overall, we considered $N_\mathrm{nets}=15$ neural networks, whose details are reported in \Cref{tab:networks}. We refer to this collection of architectures as a \textit{neural-network ensemble} and we generated six independent ensembles. The total number of independent training runs performed in this work is therefore 90. Within a given ensemble, and therefore at fixed $\lambda$ and GP hyper-parameters, each architecture is identified by an index $k=1,\cdots,15$. \Cref{tab:networks} reports, for each architecture, the number of training samples $N_\rho$, the latent-space dimension $p$, the number of layers and neurons and the total number of trainable parameters. Each architecture is trained using the Mean Squared Error (MSE) as loss function,
\begin{flalign}\label{eq:loss}
\mathcal{L}^{(k)}_\mathrm{train/val} = \frac{1}{N_\rho} \sum_{n=1}^{N_\rho}\bigg(\hat{\rho}_{\sigma,n}^{(k)}(E)-\rho_{\sigma,n}^\mathrm{true}(E)\bigg)^2\,,
\end{flalign}
where the index $n$ labels the $n$-th sample in the dataset. The quantity $\hat{\rho}_{\sigma,n}^{(k)}(E)$ denotes the prediction produced by the $k$-th network for the $n$-th sample, while $\rho_{\sigma,n}^\mathrm{true}(E)$ is the corresponding target value. The subscript ``train/val'' indicates that the same type of loss function is employed for both the training and validation datasets.
\begin{figure}
\centering
\includegraphics[width=\linewidth]{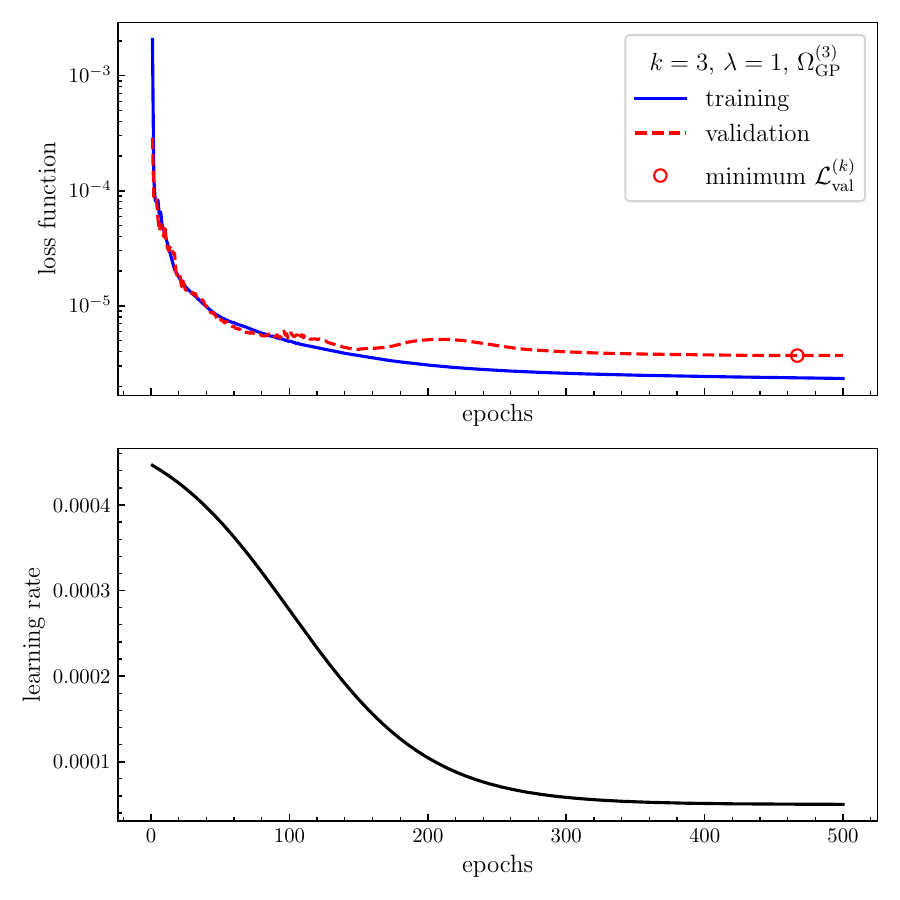}
\caption{\label{fig:loss} \emph{Top panel}: Training loss (solid blue line) and validation loss (dashed red line) as functions of the number of epochs for the neural network with index $k = 3$, $\lambda = 1$ and $\Omega_\mathrm{GP}^{(3)}$. One epoch corresponds to a single pass of the entire training set through the neural network. The circle marks the minimum of the validation loss. The network weights corresponding to this point are used for the prediction analysis. \emph{Bottom panel}: Learning rate $\eta(x)$ (\cref{eq:learning_rate}) as a function of the number of epochs.}
\end{figure}
All training runs are performed using the Mini-Batch Gradient Descent algorithm with a batch size of 32 on Google Colab \cite{googlecolab}. The neural networks are implemented using the libraries TensorFlow \cite{tensorflow2015-whitepaper} and Keras \cite{keras}. The network parameters are optimized using the Adam optimizer \cite{Adam} together with a learning-rate schedule of the form
\begin{flalign}\label{eq:learning_rate}
\eta(x)= \eta_f+\frac{\eta_i-\eta_f}{1+e^{(x-x_\eta)/\sigma_\eta}}\,,
\end{flalign}
with $\eta_i=0.5\times10^{-3}$, $\eta_f=0.5\times10^{-4}$, $\sigma_\eta=50$, and $x_\eta=100$. The function $\eta(x)$ is shown in the bottom panel of \Cref{fig:loss}. During training, we monitor the loss function at the end of each epoch, where one epoch corresponds to a complete pass through the entire training set. The loss is evaluated on both the training and validation datasets. The latter is not used to update the network parameters and serves only to assess the ability of the model to generalize to previously unseen data. The network configuration ultimately used for prediction corresponds to the epoch at which the validation loss reaches its minimum value, thereby preventing overfitting. The minimum validation loss reached by each network is reported in \Cref{tab:performances}. The top panel of \Cref{fig:loss} shows a representative example of the loss-function minimization for the network with index $k=3$, noise level $\lambda=1$ and $\Omega_\mathrm{GP}^{(3)}$. As can be seen, the training proceeds smoothly and no significant signs of overfitting are observed within the training range considered. In general, the minimum of the validation loss is reached after approximately 400 to 600 epochs.
\begin{figure*}
\centering
\includegraphics[width=\linewidth]{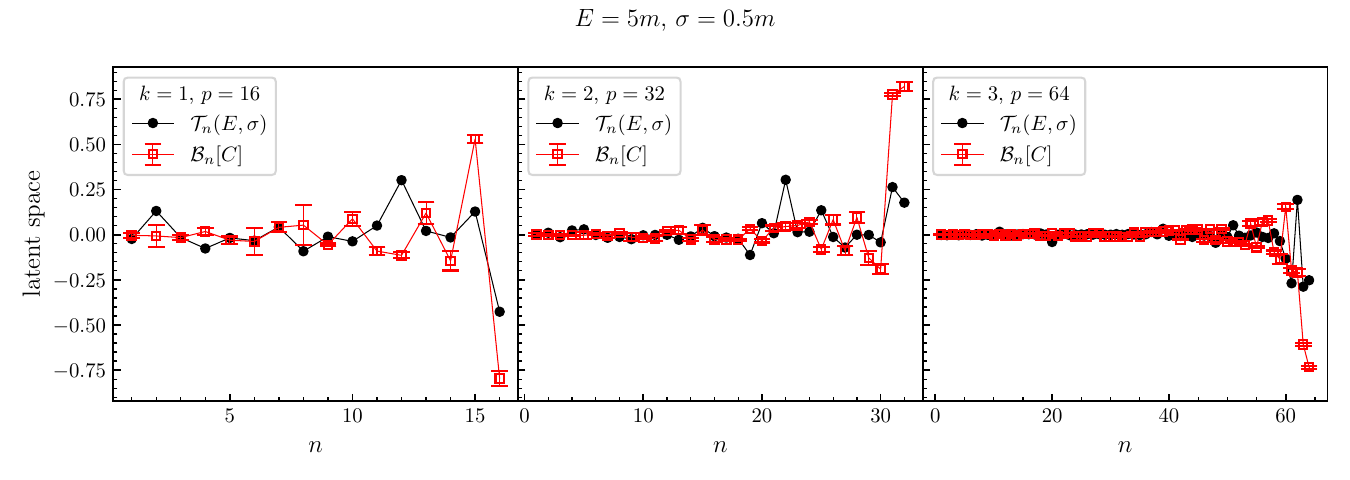}
\includegraphics[width=\linewidth]{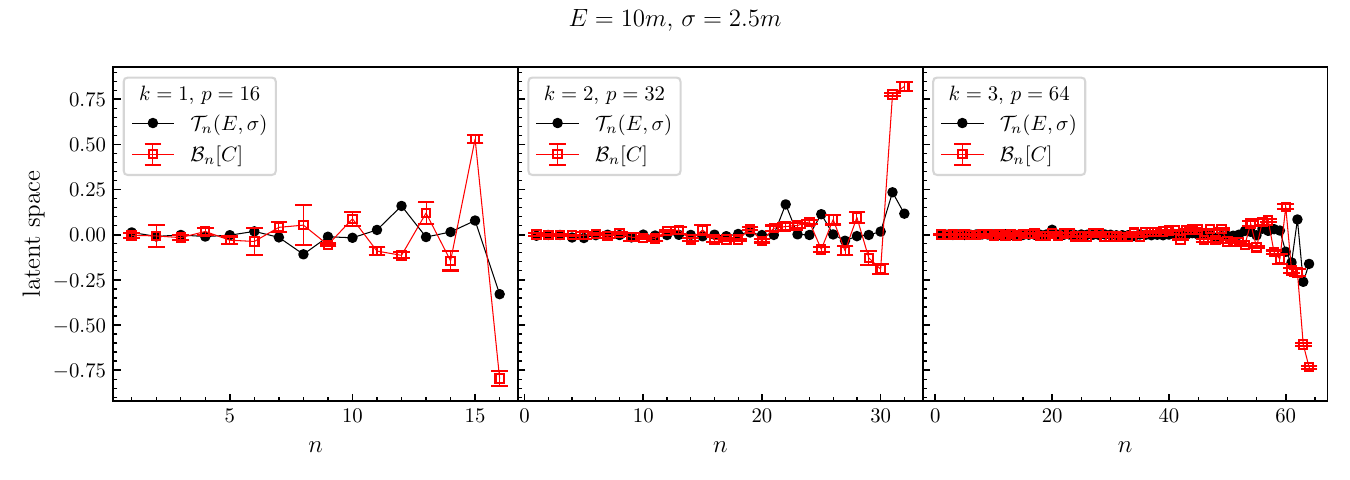}
\includegraphics[width=\linewidth]{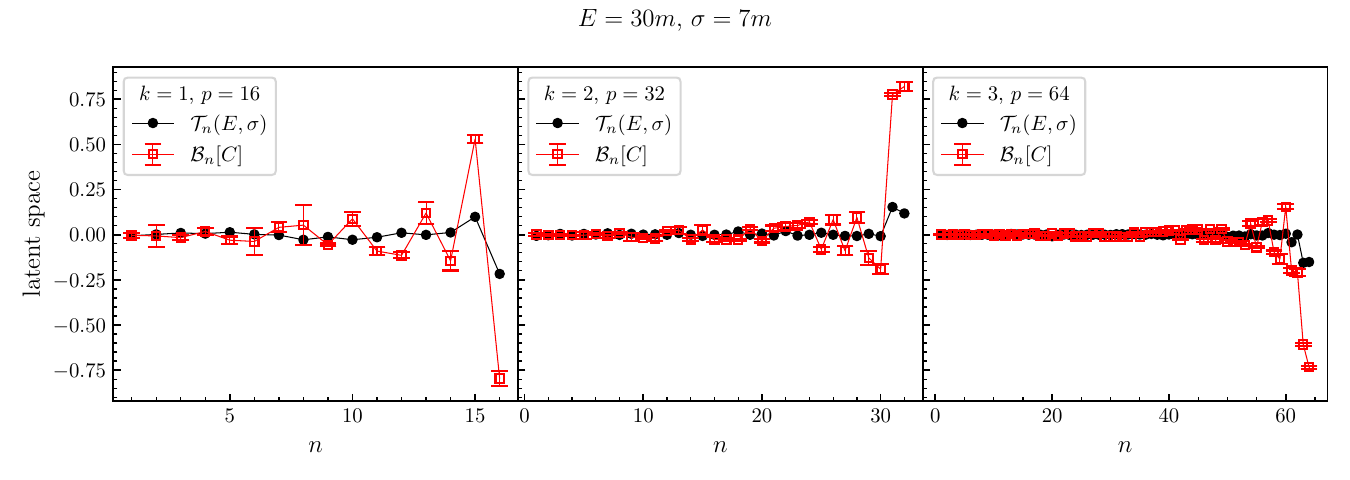}
\caption{\label{fig:latent_spaces}Branch (red points) and Trunk (black points) latent representations for $(E,\sigma)=(5m,0.5m)$ (top panels), $(E,\sigma)=(10m,2.5m)$ (middle panels), and $(E,\sigma)=(30m,7m)$ (bottom panels). In each row, we compare the trained architectures (with $\lambda=1$ and $\Omega_\mathrm{GP}^{(3)}$) labeled, according to \Cref{tab:networks}, by $k=1$, $2$, and $3$. The three architectures differ only in the dimension of the latent space, which is $p=16$, $32$, and $64$, respectively. As input to the Branch Net, we use 500 bootstrap samples of the Euclidean correlator measured on ensemble A4 (see \Cref{tab:ensembles}). The error bars represent the corresponding bootstrap uncertainties, determined from the 500 network outputs. The points are ordered according to the absolute value of the Branch Net output, from smallest to largest.}
\end{figure*}

We discuss the performance of the neural networks in \Cref{sec::systematic}, in connection with the procedure developed to estimate the systematic uncertainty. Before doing so, we present in \Cref{fig:latent_spaces} the latent representations, after training, for both the input correlator (the output of the Branch Net) and the pair $(E,\sigma)$ (the output of the Trunk Net). Different networks are shown from left to right, while different values of $(E,\sigma)\in\Omega_{E,\sigma}$ are displayed from top to bottom. As input correlator, we use $C^\mathrm{A4}$. In particular, the figure compares the outputs of the networks labeled by $k=1$, $2$, and $3$, which have the same number of hidden layers, neurons, and training samples, but differ in the dimension of the latent space. The training runs refer to $\lambda=1$ and $\Omega_\mathrm{GP}^{(3)}$. The points have been reordered according to the increasing absolute value of the Branch Net output (red points). In general, there is no way to predict a priori the behavior of the latent representations, since it depends strongly on the neural-network architecture, including the choice of activation functions. Different architectures can therefore produce substantially different latent spaces. What we observe is that both the Branch and Trunk outputs are typically oscillatory in sign and do not exhibit any obvious pattern. Their magnitude, however, remains very small and, for all the examples shown in \Cref{fig:latent_spaces}, is always below one. This is not the case with the HLT method, whose coefficients in \Cref{eq:master_equation_HLT} can become very large in magnitude. DeepONet therefore encodes the solution in a radically different way and, through its non-linearity, appears to provide an intrinsic regularization of the ill-posed nature of the problem. A second observation is that many components of the latent space are close to zero, particularly when the latent-space dimension $p$ is increased. This suggests that, at least for the task and parameter ranges considered in this work, increasing the dimension of the latent space does not lead to a noticeable improvement in the accuracy of the solution. This conclusion is also supported by the minimum validation losses reported in \Cref{tab:performances}, which are essentially identical for the three architectures. It is, of course, entirely plausible that the situation may be different for more challenging reconstruction tasks, larger domains $\Omega_{E,\sigma}$, or more general classes of training functions.

\subsection{\label{sec::systematic}Systematic errors and biases}

In this section, we explain how the ensembles of neural networks introduced in the previous section are used to make predictions on new data and to estimate the associated systematic uncertainties. It is understood that, before being provided as input to the neural network, any correlator $\vec{C}$ is preprocessed according to the procedure described in \Cref{sec::preprocessing}. Once the training has been completed, the network can provide an estimate of the smeared spectral density $\hat{\rho}_\sigma^{(k)}(E)$ at any point in the domain $\Omega_{E,\sigma}$. In order to illustrate the strategy and its performance, we restrict our attention to a discrete set of energies and smearing parameters. For the energy, predictions are computed at the following 37 values,
\begin{flalign}\label{eq:energies}
\Omega^\mathrm{pred}_E=\{4,5,6,\cdots,40\}\cdot m\,.
\end{flalign}
As for the smearing parameter, we consider the set with 22 values given by
\begin{flalign}\label{eq:sigmas}
\Omega^\mathrm{pred}_\sigma=\{0.5,& 0.6, 0.7, 0.8,0.9, 1, 1.1, 1.2, 1.3, 1.4, 1.5,\nonumber\\
&2, 2.5, 3, 3.5, 4, 4.5, 5, 5.5, 6, 6.5, 7\}\cdot m\,.
\end{flalign}
 The closely spaced values of $\sigma$ in the first row of \Cref{eq:sigmas} will be used in \Cref{sec::sigma_to_zero} to perform the $\sigma\to0$ extrapolation of  $\rho^\mathrm{O(3)}(E)$ from lattice correlators. The collective set $\Omega_{E,\sigma}^\mathrm{pred}=\Omega_E^\mathrm{pred}\times\Omega_\sigma^\mathrm{pred}$ is then a subset of 814 points in the $\Omega_{E,\sigma}$ domain.
\begin{figure}
\centering
\includegraphics[width=\linewidth]{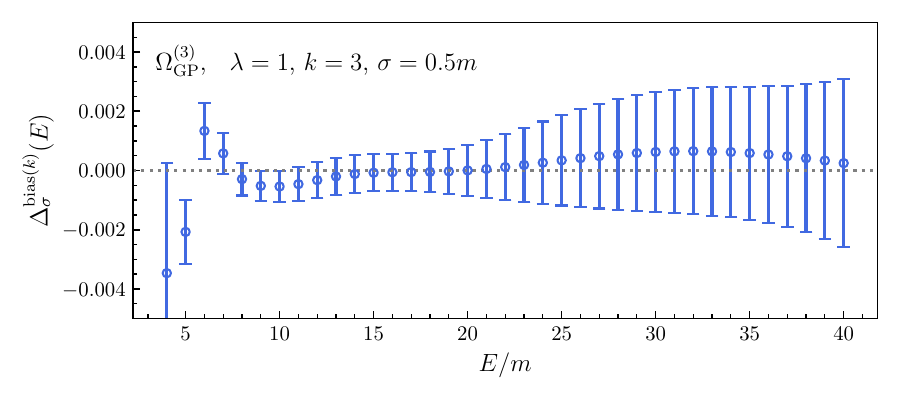}
\includegraphics[width=\linewidth]{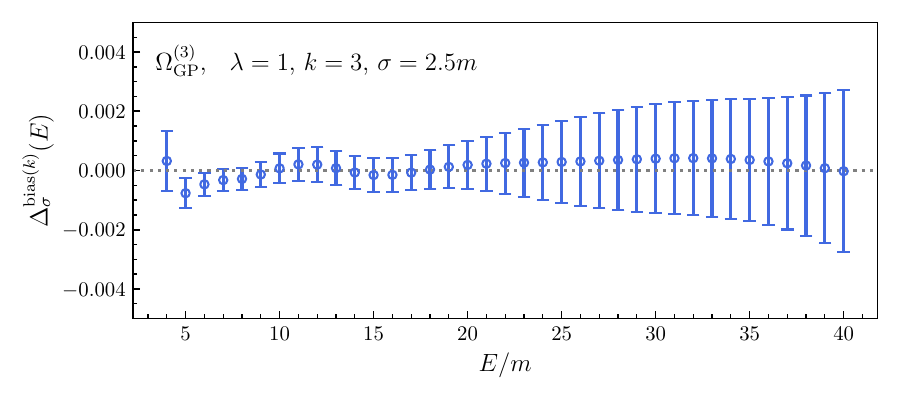}
\includegraphics[width=\linewidth]{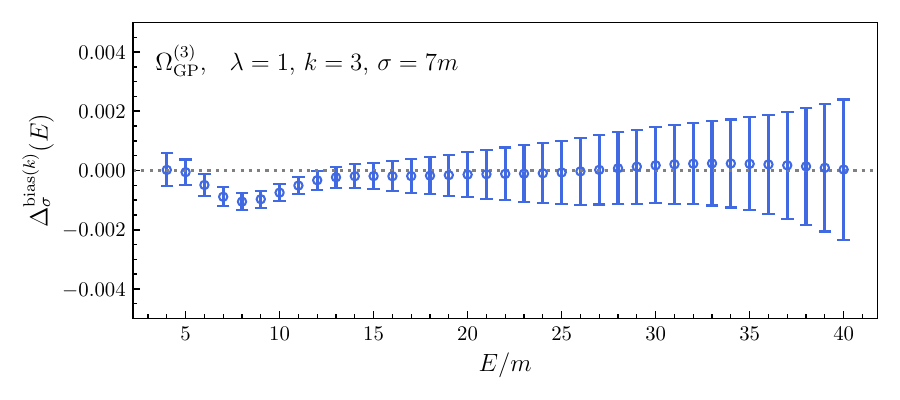}
\caption{\label{fig:bias} Global bias for the neural network with index $k=3$ in case of $\lambda=1$, $\Omega_\mathrm{GP}^{(3)}$, and, from top to bottom, for $\sigma=0.5m$, $2.5m$ and $7m$. The error bars correspond to one standard deviation calculated over $N_\mathrm{bias}=1000$ mock correlators not used during the training.}
\end{figure}

Each network in the six ensembles of \Cref{tab:performances} provides only an approximation to the target operator and, as such, its prediction cannot be exact but is inevitably affected by a systematic uncertainty. We consider two possible sources of systematic error: a global bias, independent of the specific input correlator, and an input-dependent effect. To quantify the presence of a global bias, we generate a set of $N_\mathrm{bias}=1000$ new mock samples according to the procedure described in \Cref{sec::sampling}. For consistency, biases and closure tests for the ensemble corresponding to $\Omega_\mathrm{GP}^{(i)}$ are performed by generating data from the GP with hyper-parameters $\Omega_\mathrm{GP}^{(i)}$. These samples are distinct from both the validation set and all datasets used during training. The corresponding correlators are then corrupted with noise according to \Cref{eq:noise}, using either $\lambda=1$ or $\lambda=3.5$ depending on the ensemble of networks under consideration. For each model, we collect the 1000 predictions at given $(E,\sigma)\in\Omega_{E,\sigma}^\mathrm{pred}$ and define the global bias according to
\begin{flalign}
\Delta_{\sigma}^{\mathrm{bias}(k)}(E) = \frac{1}{N_\mathrm{bias}}\sum_{n=1}^{N_\mathrm{bias}}\bigg(\rho^\mathrm{true}_{\sigma,n}(E)-\hat{\rho}^{(k)}_{\sigma,n}(E)\bigg)\,.
\end{flalign}
The associated standard deviation is computed as well in order to quantify the statistical significance of $\Delta_\sigma^\mathrm{bias}(E)$. In \Cref{fig:bias}, we show the global bias for the network with index $k=3$, $\lambda=1$, $\Omega_\mathrm{GP}^{(3)}$, and for three values of $\sigma$. As can be seen, a statistically significant bias is observed only at low energies, while no evidence for a nonzero bias is found for $E\ge10m$. Notice, however, that the magnitude of the bias is approximately two orders of magnitude smaller than the typical amplitude of the spectral functions generated from the GP and shown in \Cref{fig:function_space}. A similar behavior is observed for all the other neural networks. To avoid introducing additional notation, it is understood from this point on that the prediction of each neural network is corrected accordingly by adding the bias estimate,
\begin{flalign}
\hat{\rho}^{(k)}_{\sigma,n} \to \hat{\rho}^{(k)}_{\sigma,n} + \Delta_\sigma^\mathrm{bias(k)}(E)\,.
\end{flalign}

The second source of systematic uncertainty is more difficult to estimate, as it is associated with the intrinsic limitations of a given neural network in representing the target operator exactly. In Ref.~\cite{Buzzicotti:2023qdv}, we proposed to quantify this uncertainty by exploiting an ensemble of neural networks under the assumption that the exact solution should be recovered in the limit of infinitely large neural networks trained on infinitely large datasets. While, from a conceptual point of view, one expects the accuracy of a neural network to improve as its representational power increases, this is often not the case in practice. If the capacity of the network significantly exceeds the actual complexity of the problem, additional effort during training is required to suppress the unnecessary degrees of freedom, which can ultimately lead to a degradation of the performance. For this reason, we revisit here the strategy proposed in Ref.~\cite{Buzzicotti:2023qdv}. The basic idea is that the best network is not necessarily the largest one, but rather the one that makes the smallest error on previously unseen data. Accordingly, we assign to each network in the ensemble (at fixed $\lambda$) a weight based on the minimum value reached by the validation loss,
\begin{flalign}\label{eq:min_val}
\mathcal{L}^{(k)}_{\mathrm{val}} = \mathrm{min}\;\mathcal{L}^{(k)}_{\mathrm{val}}\big\rvert_{\mathrm{w.r.t.\;epochs}}\,.
\end{flalign}
The corresponding normalized weight is defined as
\begin{flalign}\label{eq:weight}
w^{(k)}=\frac{e^{-\alpha \mathcal{L}^{(k)}_{\mathrm{val}}}}{\sum_{k=1}^{N_\mathrm{nets}} e^{-\alpha \mathcal{L}^{(k)}_{\mathrm{val}}}}\,.
\end{flalign}
The parameter $\alpha$ is set to $10^5$ for all the network ensembles, except for $\Omega_\mathrm{GP}^{(4)}$ and $\Omega_\mathrm{GP}^{(5)}$, where it is set to $10^4$, given that the validation loss is approximately one order of magnitude larger compared to the other cases. The denominator is such that the weights within the same neural-network ensemble add up to one. The rationale behind \Cref{eq:min_val,eq:weight} is that networks achieving smaller validation losses are expected to provide more accurate approximations of the target operator and should therefore contribute more significantly to the final prediction. The quantity $\mathcal{L}^{(k)}_{\mathrm{val}}$ provides a measure of the performance of the $k$-th network on previously unseen data and therefore constitutes our primary indicator of the generalization capability and predictive accuracy of the corresponding model. The values of $\mathcal{L}^{(k)}_\mathrm{val}$ and the associated weights for each network are reported in \Cref{tab:performances}. As can be seen, at fixed training-set size $N_\rho=10^5$ and training set $\Omega_\mathrm{GP}^{(3)}$, the largest weight, corresponding to the smallest validation loss, is assigned to the network with index $k=3$ for both $\lambda=1$ and $\lambda=3.5$. Interestingly, this network is not the largest one in terms of trainable parameters, but rather occupies an intermediate position among the architectures considered. For all the neural network ensembles, the most significant improvement in performance is obtained by training the same architecture ($k=3,7,8$ and 9)  on progressively larger datasets, culminating in the best-performing model, $k=9$, which is trained on $8\times10^5$ samples. The interpretation is that, for the specific reconstruction task and parameter range considered in this work, the capacity provided by two hidden layers with 64 neurons each is already sufficient to approximate the target operator accurately. The worst performance is, in most cases, observed for models $k=14$ and $k=15$, which contain only a single hidden layer. Increasing the depth from two to three hidden layers (models with $k=10,\ldots,13$) does not lead to any significant improvement in performance.

The final prediction and its associated uncertainty are assembled as follows. Given a new correlator represented by $N_b$ replicas, corresponding for instance to different bootstrap samples, whether it is a mock correlator generated according to \Cref{eq:noise} or an actual lattice correlator, we first evaluate the statistical uncertainty at fixed $(E,\sigma)$. This is done by feeding all $N_b$ replicas to the $k$-th network and computing the bootstrap uncertainty from the corresponding $N_b$ predictions. The same set of replicas is used for each model in order to keep track of statistical correlations. In this way, we obtain the statistical uncertainty associated with the $k$-th network, denoted by $\Delta_\sigma^{\mathrm{stat}(k)}(E)$. As the final statistical uncertainty of the reconstruction, we quote the weighted average of the individual statistical uncertainties combined in quadrature,
\begin{flalign}\label{eq:ensemble_statistical}
\Delta^\mathrm{stat}_{\sigma}(E)=\sqrt{\sum_{k=1}^{N_\mathrm{nets} }w^{(k)}\Big(\Delta^{\mathrm{stat(k)}}_{\sigma}\Big)^2}\,.
\end{flalign}
Accordingly, the central value of the final prediction is obtained from the weighted average of all the predictions given by the ensemble of networks
\begin{flalign}\label{eq:ensemble_average}
\hat{\rho}_{\sigma}(E) = \sum_{k=1}^{N_\mathrm{nets}} w^{(k)}\, \hat{\rho}^{(k)}_{\sigma}(E)\;.
\end{flalign}
\Cref{eq:ensemble_average} combines the predictions of the different networks, assigning a larger weight to those associated with better performance on the validation set. Our estimate of the systematic uncertainty associated with the neural-network approximation of the target operator, which we denote by $\Delta_\sigma^\mathrm{net}(E)$, is obtained from the weighted spread of the individual predictions around the ensemble average defined in \Cref{eq:ensemble_average},
\begin{flalign}\label{eq:ensemble_systematic}
\Delta^\mathrm{net}_{\sigma}(E)=\sqrt{\sum_{k=1}^{N_\mathrm{nets}} w^{(k)}\Big(\hat{\rho}^{(k)}_{\sigma}(E)-\hat{\rho}_{\sigma}(E)\Big)^2}\,.
\end{flalign}
This uncertainty estimate automatically suppresses the impact of solutions that deviate significantly from the ensemble average while at the same time being associated with worse validation performance. Conversely, it properly accounts for fluctuations among models with comparable weights, for which the relative quality of the approximation is similar. The final uncertainty of the spectral reconstruction procedure, denoted by $\Delta_\sigma^\mathrm{rec}(E)$, is then obtained by combining the statistical and systematic contributions in quadrature,
\begin{flalign}\label{eq:ensemble_total}
\Delta^\mathrm{rec}_{\sigma}(E)=\sqrt{\big(\Delta^\mathrm{stat}_{\sigma}(E)\big)^2+\big(\Delta^\mathrm{net}_{\sigma}(E)\big)^2}\,.
\end{flalign}
The estimate of the total error is of course different from data to data and from energy (smearing parameter) to energy (smearing parameter) and the procedure has to be repeated in order to build the solution for any new correlator.

\subsection{\label{sec::closure_test}Closure test}
\begin{figure}
\centering
\includegraphics[width=\linewidth]{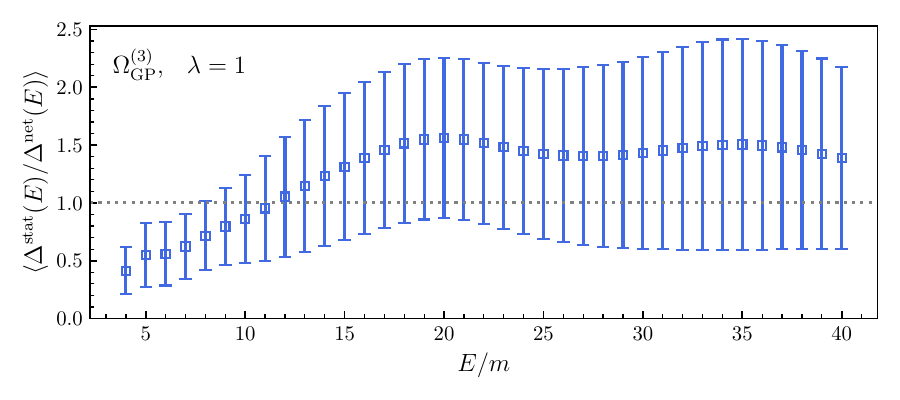}
\includegraphics[width=\linewidth]{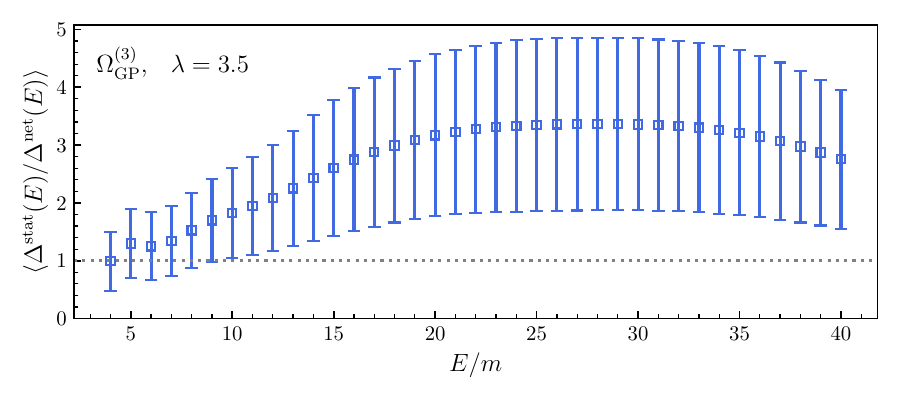}
\includegraphics[width=\linewidth]{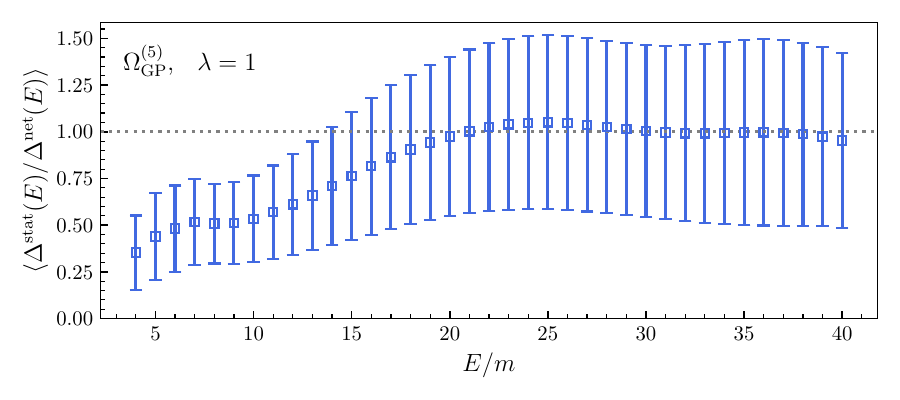}
\caption{\label{fig:ratio} Average and one standard deviation of the ratio between statistical and systematic error associated with the neural network ensemble as a function of the energy. The top panel corresponds to  $\lambda=1$ and $\Omega_\mathrm{GP}^{(3)}$, the middle panel to $\lambda=3.5$ and $\Omega_\mathrm{GP}^{(3)}$ and the bottom panel to $\lambda=1$ and $\Omega_\mathrm{GP}^{(5)}$. For each energy, the statistics are computed collectively from all reconstructions obtained from 1000 previously unseen mock data, each evaluated at the 22 values of $\Omega_\sigma^\mathrm{pred}$ listed in \cref{eq:sigmas}. }
\end{figure}
\begin{figure*}
\centering
\includegraphics[width=\linewidth]{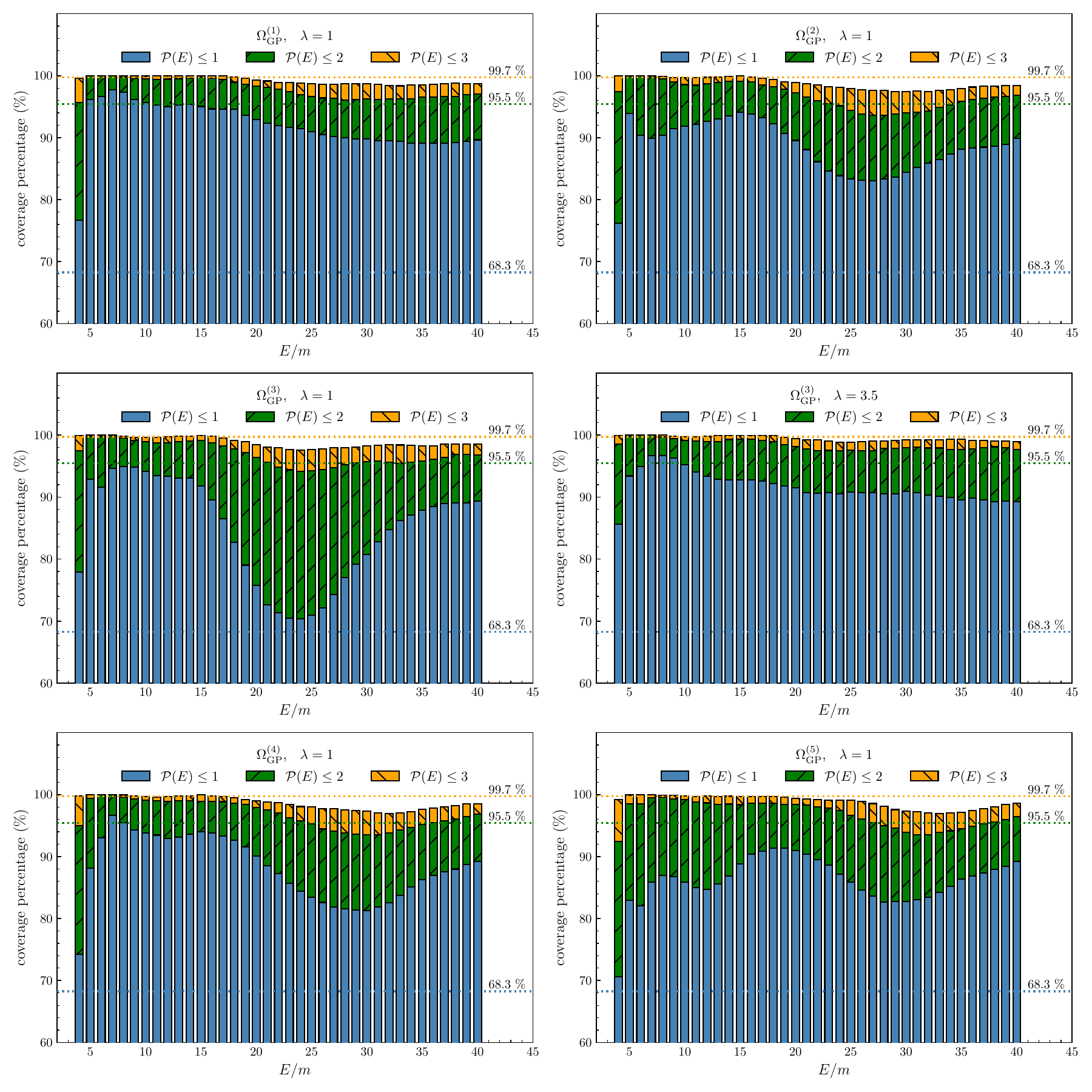}
\caption{\label{fig:pull}
Validation of the uncertainty estimate provided by the ensemble of neural networks as a function of the energy for the six combinations of $\lambda$ and $\Omega_\mathrm{GP}^{(i)}$ considered in this work. For each energy bin, the stacked bars show the fraction of mock data for which the exact result lies within one (blue), two (green), and three (orange) standard deviations of the reconstructed value. The horizontal dotted lines indicate the Gaussian expectations of 68.3\%, 95.5\%, and 99.7\%. The statistics are computed from 1000 previously unseen mock smeared spectral functions, each reconstructed at the 22 values of $\Omega^\mathrm{pred}_\sigma$ listed in \cref{eq:sigmas}.}
\end{figure*}
In order to verify that the procedure to estimate the errors described in the previous section is robust, we perform a closure test to measure the failure rate on a set of previously unseen mock data. This is a crucial step if the approach is ultimately to be employed for phenomenological applications and precision predictions. To this end, we generate 1000 new mock correlators, distinct from the training set, the validation set, and the sample used to determine the global bias. Each correlator is then corrupted with noise by generating $N_b=500$ bootstrap replicas according to \Cref{eq:noise}, randomly sampled from a set of 5000 bootstrap samples of $C^{A4}$ generated independently of those used during training. The test is performed for all the six combinations of $\lambda$ and $\Omega_\mathrm{GP}^{(i)}$ considered in this work. For each correlator, we compute the exact value $\rho_\sigma(E)$ for all the pairs in $\Omega_{E,\sigma}^\mathrm{pred}$. The correlator replicas are then fed to each model in the ensemble. For every prediction, we compute the statistical uncertainty according to \Cref{eq:ensemble_statistical}, the central value according to \Cref{eq:ensemble_average}, the systematic uncertainty according to \Cref{eq:ensemble_systematic}, and the total uncertainty according to \Cref{eq:ensemble_total}. A first interesting question is to assess the relative importance of the systematic and statistical uncertainties. To this end, at fixed energy, we collect the $1000\times22$ predictions corresponding to all mock samples and all values of $\Omega_\sigma^\mathrm{pred}$, and compute the mean and standard deviation of the ratio $\Delta^\mathrm{stat}(E)/\Delta^\mathrm{net}(E)$. The result is shown in \Cref{fig:ratio} for both $\lambda=1$ (top panel) and $\lambda=3.5$ (middle panel) in case of $\omegathree$ and for $\omegafive$ (bottom panel). As can be seen, in both cases with $\lambda=1$ the ratio is compatible with 1 over most of the energy range, indicating that the systematic uncertainty is, on average, as important as the statistical one. In the case $\lambda=3.5$, the statistical uncertainty is on average about three times larger than the systematic uncertainty, consistent with the fact that the input correlators are affected by a noise level approximately 3.5 times larger.

\begin{figure}
\centering
\includegraphics[width=\linewidth]{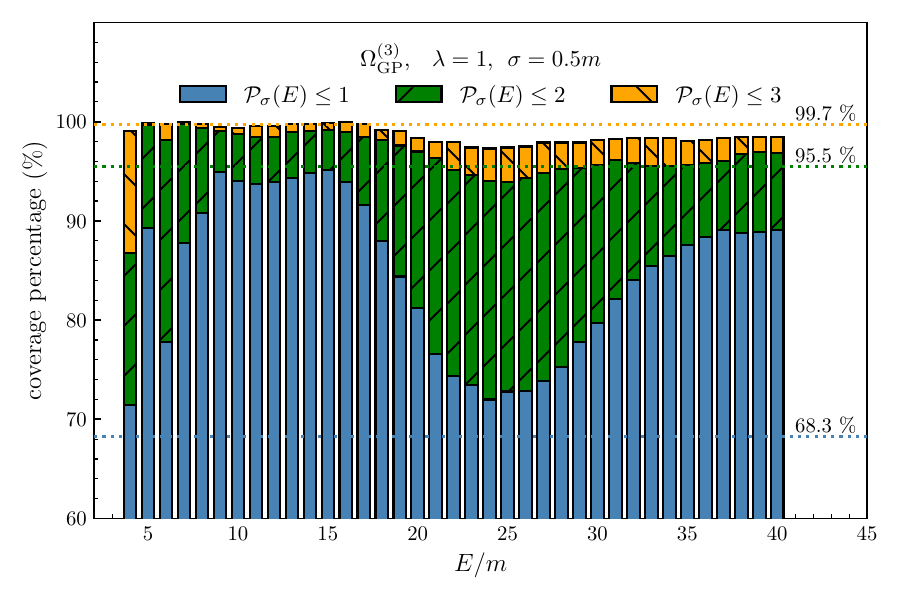}
\includegraphics[width=\linewidth]{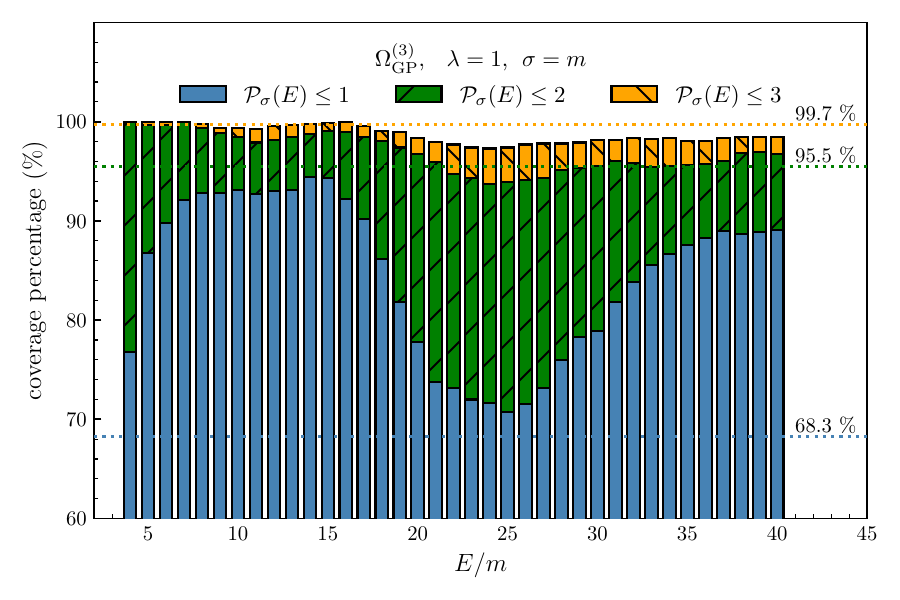}
\includegraphics[width=\linewidth]{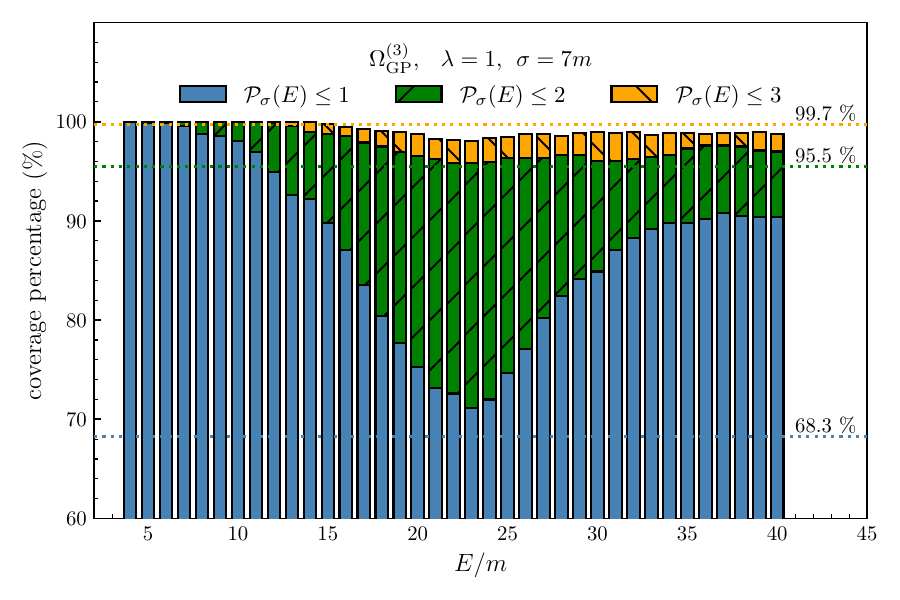}
\caption{\label{fig:pull_sigma} Coverage level of the variable $\mathcal{P}_\sigma(E)$ for $\omegathree$ and $\lambda=1$ in case of smearing parameters $\sigma=0.5m$ (top panel), $\sigma=m$ (middle panel) and $\sigma=7m$ (bottom panel). Unlike \cref{fig:pull}, each bar now contains only the 1000 values corresponding to the given $\sigma$.}
\end{figure}

To quantify and validate the procedure, we define the following pull variable 
\begin{flalign}
\mathcal{P}_{\sigma}(E)=\frac{|\hat{\rho}_{\sigma}(E)-\rho^\mathrm{true}_{\sigma,n}(E)|}{\Delta^\mathrm{rec}_{\sigma}(E)}\,,
\end{flalign}
measuring the deviation of the prediction from the true value with respect to the total error. The pull variable is calculated for all the 1000 new mock samples. At fixed energy, we then collect the resulting $1000\times22$ values of $\mathcal{P}_\sigma(E)$, we denote $\mathcal{P}(E)$ this aggregation variable, and compute the percentage of pulls lying within one, two, and three standard deviations from zero. The results of this analysis are shown in \Cref{fig:pull} for all the six neural network ensembles and compared to the expected $68.3\%$, $95.5\%$ and $99.7\%$ confidence levels. In all cases, the fraction of pulls lying within one standard deviation is always larger than the expected value of $68.3\%$ and reaches maximum values of $90\%-95\%$. Instead, the fractions of pulls within two and three standard deviations are consistent with the corresponding confidence levels expected for a Gaussian distribution. Deviations larger than three standard deviations are observed in less than $2\%$ of the cases, at intermediate/high energies. This analysis reveals that the estimate of the total uncertainty coming from the use of the ensemble of neural networks of \cref{tab:networks} is rather conservative for all the choices of $\lambda$ and of the GP hyper-parameters, indicating an overestimation of the systematic errors according to \cref{eq:ensemble_systematic}. This is presumably related to the way we determine the weights $w^{(k)}$ in \cref{eq:weight}. In particular, changing the magnitude of the parameter $\alpha$ also changes the relative magnitude of the weights across different models, suggesting that a tuning of this parameter could be performed by matching the coverage percentages to the expected Gaussian levels. As a first exploration of the method presented in this work, we decided, however, to remain on the conservative side and avoid such a technical tuning. Additionally, in \cref{fig:pull_sigma} we show the pull variable $\mathcal{P}_\sigma(E)$ at fixed values of $\sigma=0.5m$, $\sigma=m$, and $\sigma=7m$, for $\lambda=1$ and $\omegathree$. The coverage percentages of the pull variable are approximately the same across the energy range in all three cases, with a tendency to become over-conservative at low energies for the largest value of $\sigma$.

\begin{figure}
\centering
\includegraphics[width=\linewidth]{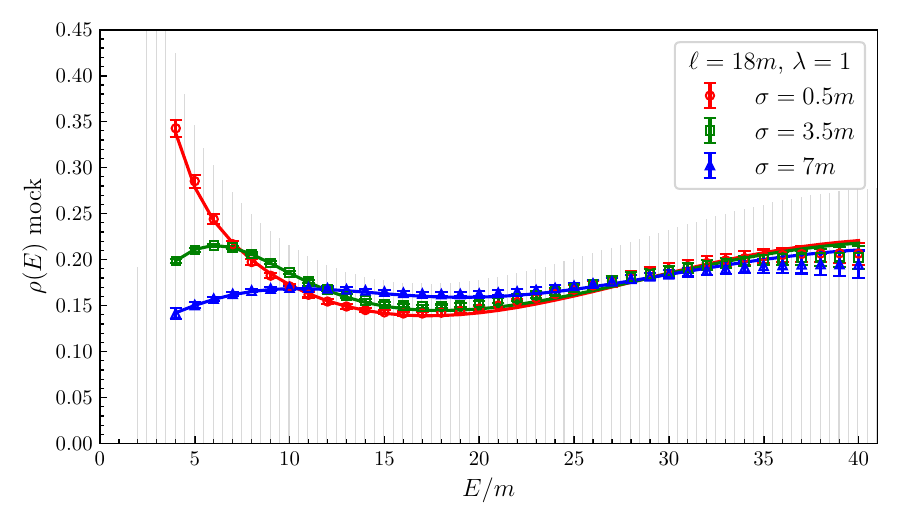}
\includegraphics[width=\linewidth]{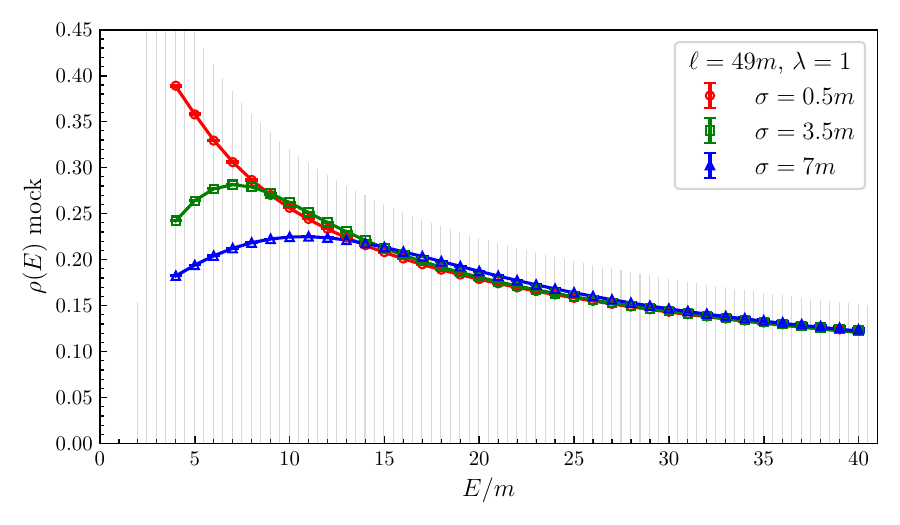}
\includegraphics[width=\linewidth]{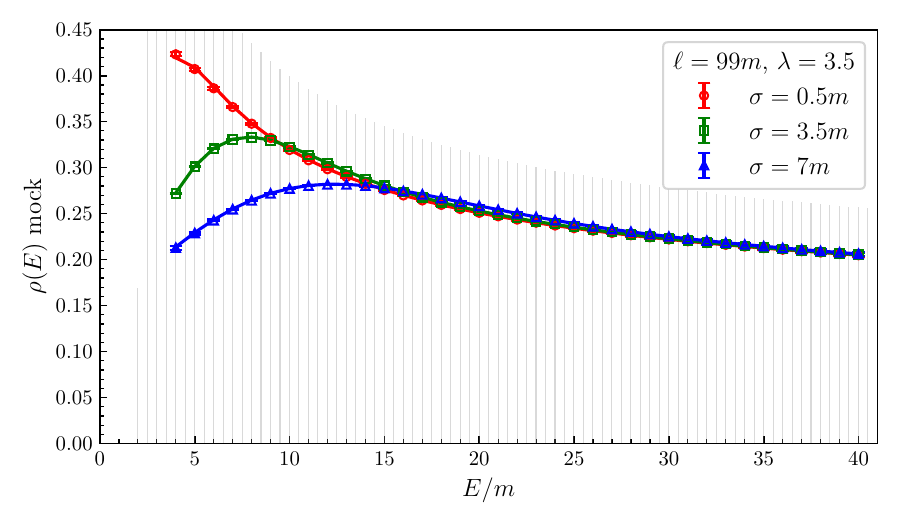}
\caption{\label{fig:mock} Examples of reconstructed smeared spectral functions for several values of $\sigma$ and for previously unseen mock data generated from $\Omega_\mathrm{GP}^{(3)}$. The error bars include both statistical and systematic uncertainties. The solid lines correspond to the true smeared spectral functions. The vertical lines are drawn in correspondence of the discrete energy spectrum and are proportional to $\rho_n$ (see \cref{sec::sampling}).}
\end{figure}

A selection of reconstructed smeared spectral densities for different values of $\sigma$, obtained from mock correlators associated with $\omegathree$, is shown in \Cref{fig:mock}, together with a comparison to the corresponding exact results. The figure includes examples generated from spectral functions with different degrees of smoothness, ranging approximately from $\ell=18m$ to $\ell=100m$, and demonstrates the overall accuracy and precision of the reconstruction procedure. Excellent agreement with the true smeared spectral densities is observed across the entire range considered.

The findings of this section provide strong validation of the proposed strategy. They demonstrate that the operator mapping a lattice Euclidean correlator to its associated smeared spectral density, defined over the continuous domain $\Omega_{E,\sigma}$, can be successfully approximated using a DeepONet architecture. Moreover, they show that the systematic uncertainty associated with this approximation can be reliably quantified through the ensemble procedure described above in presence of different levels of noise. The effectiveness of this procedure relies on the diversity of the networks entering the ensemble. More important than the total number of networks is the diversity of their expressive power. This can be achieved by varying the architecture, as we have done in this work, but also by considering different activation functions, loss functions, or other architectural choices. While there is no reason to expect the reconstruction to be reliable for spectral functions lying outside the function space represented by the GP, we show in \Cref{app:out_of_distribution} that the neural networks are in fact able to generalize, to some extent, even to out-of-distribution spectral functions. In the same appendix we also discuss the performance for out-of-domain smearing parameters. Having validated the methodology on mock data, we now turn to the reconstruction of the spectral function in a realistic setting using actual lattice correlators.

\section{\label{sec:benchmark} Results on Benchmark model}
\begin{figure}
\centering
\includegraphics[width=\linewidth]{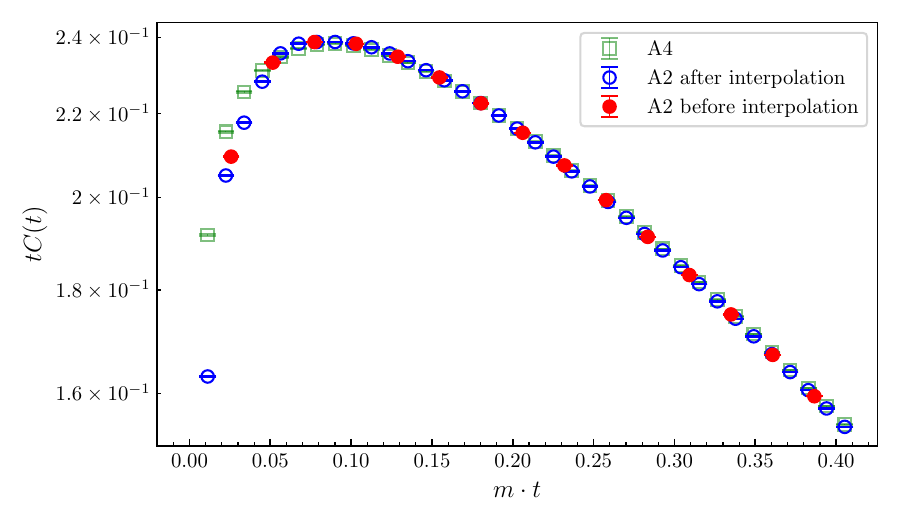}
\includegraphics[width=\linewidth]{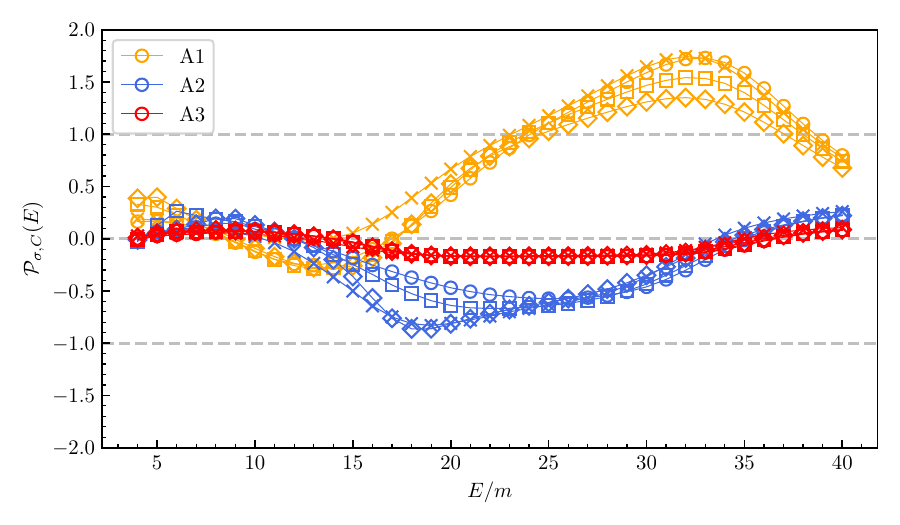}
\caption{\label{fig:interpolation} \emph{Top panel}: Comparison of the Euclidean correlation function measured on ensemble A2 before interpolation (red points) and after interpolation (blue points). The (cubic spline) interpolation is required to match the times at which the correlators in the training set have been generated. To illustrate the presence of short-distance discretization effects, we also show the correlator measured on ensemble A4. \emph{Bottom panel}: Statistical deviation of $\rhohat_\sigma(E)$, computed using a quadratic interpolation, with respect to $\rhohat_\sigma(E)$ computed using a cubic spline interpolation, for $\sigma=0.5m$ (circles), $\sigma=m$ (squares), $\sigma=2m$ (diamonds) and $\sigma=5m$ (crosses), and for ensembles A1-A3 in the case of $\omegathree$.}
\end{figure}
After validating the strategy on mock data, we finally move on to benchmarking the method using actual lattice correlators and determining the spectral density of the O(3) non-linear $\sigma$ model. The chain of operations illustrated in this section is repeated independently for the five neural network ensembles (with $\lambda=1$), which, being based on training sets obtained from GPs with different choices of the hyper-parameters, allow us to assess the impact of these choices on the final result. The neural network ensemble with $\lambda=3.5$ and $\Omega_\mathrm{GP}^{(3)}$ is used only to estimate finite-size effects. The correlators corresponding to ensembles A4, B1, and B2 can be directly provided as inputs to the neural networks, truncated at time slice 540, since they are already defined on the time grid $\Omega_t$. The correlators corresponding to ensembles A1-A3, on the other hand, must first be interpolated onto the same set of times. To obtain a smooth interpolation, we employ a standard cubic-spline ansatz using the logarithm of the correlator at the original ensemble time slices and including the information provided by $C(0)$. An example of the resulting interpolation for $C^\mathrm{A2}$ is shown in the top panel of \Cref{fig:interpolation}. Notice that this operation is performed only to match the size of the Branch Net input layer, which requires a fixed number of points. The higher density of points in case of A1-A3 ensembles after the interpolation is statistically correlated and cannot contain more physical information than what is contained in the original correlator. To check for possible systematics associated with the interpolation, since there is no unique way to perform it, we compare the reconstructed smeared spectral function with that obtained using a different way of interpolating the correlator. The value of $C(t)$ for $t\in\Omega_t$ is obtained by interpolating the logarithm of the correlator with a three-parameter quadratic fit in $m\cdot t$, with parameters determined by fitting the three values of the correlator closest to $t$. This correlator is then given as input to the ensemble of neural networks, and the reconstructed smeared spectral function is compared to that obtained using a cubic-spline interpolation. To quantify the difference, we compute the pull variable $\mathcal{P}_{\sigma,C}(E)$, representing the statistical deviation between the two reconstructions. This quantity is shown in the bottom panel of \cref{fig:interpolation} for correlators A1-A3 in the case of $\omegathree$ and for different values of $\sigma$. As can be seen, the pull variable is almost vanishing for A3 and never larger than one for A2. For ensemble A1, the coarsest one, it reaches at most 1.5 at high energies. This behavior is consistent with the fact that the systematic effect associated with the interpolation mostly affects the short-time part of the correlator, and it is expected to become more pronounced as the lattice spacing becomes coarser. Since the continuum extrapolation of $\rhohat_\sigma(E)$ (see \cref{sec::continuum_limit}) is, in most cases, compatible with the spectral function obtained on ensembles A2-A4 (with A4 not affected by any interpolation), we have decided not to assign any systematic uncertainty associated with the interpolation of the correlator. The procedure used to reconstruct the smeared spectral densities, together with their associated statistical and systematic uncertainties, is the one described in \Cref{sec::systematic}. It is applied to all points in $\Omega_{E,\sigma}^{\mathrm{pred}}$ and to each of the six lattice correlators introduced above. In the following, we indicate with $\hat{\rho}_\sigma^\mathrm{O(3)}[\mathrm{ID}]$ the reconstructed value of the smeared spectral function corresponding to the ensemble identified by ID. We now discuss the sequence of limits that must be performed in order to recover the physical spectral function and compare it with the analytic result. These are, in order, the infinite-volume limit, the continuum limit, and finally the $\sigma\to0$ limit.

\subsection{\label{sec::infinite_volume_limit}Infinite-volume limit}

When the goal is to recover the underlying spectral function, it is crucial to control Finite-Volume Effects (FVEs), since the removal of the smearing parameter can only be performed after the infinite-volume limit has been taken (see \Cref{eq:double_limit}). In the appendix of Ref.~\cite{O3}, it was shown in detail, by studying the two-particle contribution to the O(3) model, that finite-volume effects for spectral quantities smeared with kernels belonging to a specific class, which includes the Gaussian kernel considered in this work, are exponentially suppressed with the volume\footnote{See also Ref.~\cite{Bresciani:2026kjv} for a recent discussion in the context of the $R$-ratio.}. This is not always the case since many observables receive power-like finite-volume corrections. Indeed, the implementation of the formulas of Ref.~\cite{O3} shows that for the smallest value of $\sigma\in \Omega_{\sigma}^\mathrm{pred}$, we have $|\rho_{\sigma,\infty}^\mathrm{O(3)}(E)-\rho_{\sigma,L}^\mathrm{O(3)}(E)|<10^{-9}$ for all the energies in $\Omega_E^\mathrm{pred}$ and for the volume corresponding to the A4 ensemble. Even though finite-volume effects are therefore expected to be negligible, we quantify them with a data-driven approach. To this end, we compare the reconstructed smeared spectral densities $\hat{\rho}_\sigma^\mathrm{O(3)}(E)$ obtained from the A4 ensemble with those obtained from the B1 and B2 ensembles, which allow us to probe, respectively, the dependence on the spatial and temporal extent of the lattice.  We then define the two pull variables
\begin{figure}
\centering
\includegraphics[width=\linewidth]{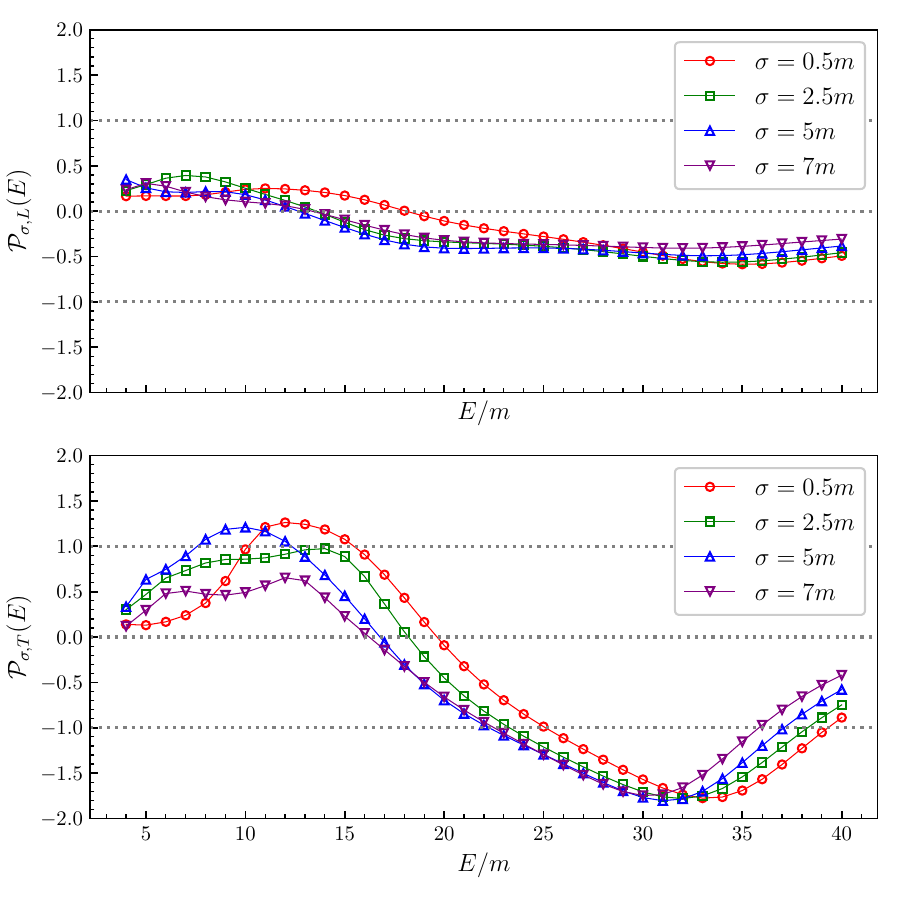}
\caption{\label{fig:pull_FVE} Pull variables for the finite-spatial (top panel) and -temporal (bottom panel) effects in case of $\Omega_\mathrm{GP}^{(3)}$. The different points refer to smearing parameters $\sigma=0.5m$ (red points), $\sigma=2.5m$ (green points), $\sigma=5m$ (blue points) and $\sigma=7m$ (purple points).}
\end{figure}
\begin{flalign}
\mathcal{P}_{\sigma,L}(E)=\frac{\hat{\rho}_\sigma^\mathrm{O(3)}(E)[\mathrm{B1}]-\hat{\rho}_\sigma^\mathrm{O(3)}(E)[\mathrm{A4}]}{\sqrt{\big(\Delta^\mathrm{rec}_\sigma(E)[\mathrm{B1}]\big)^2+\big(\Delta^\mathrm{rec}_\sigma(E)[\mathrm{A4}]\big)^2}}
\end{flalign}
and
\begin{flalign}
\mathcal{P}_{\sigma,T}(E)=\frac{\hat{\rho}_\sigma^\mathrm{O(3)}(E)[\mathrm{B2}]-\hat{\rho}_\sigma^\mathrm{O(3)}(E)[\mathrm{A4}]}{\sqrt{\big(\Delta^\mathrm{rec}_\sigma(E)[\mathrm{B2}]\big)^2+\big(\Delta^\mathrm{rec}_\sigma(E)[\mathrm{A4}]\big)^2}}\,.
\end{flalign}
The denominators are the combination in quadrature of the reconstruction errors after applying the ensemble procedure. As an estimate for a possible residual finite-spatial-extent effect we quote the quantity
\begin{flalign}
\Delta_\sigma^L(E)=\Big|\hat{\rho}_\sigma^\mathrm{O(3)}(E)[\mathrm{B1}]-\hat{\rho}_\sigma^\mathrm{O(3)}(E)[\mathrm{A4}]\Big|\mathrm{erf}\bigg(\frac{\mathcal{P}_{\sigma,L}(E)}{\sqrt{2}}\bigg)\,,
\end{flalign}
and analogously, as an estimate for the finite-temporal effect,
\begin{flalign}
\Delta_\sigma^T(E)=\Big|\hat{\rho}_\sigma^\mathrm{O(3)}(E)[\mathrm{B2}]-\hat{\rho}_\sigma^\mathrm{O(3)}(E)[\mathrm{A4}]\Big|\mathrm{erf}\bigg(\frac{\mathcal{P}_{\sigma,T}(E)}{\sqrt{2}}\bigg)\,.
\end{flalign}
The error function maps the quantities $\mathcal{P}_{\sigma,L}(E)$ and $\mathcal{P}_{\sigma,T}(E)$ into probabilistic weights. Accordingly, for $x \gg 1$, $\mathrm{erf}(x) \to 1$, while for $x \ll 1$, $\mathrm{erf}(x) \to 0$. This procedure for estimating systematic uncertainties was introduced in Ref.~\cite{Rratio} and has since become standard in our applications of spectral reconstruction. We then combine the two errors to obtain a single estimate of the FVEs,
\begin{flalign}\label{eq:FVE}
\Delta_\sigma^\mathrm{FVE}(E) = \Delta_\sigma^L(E) + \Delta_\sigma^T(E)\,.
\end{flalign}
This quantity is computed for all values of $E$ and $\sigma$ in $\Omega^\mathrm{pred}_{E,\sigma}$ and is propagated as an additional systematic uncertainty to $\hat{\rho}_\sigma^\mathrm{O(3)}(E)[\mathrm{A1}]$, $\hat{\rho}_\sigma^\mathrm{O(3)}(E)[\mathrm{A2}]$,  $\hat{\rho}_\sigma^\mathrm{O(3)}(E)[\mathrm{A3}]$, and $\hat{\rho}_\sigma^\mathrm{O(3)}(E)[\mathrm{A4}]$ by combining it in quadrature with the corresponding reconstruction uncertainties $\Delta^\mathrm{rec}_\sigma(E)$. \Cref{fig:pull_FVE} shows the quantities $\mathcal{P}_{\sigma,L}(E)$ (top panel) and $\mathcal{P}_{\sigma,T}(E)$ (bottom panel) for all energies and for a selection of values of $\sigma$ in case of $\Omega_\mathrm{GP}^{(3)}$. As can be seen, the pull variables are always smaller than two and, in most cases, smaller than one, with only a few exceptions in the energy range between $30m$ and $35m$ for $\mathcal{P}_{\sigma,T}(E)$. Such deviations are fully compatible with statistical fluctuations, especially considering that the B1 and B2 correlators are significantly noisier than A4. Similar results are also found for the reconstructions associated with different choices of the GP hyper-parameters. After including the uncertainty of \Cref{eq:FVE}, the reconstructed smeared spectral densities can be regarded as in fact extrapolated to the infinite-volume limit.

\subsection{\label{sec::continuum_limit}Continuum limit}
\begin{figure*}
\centering
\includegraphics[width=\linewidth]{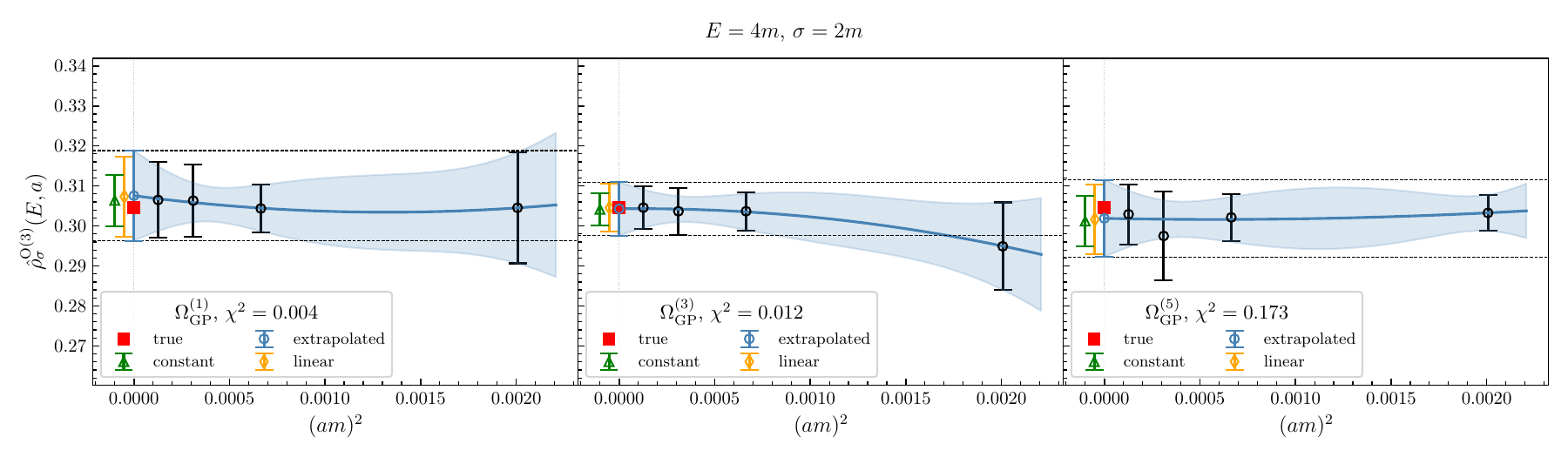}
\includegraphics[width=\linewidth]{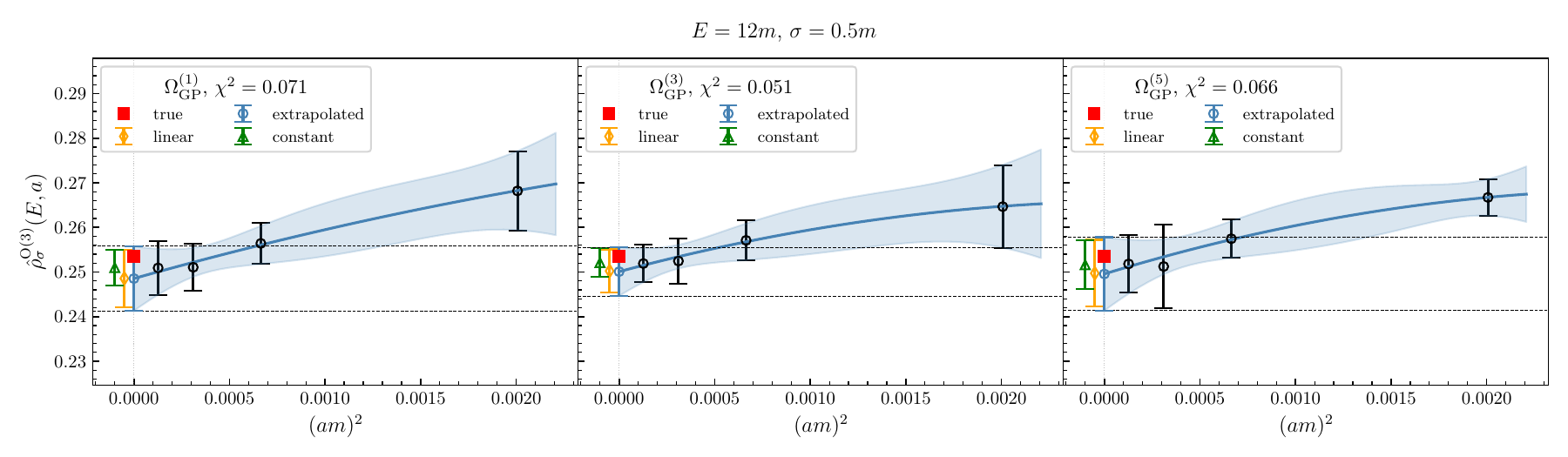}
\includegraphics[width=\linewidth]{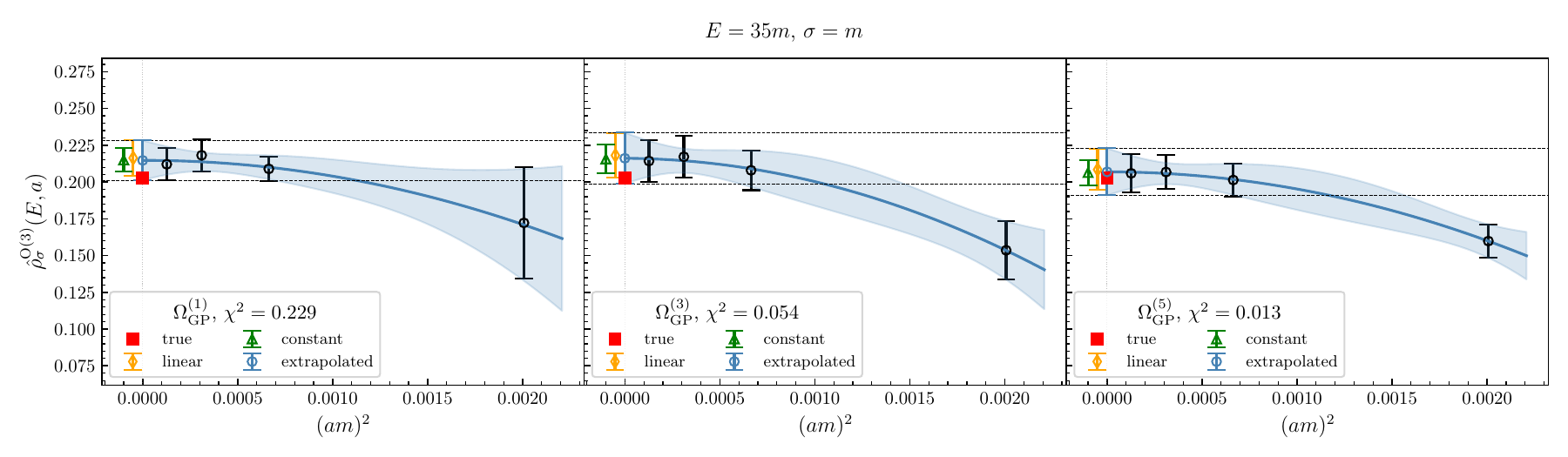}
\caption{\label{fig:continuum_limit} Continuum extrapolations for $(E,\sigma)=(4m,2m)$ (top panel), $(E,\sigma)=(12m,0.5m)$ (middle panel), and $(E,\sigma)=(35m,m)$ (bottom panel), and, from left to right, for $\omegaone$, $\omegathree$ and $\omegafive$. In all panels, the black points already include the systematics associated with the finite volume and the reconstruction. The solid line and the corresponding blue circle at $am=0$ represent our continuum extrapolation from the quadratic fit, including terms up to $(am)^4$. The $\chi^2$ (which coincides with the reduced one, since the fit admits only one degree of freedom) is written in the legend. Horizontal lines are drawn to ease the comparison between different GP hyper-parameters. The green triangle is the result of a constant fit using the two points at the finest lattice spacing. The orange diamond is the extrapolation obtained by using a linear fit through the three left-most black points. The red squares represent the expected result for $\rhohat_\sigma(E)$ obtained by applying the smearing kernel to the analytic spectral function, including the two-, four-, and six-particle contributions (see \cref{app:analytic_rho}).}
\end{figure*}
The second step towards the extraction of $\rho_\sigma^\mathrm{O(3)}(E)$ is the continuum extrapolation, which we perform using the four ensembles A1-A4. As remarked in Ref.~\cite{O3}, on the basis of Refs.~\cite{balog2009logarithmic,balog2010puzzle}, the lattice-discretized correlation function of the two-dimensional O(3) non-linear $\sigma$-model approaches the continuum slowly, with logarithmic corrections that are as sizable as the leading term $\order{a^2}$. These corrections are important, especially in the short-distance regime, when the observable to be extrapolated to the continuum is the correlator itself. Here, however, the target is the associated smeared spectral function, for which the logarithmic dependence on the lattice spacing is expected to be difficult to resolve. On the other hand, \Cref{fig:continuum_limit}, which reports a selection of continuum extrapolations corresponding to $(E,\sigma)=(4m,2m)$ (top panel), $(E,\sigma)=(12m,0.5m)$ (middle panel), and $(E,\sigma)=(35m,m)$ (bottom panel) for $\omegaone$, $\omegathree$, and $\omegafive$, shows that the reconstructed smeared spectral densities exhibit a pronounced curvature as a function of $(am)^2$ as the energy increases, meaning that the precision achieved by our reconstruction method is sufficient to resolve lattice-artifact effects. Therefore, to perform the continuum extrapolation, we consider the following reference three-parameter ansatz with terms up to $(am)^4$,
\begin{flalign}\label{eq:quadratic_fit}
\hat{\rho}_\sigma^\mathrm{O(3)}(E,am)=c_0+c_1(am)^2+c_2(am)^4\,.
\end{flalign}
The three parameters are determined by minimizing the $\chi^2$ using the four independent points corresponding to the lattice ensembles A1-A4, thus leaving one degree of freedom. The result of this fit procedure is represented by the blue curve in all panels of \Cref{fig:continuum_limit} and constitutes our final estimate for the central value and statistical error of $\hat{\rho}_\sigma^\mathrm{O(3)}(E)$ in the continuum limit. In the remainder of this subsection, we refer to this fit as \myquote{ref}. In $100\%$ of the cases, the $\chi^2$ is smaller than 0.5, highlighting a remarkable stability against fluctuations of the reconstructed smeared spectral functions obtained from statistically independent correlators. To strengthen the stability of these fits and assign a systematic error, we perform two more fits. The first is obtained by setting the coefficient $c_2$ to zero and using the three points corresponding to the ensembles A2-A4. The second uses only the coefficient $c_0$ and the points associated with the ensembles A3-A4. We refer to these fits as \myquote{lin} and \myquote{const}, respectively. The results of these fits are the orange and green points, respectively, in \cref{fig:continuum_limit}. These two extrapolations are always compatible with the one obtained from the reference fit. To estimate the systematic error, we follow a procedure similar to that used to estimate the finite-size effects and, by calling $\times\in\{\mathrm{lin},\mathrm{const}\}$, compute the quantities
\begin{flalign}
\Delta_{\sigma,a}^\times(E)=\big|\rhohat_\sigma[\times](E)-\rhohat_\sigma[\mathrm{ref}](E)\big|\erf\bigg(\frac{\mathcal{P}_{\sigma}^\times(E)}{\sqrt{2}}\bigg),
\end{flalign}
where $\mathcal{P}^\times_\sigma(E)$ is the statistical deviation between the $\times$ fit and the reference one and $\rhohat_\sigma(E)[\times]$ is the result of the continuum extrapolation obtained from the $\times$ procedure. As the systematic error associated with the continuum extrapolation, and entering the final error budget, we quote
\begin{flalign}\label{eq:syst_a}
\Delta_{\sigma,a}(E)=\max \{\Delta_{\sigma,a}^\mathrm{lin}(E),\Delta_{\sigma,a}^\mathrm{const}(E)\}.
\end{flalign}
This systematic error is already included in the extrapolated point shown in \cref{fig:continuum_limit}. As a first observation, we note that the continuum-extrapolated points associated with different choices of the GP hyper-parameters are always compatible with each other and have similar errors. This is also true for $\omegatwo$ and $\omegafour$, whose results are not shown in this section. Finally, to quantify the goodness of these extrapolations we also define the pull variable
\begin{flalign}
\mathcal{P}_{\sigma,a}(E)=\frac{\big|\hat{\rho}_\sigma^\mathrm{O(3)}(E)[\mathrm{ref}]-\hat{\rho}_\sigma^{\mathrm{O(3)}}(E)[\mathrm{A4}]\big|}{\sqrt{\big(\Delta_\sigma^\mathrm{FVE}(E)[\mathrm{A4}]\big)^2+\big(\Delta_\sigma^\mathrm{rec}(E)[\mathrm{A4}]\big)^2}}\,.
\end{flalign}
The denominator is the total error of the point at the finest lattice spacing, while the numerator is the difference between this point and the extrapolated result. The value of $\mathcal{P}_{\sigma,a}(E)$ is below 1 in $94\%$ and $86\%$ of the cases for $\omegaone$ and $\omegathree$, respectively, and always below 1.5. For all the other choices of the GP hyper-parameters, the pull variable is below 1. This is quantitative evidence that the continuum extrapolation is very well controlled with the present values of the lattice spacings.

\subsection{\label{sec::sigma_to_zero}$\sigma\to 0$ extrapolation}
\begin{figure}
\centering
\includegraphics[width=\linewidth]{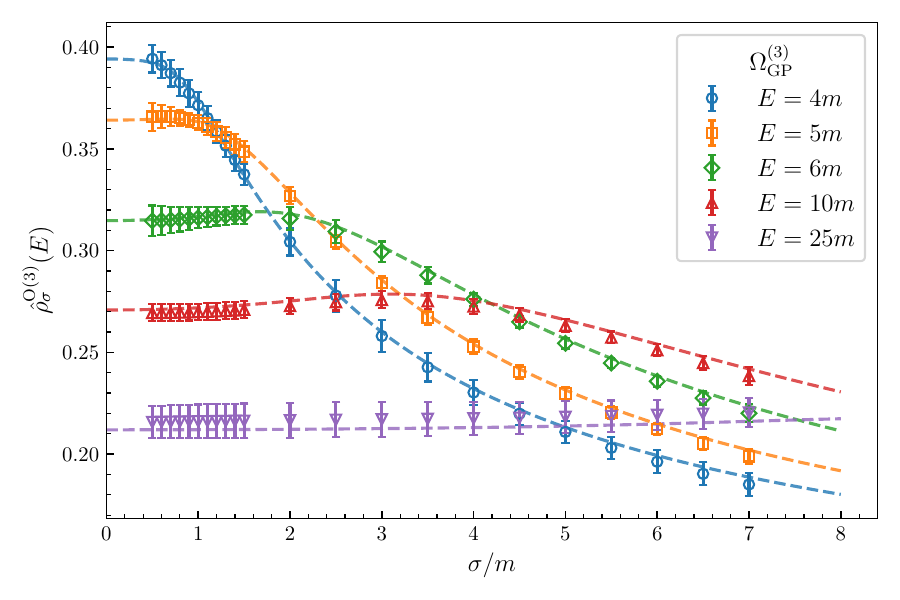}
\caption{\label{fig:sigma_trend} Examples of the reconstructed smeared spectral function, as a function of $\sigma/m$, after the continuum limit and for a selection of energies and for $\omegathree$. The dependence on $\sigma$ becomes increasingly steep as $\sigma \to 0$ and the energy approaches the threshold $2m$. By contrast, the smoothness of the underlying spectral function at high energies makes the dependence on $\sigma$ very weak in this regime. The dashed lines represent the expected result for $\rho_\sigma^\mathrm{O(3)}(E)$ (see \cref{app:analytic_rho}).}
\end{figure}

Finally, as the last step, we address the $\sigma\to0$ limit, which is required in order to recover the underlying spectral function. The removal of the smearing parameter is a delicate business, since the dependence on $\sigma$ is strongly influenced by the structure of the spectral function itself. In the presence of sharp features, such as resonances or multi-particle thresholds, controlling the extrapolation requires access to very small values of $\sigma$, whereas in energy regions where the spectral function is smooth the extrapolation is generally more affordable. This behavior is evident in \Cref{fig:sigma_trend}, where we show our determination of the reconstructed smeared spectral density $\hat{\rho}_\sigma^{\mathrm{O(3)}}(E)$, after the continuum extrapolation, for all values of $\sigma\in\Omega_\sigma^\mathrm{pred}$ and for a representative selection of energies in case of $\omegathree$. As can be seen, the dependence on $\sigma$ is very pronounced at low energies, where the spectral function rises rapidly due to the onset of the two-particle contribution. Correspondingly, the sensitivity to the smearing parameter becomes progressively weaker at higher energies, reflecting the smooth behavior of the underlying spectral density in this regime. The level of precision and control achieved by the present strategy, even at small values of $\sigma$ and across a broad range of energies, is remarkable. Such a level of precision is out of reach for the HLT method at the current level of statistical precision of the correlators. Indeed, in Ref.~\cite{O3}, the authors had to exploit the smooth high-energy behavior of the spectral function in order to perform the $\sigma\to0$ extrapolation starting from values of $\sigma$ that were significantly larger than those achieved in the present work.

\begin{figure}
\centering
\includegraphics[width=\linewidth]{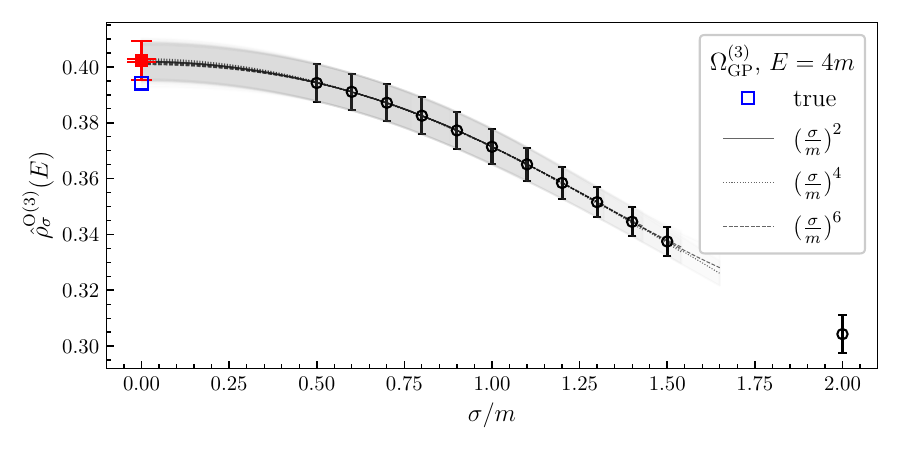}
\includegraphics[width=\linewidth]{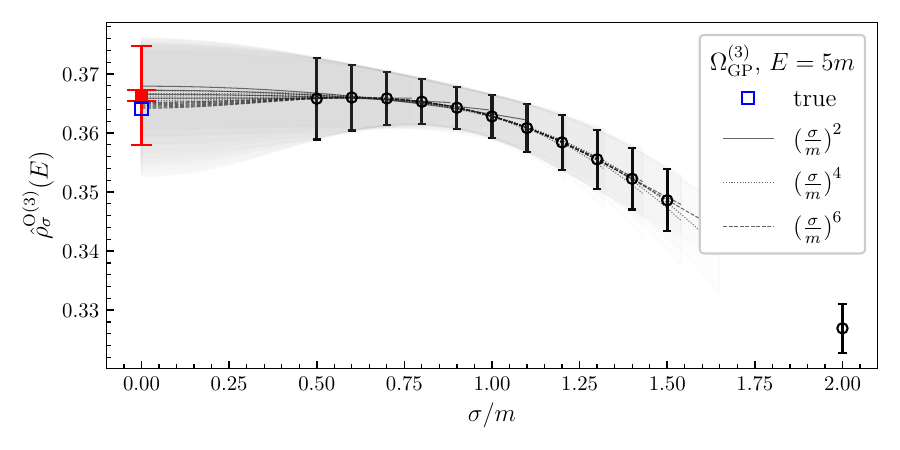}
\includegraphics[width=\linewidth]{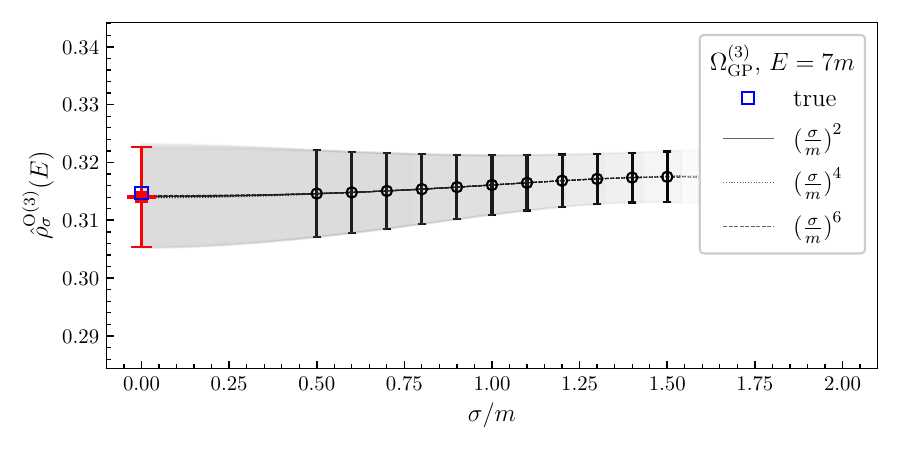}
\includegraphics[width=\linewidth]{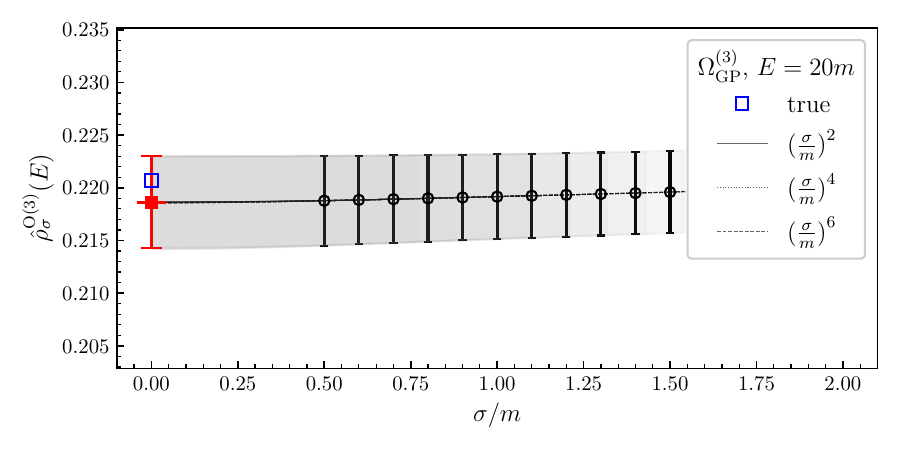}
\caption{\label{fig:sigma_extrapolations} Examples of $\sigma \to 0$ extrapolations for $E=4m$ (top panel), $E=5m$ (second panel), $E=7m$ (third panel) and $E=20m$ (last panel) for $\omegathree$. In all panels, the red point is the result at $\sigma=0$ and already includes a systematic error (represented by the largest error bar) associated with the extrapolation procedure. The solid, dashed and dotted lines are results of fits obtained by including, respectively, terms up to $(\sigma/m)^2$, $(\sigma/m)^4$ and $(\sigma/m)^6$. The blue square represents the true theoretical result for $\rho^\mathrm{O(3)}(E)$ including the two-, four- and six-particle contributions (see \cref{app:analytic_rho}).
}
\end{figure}

An extremely useful handle for controlling the $\sigma\to0$ extrapolation is provided by the asymptotic expansion of $\rho_\sigma(E)$ at small values of the smearing parameter. This result was first derived in Ref.~\cite{O3} and subsequently applied successfully to a variety of observables (see, for example, Refs.~\cite{tau1,tau2,Ds1,Bs}). For the Gaussian smearing kernel, and assuming that the underlying spectral function $\rho(E)$ is regular at energy $E$, one finds
\begin{flalign}\label{eq:asymptotic}
\rho_\sigma(E)=\rho(E)+\sum_{n=1}^{\infty} a_n\Big(\frac{\sigma}{m}\Big)^{2n}\,,
\end{flalign}
showing that the asymptotic corrections induced by the smearing depend only on even powers of the smearing parameter. This property naturally suggests the functional form of the ansatz to be used for the extrapolation. Concerning the regularity assumption on $\rho(E)$, the asymptotic expansion is, strictly speaking, not expected to hold at energies $E=2m,4m,6m,\cdots$, corresponding to the opening of multi-particle thresholds. On the other hand, the contributions beyond the two-particle sector are progressively suppressed (see \Cref{app:analytic_rho}), and the onset of each new multi-particle channel is smooth. As a result, $\rho^{\mathrm{O(3)}}(E)$ can be regarded, for all practical purposes, as a regular function at energies strictly larger than $2m$. To perform the extrapolation, we only use the reconstructed values $\hat{\rho}_\sigma^\mathrm{O(3)}(E)$ in the interval $\sigma\in[0.5m,1.5m]$. In order to obtain a robust estimate of the extrapolated value and assign a reliable systematic uncertainty, we consider three different fit ansätze motivated by \Cref{eq:asymptotic}, containing an increasing number of even powers of the smearing parameter. More specifically, the highest power included in the fit is taken to be $(\sigma/m)^2$, $(\sigma/m)^4$, and $(\sigma/m)^6$, respectively. For each fit ansatz, we perform several extrapolations by progressively extending the fit range towards larger values of $\sigma$. For the ansatz whose highest power is $(\sigma/m)^2$, we include data points up to $\sigma/m=0.7$, $0.8$, $0.9$, and $1.0$. For the ansätze whose highest powers are $(\sigma/m)^4$ and $(\sigma/m)^6$, we instead consider fit ranges extending up to $\sigma/m=1.0$, $1.1$, $1.2$, $1.3$, $1.4$, and $1.5$. In total, this procedure yields sixteen extrapolations.  The corresponding results are then combined using a Bayesian Model Average (BMA) procedure \cite{akaike1974new}. To this end, we compute the $\chi^2$ associated with each fit and assign it a weight
\begin{flalign}
\gamma_i \propto \exp\big[-(\chi_i^2+2N_i^\mathrm{params}-N_i^\mathrm{points})/2\big]\,,
\end{flalign}
where $N_i^\mathrm{params}$ and $N_i^\mathrm{points}$ denote the number of fit parameters and data points entering the $i$-th fit, respectively. The weights are subsequently normalized such that their sum equals unity. Given the extrapolated value $\hat{\rho}_i^\mathrm{O(3)}(E)$ obtained from the $i$-th fit, our final estimate of the spectral function is taken to be the weighted average
\begin{flalign}\label{eq:average_rho}
\hat{\rho}^\mathrm{O(3)}(E)=\sum_{i=1}^{16}\gamma_i\hat{\rho}_i^\mathrm{O(3)}(E)\,,
\end{flalign}
while the associated systematic uncertainty is estimated from the weighted spread of the individual extrapolations,
\begin{flalign}\label{eq:delta_sigma}
\Delta_{\sigma\to0}(E)=
\sqrt{\sum_{i=1}^{16}\gamma_i\big(\hat{\rho}_i^\mathrm{O(3)}(E)-\hat{\rho}^\mathrm{O(3)}(E)\big)^2}\,.
\end{flalign}
This uncertainty is then combined in quadrature with the statistical one, which is obtained from the bootstrap distribution associated with \Cref{eq:average_rho}. Examples of the resulting extrapolations are shown in \Cref{fig:sigma_extrapolations} in case of $\omegathree$ for several values of the energy, ranging from $E=4m$ (top panel) up to $E=20m$. Overall, the figure demonstrates a good level of control over the $\sigma\to0$ extrapolation across the entire energy range considered. The same procedure is repeated for all the five reconstructions associated with the different choices of GP hyper-parameters.


\subsection{\label{sec::final_results}Final results}

\begin{figure*}
\centering
\includegraphics[width=0.48\linewidth]{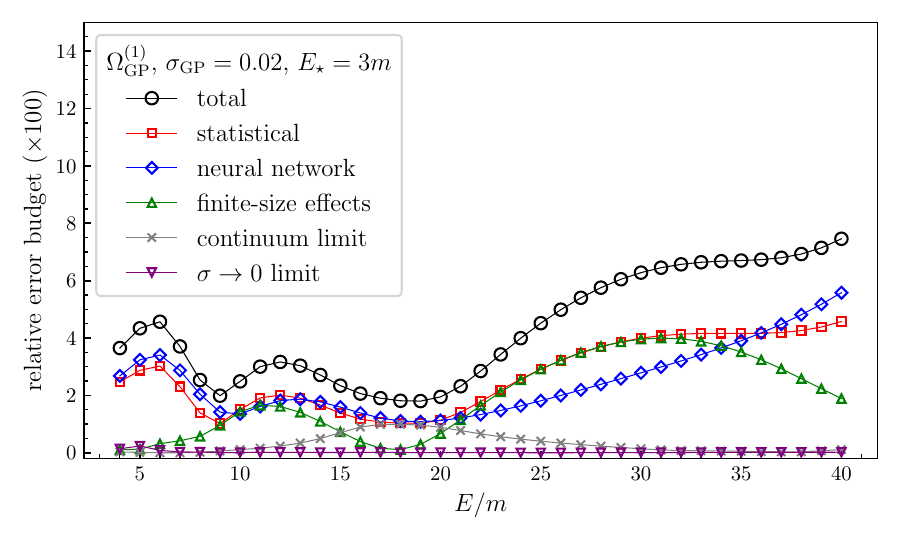}
\includegraphics[width=0.48\linewidth]{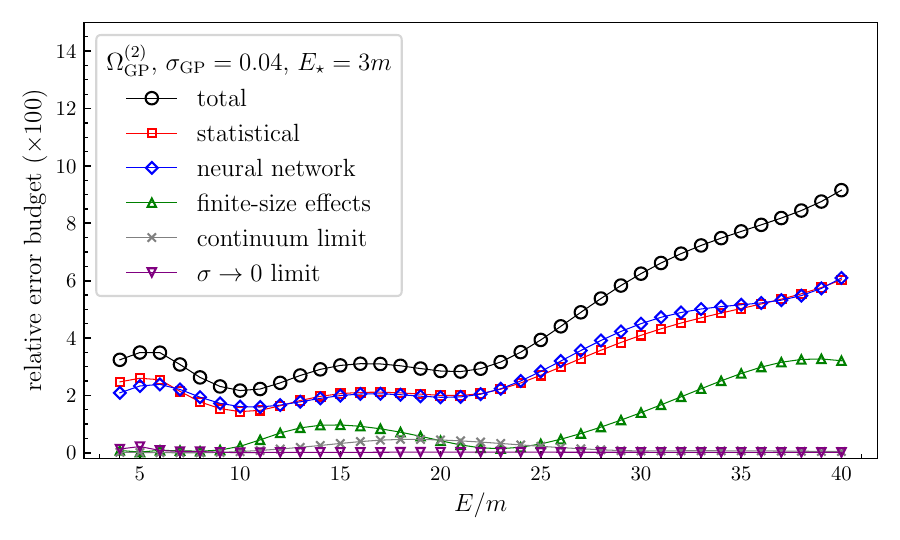}
\includegraphics[width=0.48\linewidth]{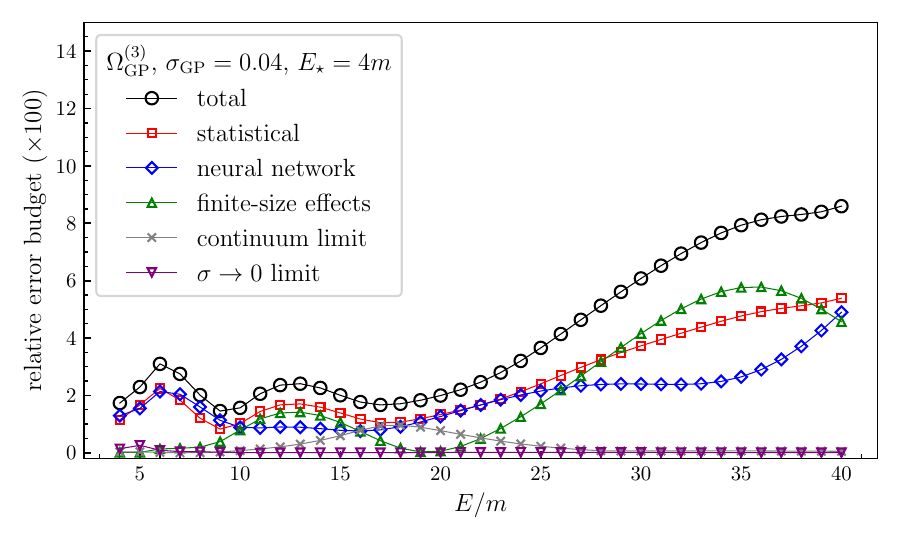}
\includegraphics[width=0.48\linewidth]{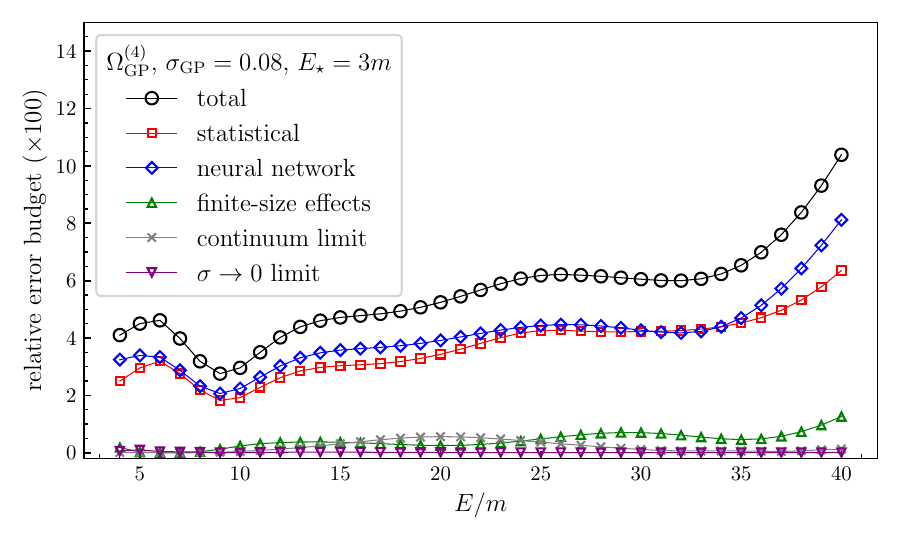}
\includegraphics[width=0.48\linewidth]{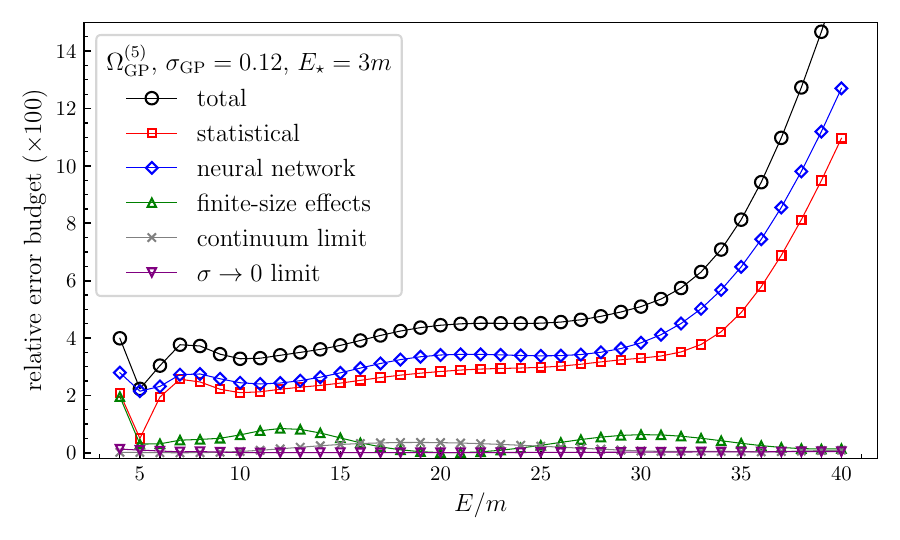}
\caption{\label{fig:error_budget} Relative error budget (multiplied by 100) of the final results obtained from the five different ensembles of neural networks produced in this work. In particular, the statistical error is in red, the systematic error derived from the ensemble of neural networks is in blue, the finite-size effect is in green, the continuum limit is in gray, and, finally, the error associated with the $\sigma \to 0$ extrapolation is in purple. All these errors reproduce the total one (black) when added in quadrature.}
\end{figure*}

In order to quantify the contribution of each source of systematic uncertainty to the total error, we construct an estimate of the error budget at fixed energy as follows. The uncertainty associated with the $\sigma\to0$ extrapolation is taken directly from \Cref{eq:delta_sigma}. For the finite-size, continuum limit and reconstruction uncertainties, we use as reference the point corresponding to the smallest value of the smearing parameter, namely the leftmost point in the panels of \Cref{fig:sigma_extrapolations}. The finite-size and continuum limit uncertainties (from \cref{eq:syst_a,eq:FVE}) are taken to be those associated with this point.  Since this point is itself obtained from the continuum-limit analysis, the reconstruction uncertainty is defined as the reconstruction error of the corresponding point at the finest lattice spacing, i.e. the leftmost point in the panels of \Cref{fig:continuum_limit}. Finally, the statistical uncertainty is determined a posteriori by requiring that all uncertainty contributions, when combined in quadrature, reproduce the total uncertainty obtained after the $\sigma\to0$ extrapolation. The resulting relative error budget is shown in \Cref{fig:error_budget} for all five reconstructions associated with the different choices of the GP hyper-parameters. In all cases, the relative total uncertainty of our final determination of $\hat{\rho}^{\mathrm{O(3)}}(E)$ is approximately $2-5\%$ for $E<20m$, increasing gradually to about $8-10\%$ at higher energies for $\Omega_\mathrm{GP}^{(1-4)}$, and up to $15\%$ for $\Omega_\mathrm{GP}^{(5)}$. This behavior is consistent with the fact that spectral reconstruction becomes progressively more challenging in the high-energy region, both because discretization effects become more pronounced and because of the intrinsic ill-conditioned nature of the inverse problem. A number of comments are in order here. From the analysis of the error budgets, it emerges that the choice of the GP hyper-parameters has the expected impact on the final result. In particular, the total error increases as the space of functions generated by the GP is progressively enlarged by relaxing the physical assumptions about the underlying spectral function. As can be seen for the cases $\omegafour$ and $\omegafive$, the increase in the error is driven by an increase in both the statistical and the systematic error estimated from the ensemble of neural networks, while the different sources of error (except for those associated with $\sigma\to0$ and $a\to0$, which are always found to be rather small) are comparable for $\Omega_\mathrm{GP}^{(1-3)}$. This pattern is entirely consistent with the intuitive expectation that, by restricting the space of possible functions, the fluctuations in the neural network predictions (both statistical and those arising from different architectures and training sets) are suppressed as well, and vice versa. However, a crucial observation is that the increase in the error when passing from $\omegaone$ to $\omegafive$ is at most a factor of 1.5-2, while the width of the GP, the parameter $\sigma_\mathrm{GP}$ (see \cref{eq:GP_parameters} or \cref{fig:error_budget}), is increased by a factor of 6. This suggests that, at least for the benchmark model considered here, the dependence of the total error on $\sigma_\mathrm{GP}$ is rather weak. Finally, the difference between $\omegatwo$ and $\omegathree$ lies only in the parameter $E_\star$, which regulates the threshold beyond which the variance of the GP is no longer suppressed. As can be seen, this parameter significantly affects the total error only at very low energy, $E=4m$ and $E=5m$, where it is about a factor of two smaller for $\omegathree$ compared to $\omegatwo$. The choice of $E_\star$ therefore has no evident impact at higher energies.

\begin{figure}
\centering
\includegraphics[width=\linewidth]{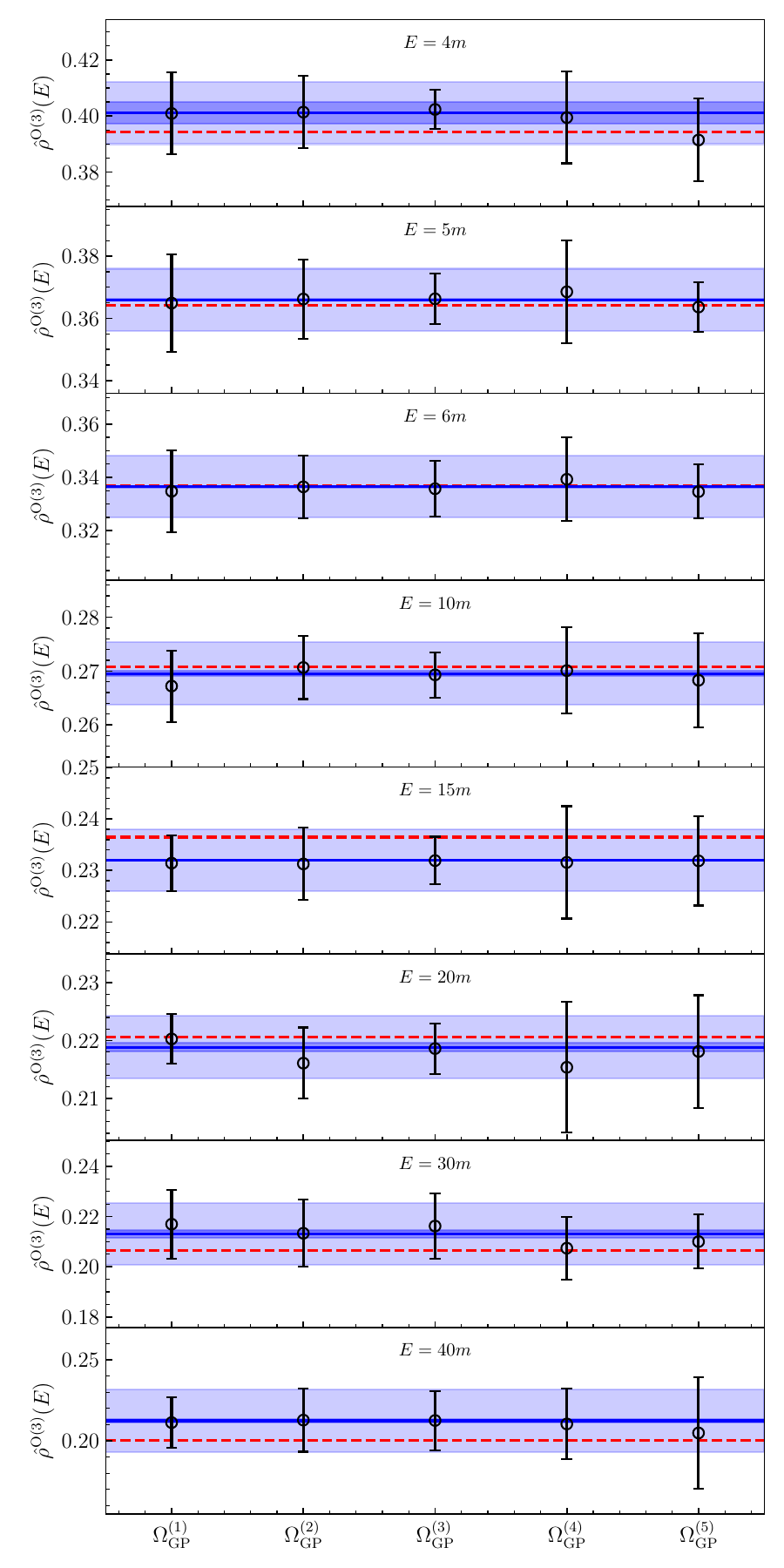}
\caption{\label{fig:GP_Comparison} Comparison of the reconstructed spectral density after $\sigma\to0$ extrapolation (black points) for different energies (from top to bottom) and for the five choices of the GP hyper-parameters $\Omega_\mathrm{GP}^{(1-5)}$ (from left to right). The horizontal blue line corresponds to our final determination of the central value from the average of the five points, with the larger band corresponding to the total error and the darker band corresponding to the systematic error, obtained as explained in the main text. The dashed red line represents the exact analytical result from the two-, four-, and six-particle contributions.}
\end{figure}

\begin{figure}
\centering
\includegraphics[width=\linewidth]{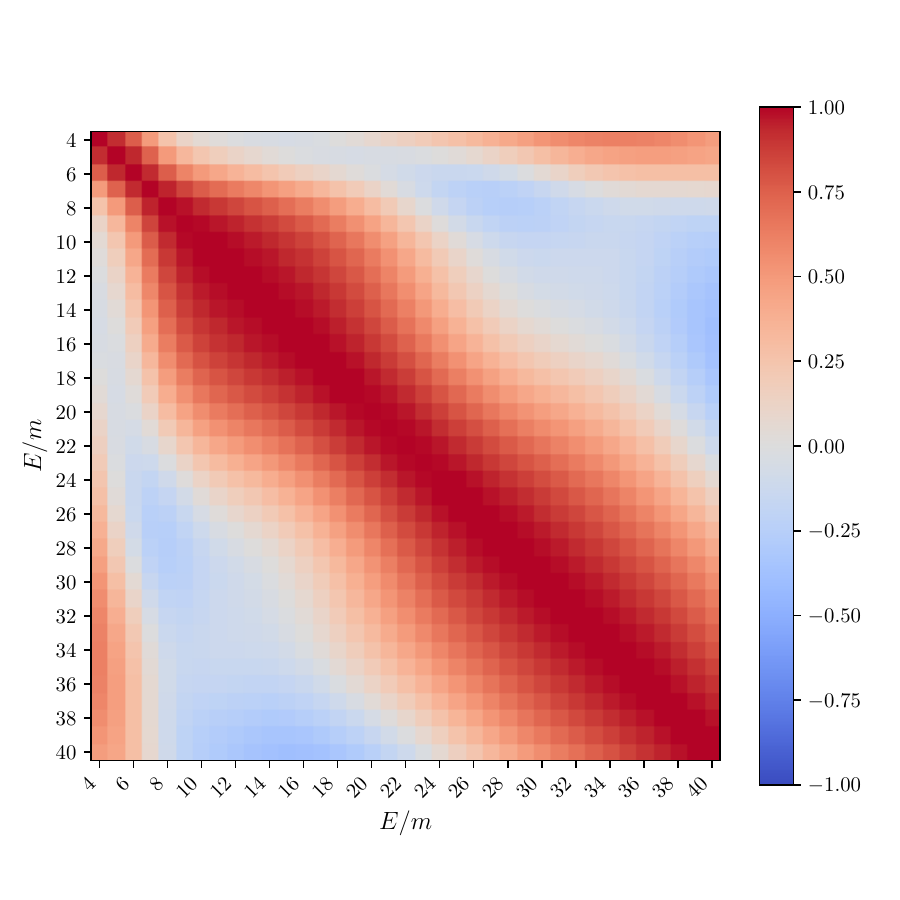}
\caption{\label{fig:correlation_matrix} Energy-energy correlation matrix of our final determination for $\rhohat(E)$ after combining the five different reconstructions.}
\end{figure}

 A complementary comparison between the five different reconstructions is shown in \cref{fig:GP_Comparison}, where the final five values of $\rhohat(E)$ corresponding to $\Omega_\mathrm{GP}^{(1-5)}$ are displayed for a selection of energies. The horizontal dashed line in the figure represents the true analytical result, as derived in \cref{app:analytic_rho}. As can be seen, the five different reconstructions are always compatible with each other and in excellent agreement with the theoretical result, with the largest deviation amounting to only $1.2$ standard deviations at $E=4m$ for $\omegathree$. The errors of the five reconstructions directly reflect the discussion above regarding the error budget. The blue band in the figure is our final determination of $\rhohat(E)$ and corresponds to a combination of the five values, obtained as follows. The central value and statistical error are obtained from the bootstrap average of the five points. The spread of the five points accounts for possible residual systematic effects induced by the prior knowledge incorporated into the training sets. To estimate it, we proceed similarly to what was done for the finite-size effects and the continuum limit, taking the maximum difference between each reconstruction and the average one, weighted by the error function of the statistical significance of such a difference. The larger band in \cref{fig:GP_Comparison} is the total error, given by the combination in quadrature of the statistical and systematic uncertainties, with the latter represented by the darker band. As can be seen, a systematic error is sizable only for $E=4m$, while it is negligible compared to the total one in all the other cases, owing to the stability of the different reconstructions. The final results are listed in \cref{tab:final_results} together with the exact values, the HLT result from Ref.~\cite{O3} and the statistical deviation of our result from the exact one. Our final determination is compatible with the true result within 0.8 standard deviations for all the energies explored in this paper. \cref{fig:correlation_matrix} instead shows a heat map of the energy-energy correlation matrix for our final determination.

\begin{table}[] 
{\setlength{\tabcolsep}{5pt}
\begin{tabular}{ccccc}
\toprule
$E/m$ \qquad& exact & DeepONet & HLT &  pull \\
\midrule
\midrule
4   &  0.394   &   0.401(10)(4)[11]  &  0.404[77]  & +0.62 \\
5   &  0.364   &   0.366(10)(0)[10]  &  0.343[46]  & +0.19 \\
6   &  0.337   &   0.337(12)(0)[12]  &  0.321[31]  & -0.04 \\
7   &  0.315   &   0.315(10)(0)[10]  &  0.321[40]  & -0.02 \\
8   &  0.297   &   0.297(8)(0)[8]    &  0.308[38]  & -0.00 \\
9   &  0.282   &   0.282(6)(0)[6]    &  0.298[42]  & -0.10 \\
10  &  0.271   &   0.270(6)(0)[6]    &  0.289[39]  & -0.22 \\
11  &  0.261   &   0.259(7)(0)[7]    &  0.282[37]  & -0.33 \\
12  &  0.253   &   0.250(7)(0)[7]    &  0.274[39]  & -0.45 \\
13  &  0.247   &   0.243(7)(0)[7]    &  0.265[40]  & -0.55 \\
14  &  0.241   &   0.237(7)(0)[7]    &  0.257[39]  & -0.65 \\
15  &  0.236   &   0.232(6)(0)[6]    &  0.250[34]  & -0.74 \\
16  &  0.232   &   0.228(5)(0)[5]    &  0.245[28]  & -0.80 \\
17  &  0.229   &   0.225(5)(0)[5]    &  0.239[23]  & -0.77 \\
18  &  0.226   &   0.223(5)(1)[5]    &  0.236[21]  & -0.66 \\
19  &  0.223   &   0.221(5)(1)[5]    &  0.235[21]  & -0.50 \\
20  &  0.221   &   0.219(5)(1)[5]    &  0.235[22]  & -0.34 \\
21  &  0.219   &   0.217(6)(1)[6]    &  0.236[24]  & -0.18 \\
22  &  0.217   &   0.216(7)(1)[7]    &  0.236[26]  & -0.05 \\
23  &  0.215   &   0.215(7)(1)[7]    &  0.236[27]  & +0.07 \\
24  &  0.213   &   0.215(8)(1)[8]    &  0.238[29]  & +0.18 \\
25  &  0.212   &   0.214(9)(1)[9]    &  0.238[32]  & +0.27 \\
26  &  0.211   &   0.214(10)(1)[10]  &  0.239[33]  & +0.34 \\
27  &  0.209   &   0.214(11)(1)[11]  &  0.239[35]  & +0.40 \\
28  &  0.208   &   0.213(11)(1)[11]  &  0.238[33]  & +0.45 \\
29  &  0.207   &   0.213(12)(1)[12]  &  0.236[32]  & +0.49 \\
30  &  0.206   &   0.213(12)(1)[12]  &  0.231[31]  & +0.53 \\
31  &  0.206   &   0.213(13)(1)[13]  &  0.226[30]  & +0.56 \\
32  &  0.205   &   0.213(13)(1)[13]  &  0.225[28]  & +0.60 \\
33  &  0.204   &   0.213(14)(1)[14]  &  0.223[26]  & +0.62 \\
34  &  0.203   &   0.213(15)(1)[15]  &  0.220[26]  & +0.65 \\
35  &  0.203   &   0.213(15)(1)[15]  &  0.217[23]  & +0.67 \\
36  &  0.202   &   0.213(16)(1)[16]  &  0.213[21]  & +0.68 \\
37  &  0.202   &   0.213(17)(1)[17]  &  0.209[17]  & +0.69 \\
38  &  0.201   &   0.213(18)(1)[18]  &  0.205[13]  & +0.69 \\
39  &  0.201   &   0.213(18)(1)[18]  &  0.201[11]  & +0.67 \\
40  &  0.200   &   0.212(19)(1)[19]  &  0.198[10]  & +0.64 \\
\bottomrule
\end{tabular}
\caption{\label{tab:final_results} Final results for the inclusive spectral function as a function of the energy. The second column is the exact analytical result, including up to six particles. The third column is our final determination, with the first error being statistical, the second systematic, and the total error in square brackets given by the combination in quadrature of the two. The fourth column contains the results obtained in Ref.~\cite{O3} using the HLT method, and only reports the total final error. The last column contains the statistical deviation of our results from the exact one.}}
\end{table}
\begin{figure*}[t]
\centering
\includegraphics[width=0.7\linewidth]{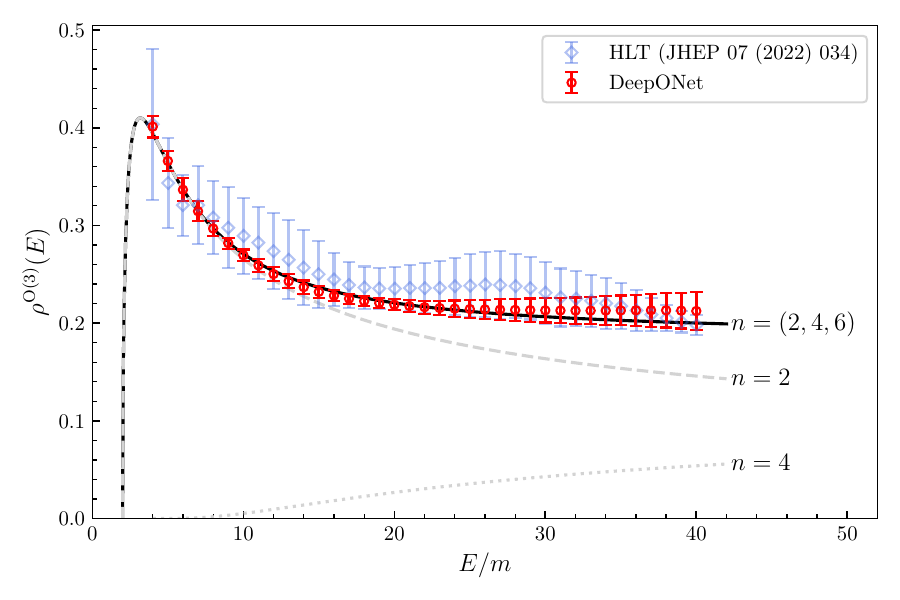}
\caption{\label{fig:final_plot} Our final result (red circles) for the spectral density of the 1+1-dimensional O(3) non-linear $\sigma$-model compared to the HLT method from Ref.~\cite{O3} (transparent blue diamonds) and to the exact two- (dashed line), four- (dotted line) and two-, four-, and six-particle contributions (solid line) to the inclusive spectral density. Our final result is obtained from the combination of five different reconstructions.}
\end{figure*}

In \Cref{fig:final_plot}, we compare our final determination of $\hat{\rho}^{\mathrm{O(3)}}(E)$ (red points) with the exact analytic prediction. The latter is shown separately for the two-particle contribution (dashed line), the four-particle contribution (dotted line), and the sum of the two-, four-, and six-particle contributions (solid line). Higher multi-particle contributions are entirely negligible on the scale of the figure. The comparison also includes the corresponding determination reported in Ref.~\cite{O3}\footnote{The central values and error bars have been extracted directly from the right panel of Fig. 9 in the published version of Ref.~\cite{O3}, since the original data of that analysis are not currently available.}, obtained using the HLT method after assessing the reconstruction systematics, the finite-size effects, the continuum limit, and the $\sigma\to0$ extrapolation. However, the analysis performed in this paper and the one performed in Ref.~\cite{O3} started from the same raw correlators (apart from the A3 ensemble, which has been regenerated for the present work), but used different bootstrap resampling, and the estimate of the systematics at each step of the analysis is also slightly different. For this reason, a correlated comparison between the two methods is not possible at present, and repeating the HLT analysis is well beyond the scope of this paper, which instead focuses on benchmarking a DeepONet-based strategy. We leave a correlated comparison, including also other spectral reconstruction methods, for a future review work. As also pointed out by the authors of Ref.~\cite{O3}, the spectral function reconstructed using the HLT method seems to exhibit a sensitivity to some structure in the spectrum for $E\gtrsim20m$. However, such a behavior is absent from the analytic prediction and is not exhibited by the DeepONet reconstruction, which, except for one ensemble, employs the same raw input data as the HLT analysis. Moreover, the spectral functions included in the training sets (see \cref{fig:function_space}) are allowed to fluctuate enough for an unpredicted contribution to the spectral function to be captured by the DeepONet. Since no such contribution is observed, we conclude that the deviation of the HLT result from the analytic one (which is, in any case, below one standard deviation) is more likely related to an artifact of the way the HLT estimator has been tuned and of the way the subsequent analysis (continuum and smearing-parameter limits) has been carried out, than to a physical spectral structure not predicted by the analytic result. Having said that, in terms of absolute errors, what immediately catches the eye is the remarkable reduction in the total uncertainty of the DeepONet determination compared to the HLT method, reaching a factor larger than five at low energies. This concludes the validation of the method for real lattice correlators and sets the stage for future applications of operator learning in four-dimensional lattice QCD with more phenomenologically relevant observables.

\section{\label{sec:conclusion} conclusions and outlooks}

In this work, we have reformulated the problem of spectral reconstruction from Euclidean correlators within the framework of Operator Learning. By exploiting DeepONet architectures, we have developed a supervised-learning strategy capable of approximating the map from Euclidean correlation functions to smeared spectral densities while providing a quantitative estimate of the associated systematic uncertainty through an ensemble of independently trained neural networks. In this work, we have chosen a Gaussian as the smearing kernel, but different kernels can easily be employed as well. Compared to our previous machine-learning approach \cite{Buzzicotti:2023qdv}, the present formulation represents a substantial conceptual and practical improvement. Rather than learning a fixed map between vectors, the network approximates the target operator itself, allowing predictions for values of the energy and smearing parameter arbitrarily chosen within the training domain. This makes the method considerably more flexible and significantly broadens its range of applicability. From a computational perspective, the proposed strategy is also considerably less expensive. Although training an ensemble of neural networks may appear computationally demanding, all the training runs performed in this work were completed within a few days using the free Google Colab platform. On standard High Performance Computing (HPC) facilities, the same workload could readily be reduced to only a few wall-clock hours.

The training set is generated by incorporating prior physical information through a novel application of Gaussian Processes in order to restrict the space of solutions, and the reconstruction has been tested under different levels of prior knowledge. The proposed strategy was first validated on previously unseen noisy mock data, corrupted with different levels of realistic noise, and subsequently applied to the reconstruction of the inclusive spectral density in the two-dimensional O(3) non-linear $\sigma$-model. After performing the infinite-volume, continuum, and $\sigma\to0$ extrapolations, and combining the results associated with different levels of prior information, the reconstructed spectral function is found to be in excellent agreement with the exact analytic result. Moreover, when compared with the current state-of-the-art HLT method of Ref.~\cite{HLT1} using the same lattice correlators \cite{O3}, the present approach achieves a substantial reduction in the total uncertainty. Even though we have not yet performed a fully model-independent study in the spirit of the discussion in \Cref{app:GP}, in a preliminary study in that direction we have found a weak dependence of the final result on the width of the GP used to generate the training set, while keeping the mean function and the degree of smoothness fixed. It is certainly possible that the assumption of smoothness plays a crucial role in the reduction of the total error. While this assumption is well motivated for the physical system considered here, it is not, in general, guaranteed to be an exploitable feature when no information on the underlying spectrum is available, and therefore the remarkable level of precision achieved in this work should not be expected to hold universally. This final consideration naturally motivates the application of the proposed strategy to more challenging and phenomenologically relevant observables, with the hadronic $R$-ratio, defined as the ratio of the $e^+e^-$ annihilation cross section into hadrons to that into muons, being an ideal next step. The $R$-ratio smeared with a Gaussian kernel has already been determined in Ref.~\cite{Rratio} using the HLT method, although a phenomenologically useful precision could only be achieved for relatively large values of the smearing parameter, $\sigma\sim4m_\pi$. An ongoing analysis based on lattice correlators with significantly improved statistical precision aims at reducing this value to $\sigma\sim2m_\pi$. It will therefore be particularly interesting to assess the improvement that can be achieved by applying the strategy developed in this work in conjunction with physics-informed training sets constructed from the well-known theoretical and phenomenological behavior of the $R$-ratio at both low and high energies.

As a further outlook, we propose to revisit the strategy used to estimate the systematic uncertainty through the use of a Gaussian Negative Log Likelihood (GNLL) loss function, which, compared to the MSE given in \Cref{eq:loss}, reads
\begin{flalign}\label{eq:GNLL}
\mathcal{L}=\frac{1}{N_\rho} \sum_{n=1}^{N_\rho}\frac{1}{2}\bigg[\frac{\big(\hat{\rho}_{\sigma,n}(E)-\rho_{\sigma,n}^\mathrm{true}(E)\big)^2}{(\Delta_{\sigma,n}^\mathrm{net}(E))^2}+\log(\Delta_{\sigma,n}^\mathrm{net}(E))^2\bigg]\,.
\end{flalign}
GNLL losses have long been known in the machine-learning literature (see, for instance, Ref.~\cite{nix1994estimating}). Compared to the standard MSE loss function, the output of the network is not just the central value $\hat{\rho}_\sigma(E)$, but rather a probability distribution, with $\Delta_{\sigma}^\mathrm{net}(E)^2$ representing the variance, which is itself predicted by the neural network. The DeepONet presented here can indeed be easily modified to output two quantities in the latent space instead of one. The quantity $\Delta_\sigma^\mathrm{net}(E)$ can then be quoted as systematic together with the statistical uncertainty. The meaning of \Cref{eq:GNLL} is quite intuitive: in order to minimize $\mathcal{L}$, the neural network can accommodate large deviations from the true result by producing large values of $\Delta_\sigma^\mathrm{net}(E)$, but at the same time it forbids arbitrarily large values of $\Delta_\sigma^\mathrm{net}(E)$ through the logarithmic term, thereby finding the optimal balance between accuracy and systematic uncertainty. The key advantage is that the predicted systematic uncertainty $\Delta_\sigma^\mathrm{net}(E)$ depends on the input correlator as well as on the choice of $(\sigma,E)$. The neural network therefore automatically optimizes this uncertainty based on the data, assigning larger errors where they are actually needed and smaller ones where the achieved accuracy is sufficient. Preliminary studies using this loss function have shown that it can be as effective as the ensemble of networks presented in this work, while offering an additional improvement in the efficiency of the machine-learning-based approach, given that only a single network has to be trained. We leave a more detailed investigation, aimed at providing a quantitative assessment of this statement, to future work.

As a final remark, this work represents the first application of DeepONet architectures to lattice field theory, and their potential is not restricted to the problem of spectral reconstruction. The strategy introduced here can naturally be exported to a broad class of problems arising in lattice field theory, with the potential to contribute to the development of next-generation lattice numerical methods based on DeepONets and Operator Learning.

\section*{Data availability}
The materials related to the implementation of the neural network, including the source code, as well as the analysis and analytical results for the benchmark model, are available from the author upon request. The lattice correlator data are available upon request from the authors of \cite{O3}.

\section*{Acknowledgements}
I am very grateful to N.~Tantalo for supporting this work throughout its development and for valuable comments on a previous version of this manuscript, which greatly improved its presentation. I warmly thank J.~Bulava, M.~T.~Hansen, and A.~Patella for granting access to the data, without which this benchmark would not have been possible, and for providing comments on a previous version of the manuscript. I also thank M.~Buzzicotti for reading a preliminary version of this work. I acknowledge enjoyable discussions with A.~C. and N.~A. on machine learning and other more speculative ideas.

\FloatBarrier

\appendix

\section{\label{app:UAT} Universal Approximation Theorem}

In this appendix, we report two Universal Approximation Theorems, as stated in Ref.~\cite{UAT}, where the corresponding proofs can also be found. These theorems provide the theoretical foundation for the success of DeepONet architectures as universal approximators. Before presenting the theorems, Ref.~\cite{UAT} identifies the class of activation functions for which the results hold. The main finding is that the activation function must be a Tauber-Wiener (TW) function, which, loosely speaking, means that it can be any continuous \textit{non-polynomial} function. The first theorem applies to linear or non-linear continuous functionals, whereas the second applies to operators, which are the objects of interest in the present work.

\emph{Theorem 1}: Suppose that $g$ is a TW function, $X$ is a Banach space, $K\subseteq X$ is a compact set, $V$ is a compact subset of $C(K)$, and $f$ is a continuous functional defined on $V$. Then, for any $\varepsilon>0$, there exist a positive integer $N$, $m$ points $x_1,\cdots,x_m\in K$, and real constants $c_i$, $\theta_i$, $\xi_{ij}$, with $i=1,\cdots,N$ and $j=1,\cdots,m$, such that
\begin{flalign}
\bigg|f(u)-\sum_{i=1}^{N}c_i g\bigg(\sum_{j=1}^{m}\xi_{ij}u(x_j)+\theta_i\bigg) \bigg| < \varepsilon
\end{flalign}
holds for all $u\in V$.

\emph{Theorem 2}: Suppose that $g$ is a TW function, $X$ is a Banach space, $K_1\subseteq X$ and $K_2\subseteq \mathbb{R}^n$ are compact sets in $X$ and $\mathbb{R}^n$, respectively, $V$ is a compact subset of $C(K_1)$, and $G$ is a non-linear continuous operator, which maps $V$ into $C(K_2)$, then, for any $\varepsilon>0$, there exist positive integers $M$, $N$, and $m$, constants $c_i^k$, $\zeta_k$, $\theta_i^k$, $\xi^k_{ij}\in \mathbb{R}$, and points $\omega_k\in\mathbb{R}^n$, $x_j\in K_1$, with $i=1,\cdots,M$, $k=1,\cdots,N$, and $j=1,\cdots,m$, such that
\begin{flalign}
\bigg|G(u)(y) -\sum_{k=1}^{N}\sum_{i=1}^{M}
c_i^k& g\bigg(\sum_{j=1}^{m}\xi_{ij}^k u(x_j)+\theta_i^k\bigg)\\
&\cdot g(\omega_k\cdot y+\zeta_k)\bigg| < \varepsilon
\end{flalign}
holds for all $u\in V$ and $y\in K_2$.

In the two statements, $C(K)$ is the Banach space of all continuous functions defined in $K$, with norm $||f||_{C(K)}=\max_{x\in K}|f(x)|$.
\section{\label{app:analytic_rho} Analytic results for the spectral function}

The O(3) non-linear $\sigma$-model is exactly integrable, meaning that the spectral density that we have used in this work as a benchmark system can be determined analytically, without relying on lattice techniques. As anticipated in \cref{eq:rho_O3}, the total inclusive spectral density can be decomposed into separate contributions associated with the propagation of multi-particle states. The $n$-multi-particle states only contain an even number of particles with an energy threshold fixed by $n\cdot m$, so that the total spectral function is 
\begin{flalign}
\rho^\mathrm{O(3)}&(E)=\theta(E-2m)\rho_2^\mathrm{O(3)}(E)\\\nonumber
&+\theta(E-4m)\rho_4^\mathrm{O(3)}(E)+\theta(E-6m)\rho_6^\mathrm{O(3)}(E)+\cdots \,.
\end{flalign}
Each contribution $\rho_n^\mathrm{O(3)}(E)$ can be expressed in terms of form factors parametrized by kinematical variables. Explicit expressions for the two-, four-, and six-particle form factors were derived in Ref.~\cite{O3BN}. The formulas required to compute the analytic spectral function are collected and summarized in the appendix of Ref.~\cite{O3}, and we do not repeat them here\footnote{Appendix A of the published version of Ref.~\cite{O3} contains at least two typographical errors which, if left uncorrected, prevent the correct reconstruction of the spectral function. Comparison with Ref.~\cite{O3BN} shows that: (i) Eq.~(A.7) is missing a square on the absolute value of the form-factor parametrization, and (ii) the second-to-last term in Eq.~(A.9) should read $-448\pi^4\tau_2$ instead of $-448\pi^2\tau_2$. The numerical implementation used in Ref.~\cite{O3} is nevertheless correct.}. The two-particle contribution, given in \cref{eq:rho2}, is the only one that can be written in closed form. The four- and six-particle contributions instead require a numerical evaluation of the phase-space integrals, which we have reimplemented in this work. \Cref{tab:analytic_rho} reports the values we obtained for the two-, four-, and six-particle contributions at energies from $2m$ to $45m$ in steps of $m$. To compute the smeared spectral function used in \cref{sec::continuum_limit,sec::sigma_to_zero}, we also generated a much finer energy grid over a significantly wider energy range and applied the smearing kernel by evaluating the convolution integral numerically using Simpson’s rule. As can be seen from \cref{tab:analytic_rho}, the six-particle contribution is approximately two orders of magnitude smaller than the sum of the two- and four-particle contributions. Contributions from higher multi-particle sectors are expected to be even more suppressed (the eight-particle contribution has never been determined) and are therefore completely negligible for the purposes of this work.

\begin{table}[]
\begin{tabular}{cccc}
\toprule
$E/m$ \qquad& $n=2$ $(\times 10)$ & $n=4$ $(\times 10^{2})$ & $n=6$ $(\times 10^{4})$ \\
\midrule
\midrule
2  &  0.0000  &  0.0000    &   0.0000 \\ 
3  &  4.0850  &  0.0000    &   0.0000 \\ 
4  &  3.9418  &  0.0000    &   0.0000 \\ 
5  &  3.6409  &  0.0008    &   0.0000 \\ 
6  &  3.3679  &  0.0173    &   0.0000 \\ 
7  &  3.1397  &  0.0750    &   0.0000 \\ 
8  &  2.9501  &  0.1823    &   0.0000 \\ 
9  &  2.7913  &  0.3328    &   0.0000 \\ 
10  &  2.6565  &  0.5157    &   0.0000 \\ 
11  &  2.5407  &  0.7205    &   0.0001 \\ 
12  &  2.4400  &  0.9386    &   0.0005 \\ 
13  &  2.3516  &  1.1636    &   0.0017 \\ 
14  &  2.2732  &  1.3907    &   0.0043 \\ 
15  &  2.2032  &  1.6167    &   0.0093 \\ 
16  &  2.1402  &  1.8391    &   0.0177 \\ 
17  &  2.0831  &  2.0566    &   0.0309 \\ 
18  &  2.0310  &  2.2681    &   0.0500 \\ 
19  &  1.9834  &  2.4730    &   0.0762 \\ 
20  &  1.9395  &  2.6710    &   0.1107 \\ 
21  &  1.8990  &  2.8620    &   0.1546 \\ 
22  &  1.8614  &  3.0460    &   0.2087 \\ 
23  &  1.8264  &  3.2231    &   0.2739 \\ 
24  &  1.7938  &  3.3935    &   0.3507 \\ 
25  &  1.7632  &  3.5573    &   0.4397 \\ 
26  &  1.7345  &  3.7149    &   0.5413 \\ 
27  &  1.7074  &  3.8665    &   0.6558 \\ 
28  &  1.6819  &  4.0123    &   0.7833 \\ 
29  &  1.6578  &  4.1526    &   0.9239 \\ 
30  &  1.6350  &  4.2877    &   1.0777 \\ 
31  &  1.6133  &  4.4177    &   1.2445 \\ 
32  &  1.5926  &  4.5430    &   1.4243 \\ 
33  &  1.5730  &  4.6637    &   1.6169 \\ 
34  &  1.5543  &  4.7801    &   1.8221 \\ 
35  &  1.5364  &  4.8924    &   2.0395 \\ 
36  &  1.5192  &  5.0008    &   2.2690 \\ 
37  &  1.5029  &  5.1055    &   2.5102 \\ 
38  &  1.4871  &  5.2065    &   2.7629 \\ 
39  &  1.4720  &  5.3042    &   3.0266 \\ 
40  &  1.4575  &  5.3987    &   3.3012 \\ 
41  &  1.4436  &  5.4901    &   3.5861 \\ 
42  &  1.4301  &  5.5786    &   3.8810 \\ 
43  &  1.4172  &  5.6643    &   4.1857 \\ 
44  &  1.4047  &  5.7473    &   4.4998 \\ 
45  &  1.3926  &  5.8277    &   4.8228 \\ 
\bottomrule
\end{tabular}
\caption{\label{tab:analytic_rho} Analytic results for the $n=2$, $4$, and $6$-particle contributions to the spectral function in the $1+1$-dimensional O(3) non-linear $\sigma$ model. Note the suppression factors $10^{-1}$, $10^{-2}$, and $10^{-4}$ by which the corresponding numbers must be multiplied. These numbers have been used to produce \cref{fig:final_plot}.}
\end{table}

\section{\label{app:GP} Discussion on Gaussian Processes  and model independence}

In this appendix, we briefly discuss how model independence can be achieved when Gaussian Processes are employed to generate the training set, or, equivalently, how the dependence of the reconstruction on the specific prior used to generate the training data can be systematically assessed. The possibility of constructing a genuinely model-independent machine-learning strategy for spectral reconstruction was investigated in our previous work \cite{Buzzicotti:2023qdv}. There, instead of generating spectral functions by sampling from a probability distribution, the training set was constructed by expanding generic spectral functions on a functional basis. In particular, we employed Chebyshev polynomials and represented a generic spectral function as
\begin{flalign}\label{eq:chebyshev}
\rho(E)=\sum_{n=0}^{N_\mathrm{B}}c_n B_n(E)\,.
\end{flalign}
Here, $B_n(E)=T_n\big(x(E)\big)$, where $T_n$ denotes the Chebyshev polynomial of degree $n$ and $x(E)$ is a suitable mapping from the energy interval of interest onto the domain $[-1,1]$, on which the Chebyshev polynomials are defined. The coefficients $c_n$ are then generated randomly. The degree of model independence is controlled by the dimension of the basis, $N_\mathrm{B}$. Small values of $N_\mathrm{B}$ only allow the generation of very smooth spectral functions, whereas in the limit $N_\mathrm{B}\rightarrow\infty$ any sufficiently regular function can be represented. By generating several training sets with progressively increasing values of $N_\mathrm{B}$, it is therefore possible to quantify the dependence of the reconstruction on the complexity of the training set and, ultimately, to study numerically the $N_\mathrm{B}\rightarrow\infty$ limit. Remarkably, the resulting strategy was found to generalize successfully also to spectral functions containing structures, such as resonances, that are not explicitly encoded in the expansion of \Cref{eq:chebyshev}.

Within the framework of Gaussian Processes (GPs), spectral functions are no longer represented through a functional basis but are instead sampled from a Gaussian probability distribution\footnote{Although the probability distribution of a GP is Gaussian, each sample can be mapped to a uniform distribution through the transformation $\rho(E)\mapsto\Phi\big[(\rho(E)-\mu(E))/\sigma_{\mathrm{GP}}\big]$, where $\Phi$ denotes the cumulative distribution function of a standard Gaussian. This transformation leaves unchanged the correlations encoded in the covariance kernel.} defined over the space of functions,
\begin{flalign}
\rho(E)\sim     \mathcal{GP}\big(\mu(E),\mathcal{K}(E,E')\big)\,.
\end{flalign}
The function $\mu(E)$ specifies the mean of the distribution, while the covariance kernel $\mathcal{K}(E,E')$ determines both the local variance of the sampled functions and the correlations between their values at different energies. To simplify the discussion, let us consider the stationary covariance kernel\footnote{This covariance function is commonly referred to as the \textit{squared-exponential kernel} and induces sample functions that are $C^\infty$. Many other covariance kernels have been proposed. Of particular interest is the \textit{Matérn class}, which allows one to control the degree of differentiability of the sampled functions (see, for instance, Refs.~\cite{valentine2020gaussian,williams2006gaussian}).}
\begin{flalign}
\mathcal{K}(E,E')=\sigma^2_\mathrm{GP}\exp\left(-\frac{(E-E')^2}{2\ell^2}\right) \;.
\end{flalign}
Usually $\sigma_\mathrm{GP}$, $\ell$ and $\mu(E)$ are referred to as \textit{hyper-parameters}. The parameter $\sigma_{\mathrm{GP}}$ controls the local variance of the distribution, since $\mathcal{K}(E,E)=\sigma_{\mathrm{GP}}^2$, while the length scale $\ell$ determines how rapidly correlations decay with the energy separation. For example, two points separated by a distance $\ell$ have a correlation coefficient $\mathcal{K}(E,E+\ell)/\sigma_{\mathrm{GP}}^2=e^{-0.5}\simeq0.61$, whereas at a separation of $2\ell$ the correlation drops to $\mathcal{K}(E,E+2\ell)/\sigma_{\mathrm{GP}}^2=e^{-2}\simeq0.14$, so that the two points can be regarded as almost uncorrelated. The correlation length then determines the degree of smoothness of the spectral functions, as $N_\mathrm{B}$ did in the previous approach.

Within this probabilistic framework, the notion of model independence is recovered in the limit in which the probability distribution samples the entire space of spectral functions with any degree of smoothness. In practice, a fully model-independent strategy can be approached by choosing a trivial mean function, e.g. $\mu(E)=0$, together with a sufficiently large value of $\sigma_{\mathrm{GP}}$, so that the generated samples span a broad range of amplitudes. A numerical example of this study is provided in \cref{fig:GP_Comparison}, where we compare the reconstruction of the spectral density by gradually increasing the parameter $\sigma_\mathrm{GP}$ at fixed mean function. The remaining hyper-parameter controlling the complexity of the sampled functions is the correlation length $\ell$. A systematic strategy therefore consists in generating a sequence of training sets corresponding to progressively smaller values of $\ell$, together with the associated ensembles of neural networks required to estimate the reconstruction uncertainty. For a given input correlator, the reconstructed spectral densities obtained from the different training sets can then be compared as a function of the value of $\ell$ employed during training. Once the reconstructed solution becomes stable within the estimated uncertainties under further reductions of $\ell$, one may conclude that the reconstruction no longer depends on the specific choice of GP hyper-parameters and can therefore be regarded, for all practical purposes, as model independent. A similar procedure can be followed when generating the training set from prior knowledge by prescribing the functional form of $\mu(E)$ together with a suitable value of $\sigma_{\mathrm{GP}}$. Generating multiple training sets while progressively decreasing the correlation length $\ell$ and/or increasing $\sigma_{\mathrm{GP}}$ makes it possible to systematically assess and quantify the sensitivity of the predictions to the GP hyper-parameters. In both cases, however, an important simplification can be exploited. The target output is not the spectral function itself, but rather its smeared counterpart. Consequently, an additional degree of smoothness is already imposed by the smearing width $\sigma$. Since the smearing operation effectively washes out fluctuations occurring on length scales smaller than $\sigma$, it is reasonable to assume that a correlation length $\ell\simeq\sigma$ is sufficient to capture all the relevant features of the target function. On this basis, the threshold for declaring model independence is expected to be reached quite fast for large values of the smearing parameter, whereas the problem becomes increasingly challenging as $\sigma$ is decreased.

Numerical evidence supporting the feasibility of this approach was presented in Ref.~\cite{Buzzicotti:2023qdv}. Since the arguments discussed here do not introduce any additional conceptual difficulties, the same conclusions are expected to hold within the framework proposed in the present work. Nevertheless, a dedicated numerical verification remains an essential task to be addressed in future work. Such a validation would require substantially greater computational resources than those employed in the present study. Indeed, the increased complexity and information content of the training sets would likely require neural networks with a larger representational capacity, i.e., models with a larger number of trainable parameters. In this regard, access to HPC resources would make this program considerably more affordable within reasonable time scales.

\bigskip
\bigskip

\section{\label{app:out_of_distribution} Performance for out-of-distribution functions and out-of-domain parameters}
\begin{figure}
\centering
\includegraphics[width=\linewidth]{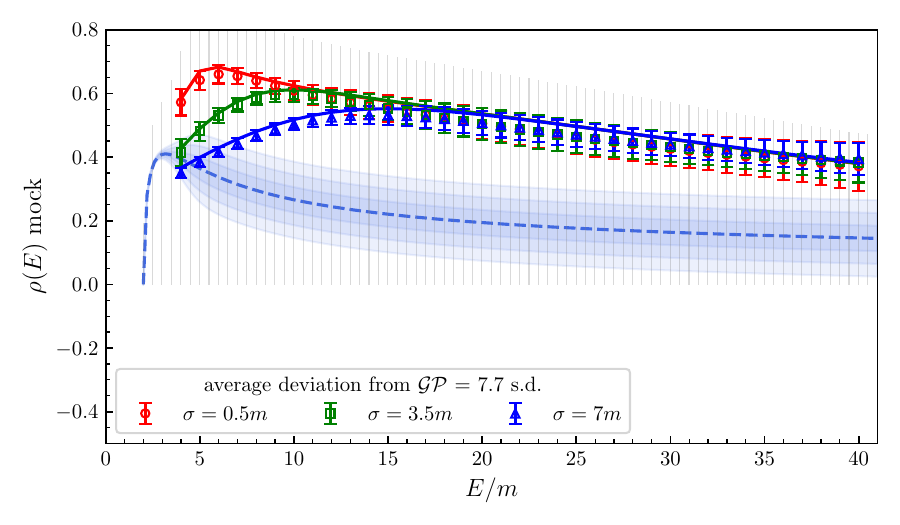}
\includegraphics[width=\linewidth]{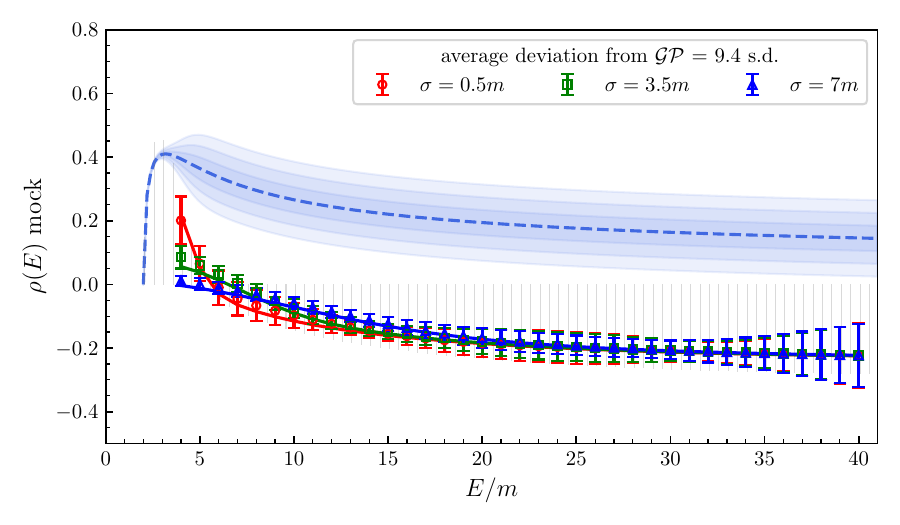}
\caption{\label{fig:out_of_distribution} Examples of reconstructions for spectral functions lying outside the distribution given by the GP ($\omegathree$) used to generate the training set, shown for different values of the smearing parameter. The blue dashed line represents the mean of the GP, while the blue bands indicate the one-, two-, and three-standard-deviation uncertainty regions. In the top panel, the unsmeared spectral function lies, on average, nearly 8 standard deviations (s.d.) above the GP mean. In the bottom panel, by contrast, it lies more than 9 s.d. below the GP mean and even receives negative contributions.}
\end{figure}
\begin{figure}
\centering
\includegraphics[width=\linewidth]{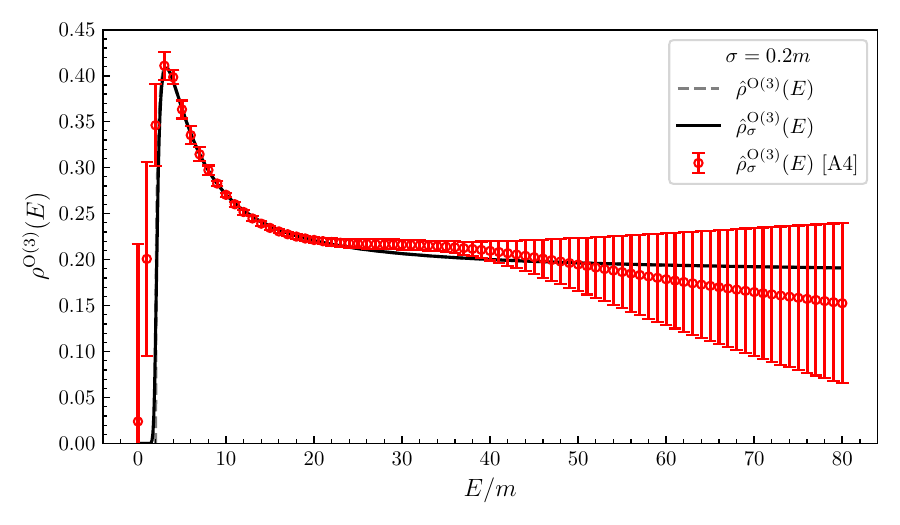}
\includegraphics[width=\linewidth]{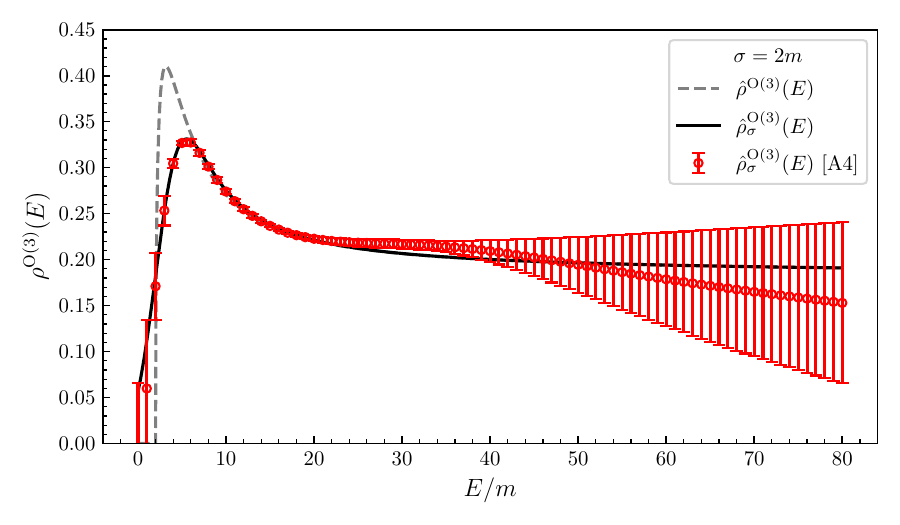}
\includegraphics[width=\linewidth]{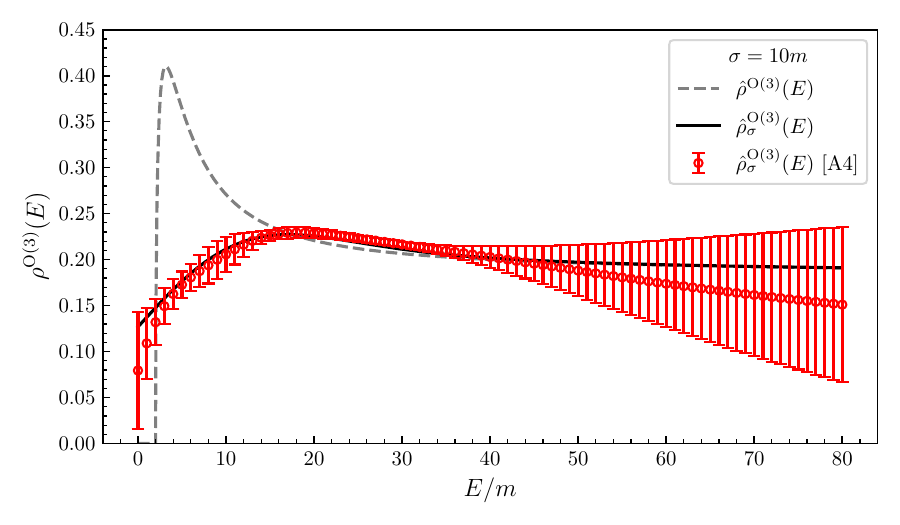}
\includegraphics[width=\linewidth]{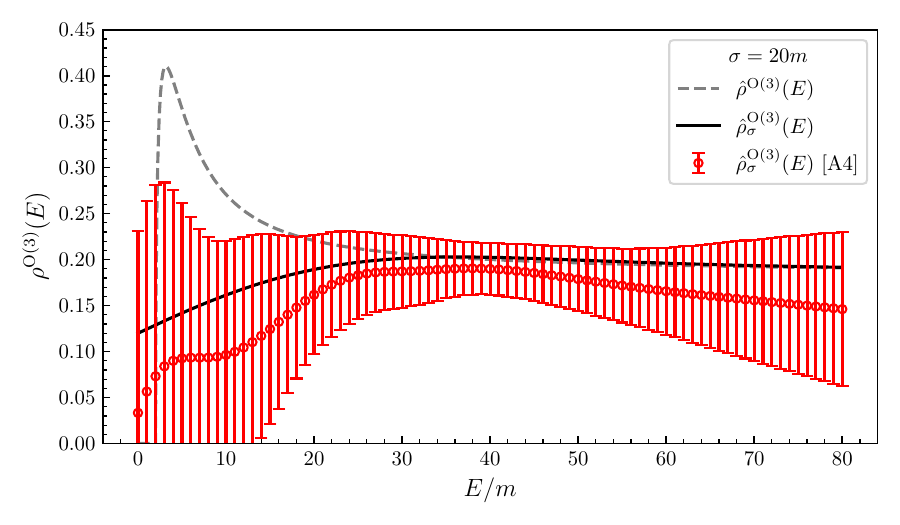}
\caption{\label{fig:out_of_range} Examples of reconstructions (red points) for choices of parameters in and outside the domain $\Omega_{E,\sigma}$. The input correlator is the lattice one measured on the ensemble A4. The exact smeared (black solid) and unsmeared (gray dashed) spectral function of the $O(3)$ model are shown for reference.}
\end{figure}

In general, when training neural networks using data defined only in a restricted space, there is no expectation that it can reliably predict solutions for new data lying outside this space. This is the case in the present work since the training set contains only spectral functions generated within the GP according to \cref{sec::dataset}. Indeed, the validation of the procedure presented in \cref{sec::closure_test} has been carried out by generating new data in the same space of functions as those used to generate the training set. It is interesting, however, to test our procedure in the presence of data that do not belong to the same function space used to generate the training set or for choices of $E$ and $\sigma$ lying outside the domain $\Omega_{E,\sigma}$. To this end, we start generating out-of-distribution spectral functions by enlarging $\sigma_\mathrm{GP}$ by a factor of five with respect to the value used to generate the training set given by $\omegathree$, while still using the two-particle contribution to the O(3) non-linear $\sigma$ model as the mean function. This defines a new probability distribution over the space of functions, admitting functions that differ much more significantly from the mean than those drawn from the GP used to generate the training set, that is the one in the middle panel of \cref{fig:function_space}. To quantify the deviation from the original GP, we compute the average deviation of the unsmeared spectral functions from the original GP mean. Using this procedure, we selected two spectral functions corresponding to two extreme cases: one whose unsmeared spectral function lies, on average, more than 7 standard deviations above the original GP mean, and another whose unsmeared spectral function lies more than 9 standard deviations below it (and which even receives negative contributions). We then generated random noise with $\lambda=1$ and reconstructed the corresponding smeared spectral functions using the ensemble of networks described in the main text. The results are presented in \Cref{fig:out_of_distribution}, where the unsmeared spectral functions (represented by the vertical lines) are also compared with the GP probability distribution used to generate the training set. As can be seen, the reconstructed smeared spectral functions are, in both cases, compatible with the true results within the quoted total uncertainty. This finding is not obvious since, in terms of the original GP, the probability of functions similar to these appearing in the training set is effectively zero. The fact that the exact result is recovered within uncertainties highlights that the neural network is learning properties of the target operator that are, at least partially, independent of the specific training data. While the extent to which our procedure can generalize to spectral functions lying far outside the training distribution is beyond the scope of this work, the investigation presented in this appendix further strengthens the conclusions drawn from \Cref{fig:final_plot} and highlights the remarkable ability of neural networks to generalize to previously unseen data. This, in turn, reinforces their potential as robust tools for performing spectral reconstructions in the presence of unseen data.

As a second scenario, we test the ensemble of neural networks associated with $\omegathree$ and $\lambda=1$ by reconstructing the smeared spectral function from the correlator computed on the ensemble A4, for values of the energy and the smearing parameter outside those used during training. In particular, we reconstruct $\rhohat_\sigma(E)$ for energies in the range $[0,80m]$, which is much larger than the range $[4m,40m]$ used during training and validation. The systematic error estimated from the ensemble of neural networks is included in the total error, and the different results are shown in \cref{fig:out_of_range} for values of $\sigma=0.2m$ (top panel) up to $\sigma=20m$ (bottom panel). For reference, the reconstructions are compared to the exact smeared spectral function for the O(3) model, even though the infinite-volume and continuum limits are not taken here. As can be seen, in all four plots the total error of $\rhohat_\sigma(E)$ grows rapidly as soon as the prediction energy falls outside $\Omega_E$. The same happens when $\sigma$ is much larger than the values contained in $\Omega_\sigma$, as is evident in the last plot of the figure. Unlike the out-of-distribution case, we conclude that reconstruction for out-of-domain parameters compared to those used during training is not recommended, and that, to obtain reliable results with usable errors, the training should be repeated by extending the training set to the region of interest. We leave for a future investigation the reconstruction of the spectral function for energies below the four-particle threshold, which is particularly challenging due to the rapid changes in the underlying spectral function. As noted in Ref.~\cite{O3}, the errors coming from the HLT method do not make the $\sigma \to 0$ extrapolation possible in the low-energy region, while the promising precision achieved by the DeepONet strategy may enable a controlled reconstruction for energies close to the two-particle threshold.

\bibliography{references}

\end{document}